\documentclass[pra,showpacs,groupedaddress,amssymb,twocolumn,longbibliography,nofootinbib]{revtex4-1}
\usepackage{graphicx}
\usepackage{amsmath}
\usepackage{amssymb}
\usepackage[usenames]{color}
\usepackage{hyperref}
\hypersetup{
colorlinks=true,       % false: boxed links; true: colored links
linkcolor=cyan,          % color of internal links
citecolor=magenta,        % color of links to bibliography
filecolor=magenta,      % color of file links
urlcolor=cyan,           % color of external links
runcolor=cyan
}

\usepackage{bm}
\usepackage{threeparttable}
\usepackage{subfigure}
\usepackage{upgreek }
\usepackage{marginnote}
\usepackage{cancel}

\newcommand{\beq}{\begin{equation}}
\newcommand{\eeq}{\end{equation}}
\newcommand{\beqnn}{\begin{equation*}}
\newcommand{\eeqnn}{\end{equation*}}
\newcommand{\bea}{\begin{eqnarray}}
\newcommand{\eea}{\end{eqnarray}}
\newcommand{\beann}{\begin{eqnarray*}}
\newcommand{\eeann}{\end{eqnarray*}}
\newcommand{\bes} {\begin{subequations}}
\newcommand{\ees} {\end{subequations}}

\newcommand{\braket}[2]{\langle #1 | #2\rangle}
\newcommand{\ket}[1]{ | #1\rangle}
\newcommand{\bra}[1]{\langle #1 | }

\newcommand{\ignore}[1]{}

\begin{document}
\title{Role of Non-stoquastic Catalysts in Quantum Adiabatic Optimization}
\author{Tameem Albash}
\affiliation{Information Sciences Institute, University of Southern California, Marina del Rey, California 90292, USA}
\affiliation{Department of Physics and Astronomy, University of Southern California, Los Angeles, California 90089, USA}
\affiliation{Center for Quantum Information Science \& Technology, University of Southern California, Los Angeles, California 90089, USA}

\begin{abstract}
The viability of non-stoquastic catalyst Hamiltonians to deliver consistent quantum speedups in quantum adiabatic optimization remains an open question. The infinite-range ferromagnetic $p$-spin model is a rare example exhibiting an exponential advantage for non-stoquastic catalysts over its stoquastic counterpart. 
We revisit this model and note how the incremental changes in the ground state wavefunction give an indication of how the non-stoquastic catalyst provides an advantage. We then construct two new examples that exhibit an advantage for non-stoquastic catalysts over stoquastic catalysts.  The first is another infinite range model that is only 2-local but also exhibits an exponential advantage, and the second is a geometrically local Ising example that exhibits a growing advantage up to the maximum system size we study.
%using a non-stoquastic catalyst over a stoquastic one
\end{abstract}

\maketitle
%%%%%%%%%%%%%%%%%%%%%%%%%%%%%%%
\section{Introduction}
%{\bf Introduction}.--- 
Optimization problems appear across a wide range of disciplines, and the development of new algorithms to tackle them is an active area of research.  Quantum algorithms hope to offer advantages over classical algorithms by capitalizing on non-classical phenomena to reach the desired solution faster. One such example is the Quantum Adiabatic Optimization (QAO) algorithm \cite{finnila_quantum_1994,Brooke1999,kadowaki_quantum_1998,farhi_quantum_2000,Santoro}, a heuristic quantum algorithm based on the adiabatic paradigm of quantum computation \cite{aharonov_adiabatic_2007}.  The algorithm uses a time-dependent Hamiltonian that interpolates from a Hamiltonian $H_{\mathrm{initial}}$ with an easily prepared ground state to a Hamiltonian $H_{\mathrm{final}}$, whose ground state encodes the solution to the optimization problem. By initializing the system in the ground state of $H_{\mathrm{initial}}$ and using an interpolation time $t_f$ that satisfies the adiabatic condition, the algorithm is guaranteed by the adiabatic theorem \cite{Kato:50,Teufel:book,Nenciu:93,Avron:99,Hagedorn:2002kx,Reichardt:2004,Ambainis:04,jansen:07,lidar:102106,PhysRevA.77.042319,Cheung:2011aa,Elgart:2012fk,Ge:2015wo} to reach the ground state of $H_{\mathrm{final}}$ with high probability.  Thus, the scaling with system size of the algorithm's runtime $t_f$ is given in terms of the adiabatic condition, which in turn is given in terms of an inverse power of the minimum ground state energy gap $\Delta_{\mathrm{min}}$ along the interpolation path \cite{jansen:07}.
%in turn is given in terms of is given in terms of the scaling of the minimum ground state energy gap along the interpolation path.

The standard implementation of the QAO algorithm implements a time-dependent stoquastic Hamiltonian \cite{Bravyi:QIC08,Bravyi:2009sp}.  The partition function associated with a stoquastic Hamiltonian can always be decomposed into a sum of positive weights that can be used in a Markov process, and quantum Monte Carlo algorithms can be used to try to emulate the QAO algorithm.  While no proof exists (for counter-examples, see Refs.~\cite{Hastings:2013kk,Jarret:16}), this is often cited as a strong indication that stoquastic Hamiltonians may not be sufficiently rich to generate quantum speedups over classical algorithms for generic optimization problems. 

%% One approach to circumvent this problem has been to use different driver Hamiltonians \cite{} or the introduction of richer interactions using `catalyst' Hamiltonians \cite{}, with several studies using the latter on random Ising optimization problems \cite{crosson2014different,PhysRevB.95.184416}, using both stoquastic and non-stoquastic interactions.
The introduction of more exotic interactions via intermediate `catalyst' Hamiltonians \cite{FarhiAQC:02} is one way to enrich the QAO algorithm. (We note that another approach is to use different initial `driver' Hamiltonians \cite{2002quant.ph..8189B,PhysRevA.93.062312,PhysRevApplied.5.034007}). The use of catalyst Hamiltonians has been pursued in several studies on random Ising optimization problems \cite{crosson2014different,PhysRevB.95.184416}, using both stoquastic and non-stoquastic interactions. While the latter are known to be necessary to perform universal adiabatic quantum computation \cite{aharonov_adiabatic_2007}, these studies have shown that \emph{typically} stoquastic catalysts outperform non-stoquastic ones, i.e. stoquastic catalysts tend to make the minimum gap along the evolution larger than non-stoquastic catalysts. 

 However, rare cases exist where the non-stoquastic catalyst can raise the gap more than the stoquastic catalyst.  Perhaps the most striking example of this is the case of the infinite-range ferromagnetic $p$-spin models \cite{Seki:2012,Seoane:2012uq,10.3389/fict.2017.00002,PhysRevA.95.042321}, where it was observed that for certain parameter choices the non-stoquastic catalyst can change a first order phase transition to a second order one for odd $p>3$.  A general mechanism for this enhancement was provided in Ref.~\cite{2018arXiv180607602D} in terms of a quantum Rayleigh limit, whereby the ground state profile coalesces from bimodal to unimodal.  This beyond mean-field treatment also demonstrated the enhancement for the $p=3$ case.

In this work, we revisit the infinite-range ferromagnetic $p$-spin models and study them at finite system size.  By studying the behavior of the ground state wavefunction along the interpolation, we are able to gain a new understanding of why the case with a non-stoquastic catalyst exhibits such a stark difference relative to its stoquastic counterpart.  We then construct two new examples that also exhibit an advantage for non-stoquastic catalysts over their stoquastic counterparts.  The first example is based on the prototypical large-spin tunneling problem and corresponds to an infinite-range 2-local model.  It exhibits a similar exponential advantage for the non-stoquastic catalyst over its stoquastic counterpart as for the $p$-spin model. The second example is a geometrically local Ising example that exhibits a growing advantage for the non-stoquastic catalyst over its stoquastic counterpart, at least up to the maximum size of 24 qubits that we study.
%
%%%%%%%%%%%%%%%%%%%%%%%%%%%%%%%%%%
\section{Infinite-range ferromagnetic $p$-spin models} \label{Sec:pSpin}
%{\bf Infinite-range ferromagnetic $p$-spin models}.---
In order to solve the infinite-range ferromagnetic $p$-spin model with QAO, we take an interpolating Hamiltonian acting on $n$ qubits of the form \cite{Seki:2012,crosson2014different}:
\begin{eqnarray} \label{eqt:pSpinH}
H_\lambda(s) &=& -(1-s) \sum_{i=1}^n \sigma_i^x - \frac{s}{n^{p-1}} \left(\sum_{i=1}^n \sigma_i^z \right)^p \nonumber \\
&& + \lambda \frac{s (1-s)}{n} \left( \sum_{i=1}^n \sigma_i^x \right)^2 \ ,
\end{eqnarray}
where $s \in [0,1]$ is our dimensionless interpolating parameter and $\sigma_i^{x},\sigma_i^z$ are the Pauli $x$ and $z$ operators acting on the $i$-th qubit. We take the parameter $\lambda$ to be constant during the evolution, but its value can be optimized at each size in order to maximize the minimum gap encountered during the interpolation.  This optimized choice of $\lambda$, which we denote by $\lambda_{\mathrm{opt}}$ (we suppress the $n$ dependence) defines an optimal interpolating protocol. This is analogous to the optimal path defined in Ref.~\cite{2018arXiv180607602D}. The first term in Eq.~\eqref{eqt:pSpinH} is the `driver'  transverse field Hamiltonian, which is the only term that is non-zero at $s  = 0$.  The second term is the infinite-range ferromagnetic $p$-spin model Hamiltonian, which represents the optimization problem that we wish to solve and is the only term that is non-zero at $s = 1$.  The last term is the catalyst Hamiltonian, which is non-zero for $s \neq 0$ and $\lambda \neq 0$.  For $\lambda \leq 0$, all off-diagonal terms in the matrix representation of $H_\lambda(s)$ in the computational basis\footnote{The computational basis is given by the single qubit basis states \protect{$\left\{\ket{0}, \ket{1} \right\}$} satisfying \protect{$\sigma^z \ket{0} = \ket{0}$} and \protect{$\sigma^z \ket{1} = - \ket{1}$}.} are negative, and we say that the Hamiltonian is stoquastic~\cite{Bravyi:QIC08,Bravyi:2009sp}.  For $\lambda > 0$, the catalyst Hamiltonian introduces positive off-diagonal elements in the matrix representation of $H_\lambda(s)$, which suggests that $H_\lambda(s)$ is non-stoquastic. However, there is a simply single-qubit transformation that makes the Hamiltonian stoquastic\footnote{Applying a Hadamard transformation, which rotates the computational basis from \protect{$\ket{0},\ket{1}$} to \protect{$\ket{+},\ket{-}$}, takes \protect{$\sigma^x_i \to \tau^z_i, \sigma^z_i \to \tau^x_i$}, and the resulting Hamiltonian is stoquastic in the new basis.}.  Nevertheless, while not strictly non-stoquastic, the ground state of $H_{\lambda}(s)$ with $\lambda > 0$ in the computational basis can have both positive and negative amplitudes.

The Hamiltonian in Eq.~\eqref{eqt:pSpinH} enjoys several symmetries.  First, it is invariant under the permutation of any group of qubits.  Because the ground state of $H_\lambda(0)$ is the uniform superposition state, the QAO algorithm starts within the subspace spanned by the completely symmetric states, which we denote by $\mathcal{S}$, and the evolution under $H_{\lambda}(s)$ cannot take the state out of this subspace.  A convenient basis for this subspace is given by the Dicke states \cite{Dicke:54}, which we denote by $\ket{S,M}$ with $S = n/2$ and $M = -n/2, -n/2+1, \dots, n/2$, such that $\frac{1}{2} \sum_{i=1}^n \sigma_i^z \ket{S,M} = M \ket{S,M}$.  Second, for $p$ even
%~\footnote{While the case of $p$ even was not explicitly studied in Ref.~\cite{Seki:2012}, it is a convenient starting point for our analysis.}
the Hamiltonian is also invariant under the transformation by $P = \prod_{i=1}^n \sigma_i^x$, and the ground state of $H_\lambda(0)$ has eigenvalue $1$ under this operator.  The evolution is then restricted to the subspace $\mathcal{S}'$ spanned by the linear combination of completely symmetric states with eigenvalue $1$ under $P$.  The runtime scaling for the QAO algorithm is thus given by the minimum ground state energy gap $\Delta_{\mathrm{min}}'$ in the $\mathcal{S}'$ subspace.

\begin{figure}[htbp]
   \centering
   \subfigure[]{\includegraphics[width=0.48\columnwidth]{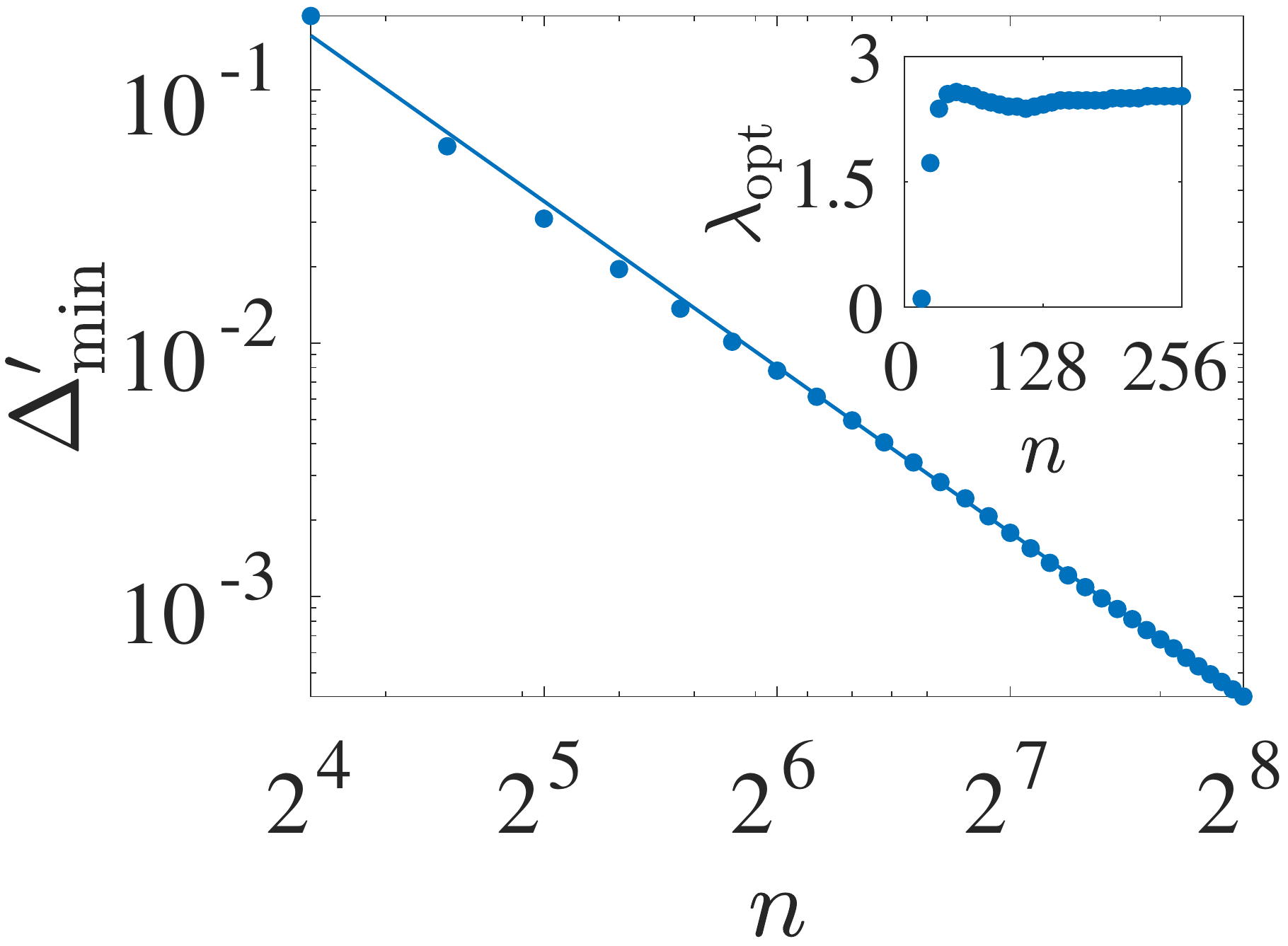} }
    \subfigure[]{\includegraphics[width=0.48\columnwidth]{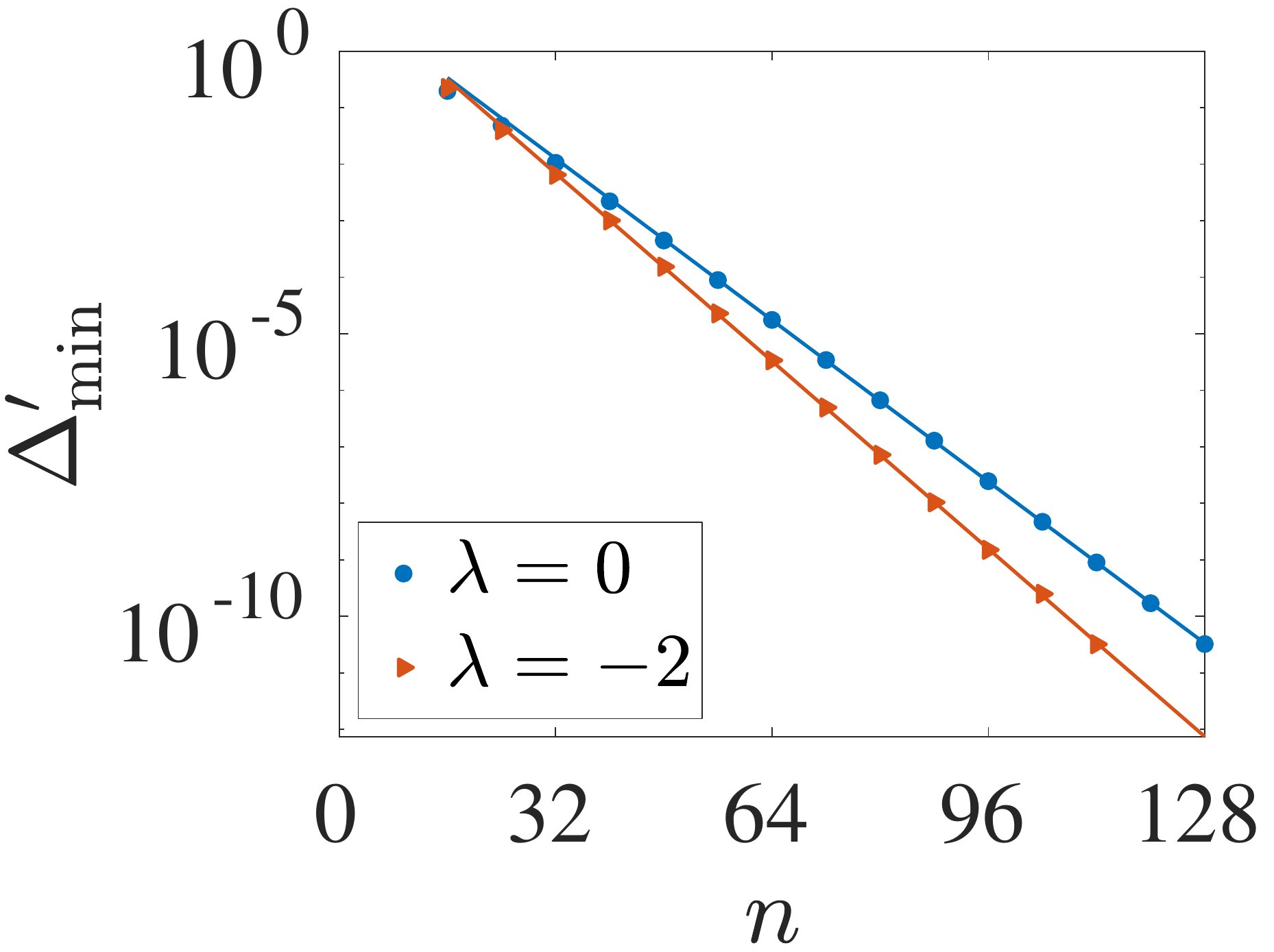} }
   \caption{Scaling of the minimum gap for the infinite-range ferromagnetic $(p = 6)$-spin model with (a) a non-stoquastic catalyst with optimized $\lambda$, (b) a stoquastic catalyst with $\lambda = 0$ and $\lambda = -2$.  In (a), the solid line corresponds to a fit of $\sim~n^{-2.17}$. The inset shows the optimized values for $\lambda$ used.  The error bars, which are not visible because they are on the size of the data points, correspond to our uncertainty in the exact optimum value of $\lambda$. In (b), the solid line  corresponds to a  fit of $ \sim~\exp (-0.21n)$ and $\sim~\exp(-0.24n)$ respectively.}
   \label{fig:MF_GapScaling}
\end{figure}

%For simplicity, we consider an interpolating path $H_\alpha(0,0) \rightarrow H_\alpha(0, \lambda_\ast) \rightarrow H_\alpha(1,\lambda_\ast) \rightarrow H_\alpha(1,1)$, and we begin with the case of $p$ even.  
We show in Fig.~\ref{fig:MF_GapScaling} results for $p=6$, where in the case of the non-stoquastic Hamiltonian ($\lambda >0$ in Eq.~\eqref{eqt:pSpinH}) the scaling behavior of $\Delta_{\mathrm{min}}'$ can be polynomial if $\lambda$ is chosen to be sufficiently large,
% or exponential if $\lambda_\ast$ is too large,
 in agreement with the conclusions of Ref.~\cite{Seki:2012}.  This is to be contrasted with the stoquastic case ($\lambda \leq 0$) where the gap scaling is always exponential.  (We find that $\lambda = 0$ maximizes the minimum gap for the stoquastic case.)

\begin{figure}[htbp]
   \centering
      \subfigure[]{\includegraphics[width=0.48\columnwidth]{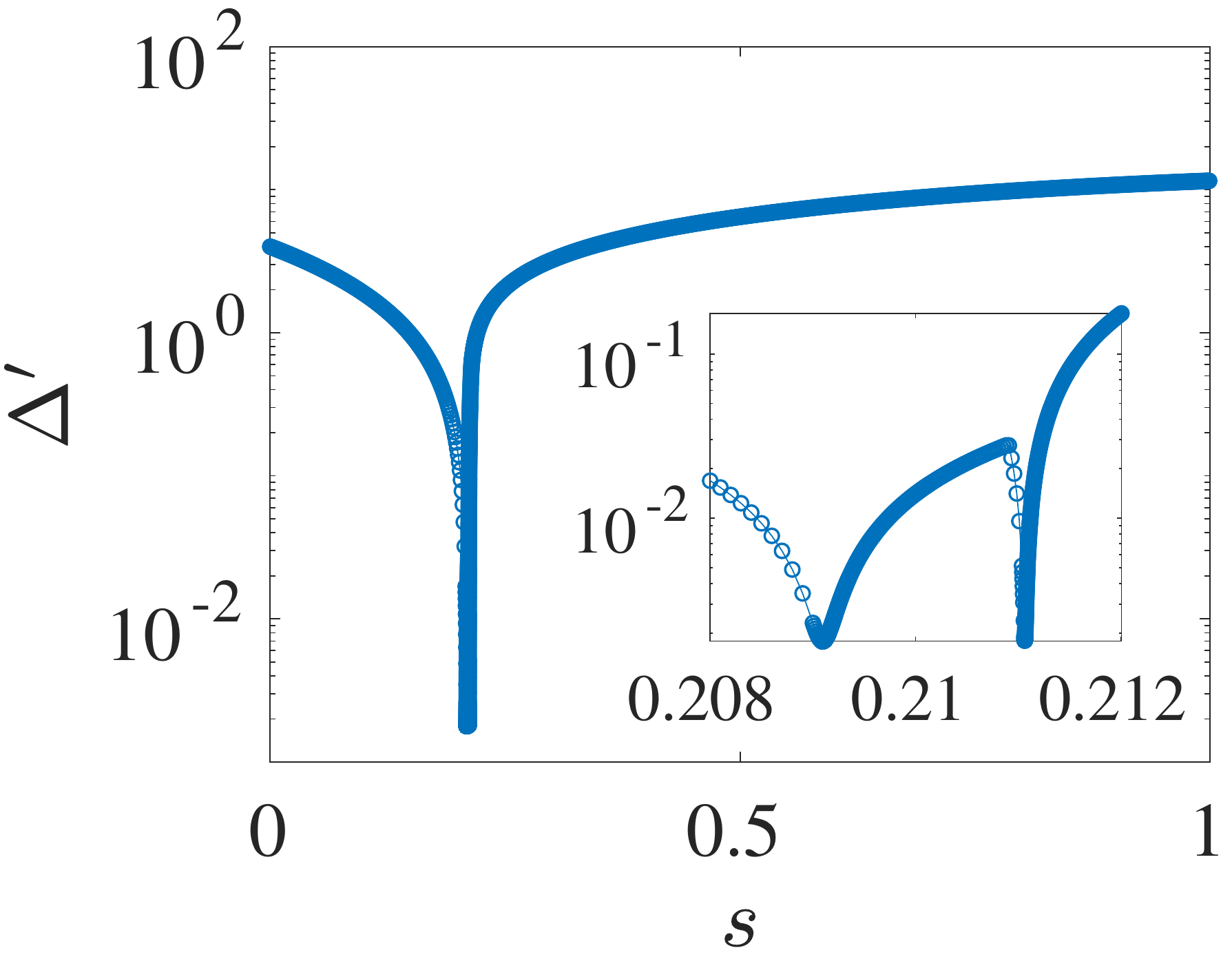} \label{fig:MF_GS1A}}
            \subfigure[]{\includegraphics[width=0.46\columnwidth]{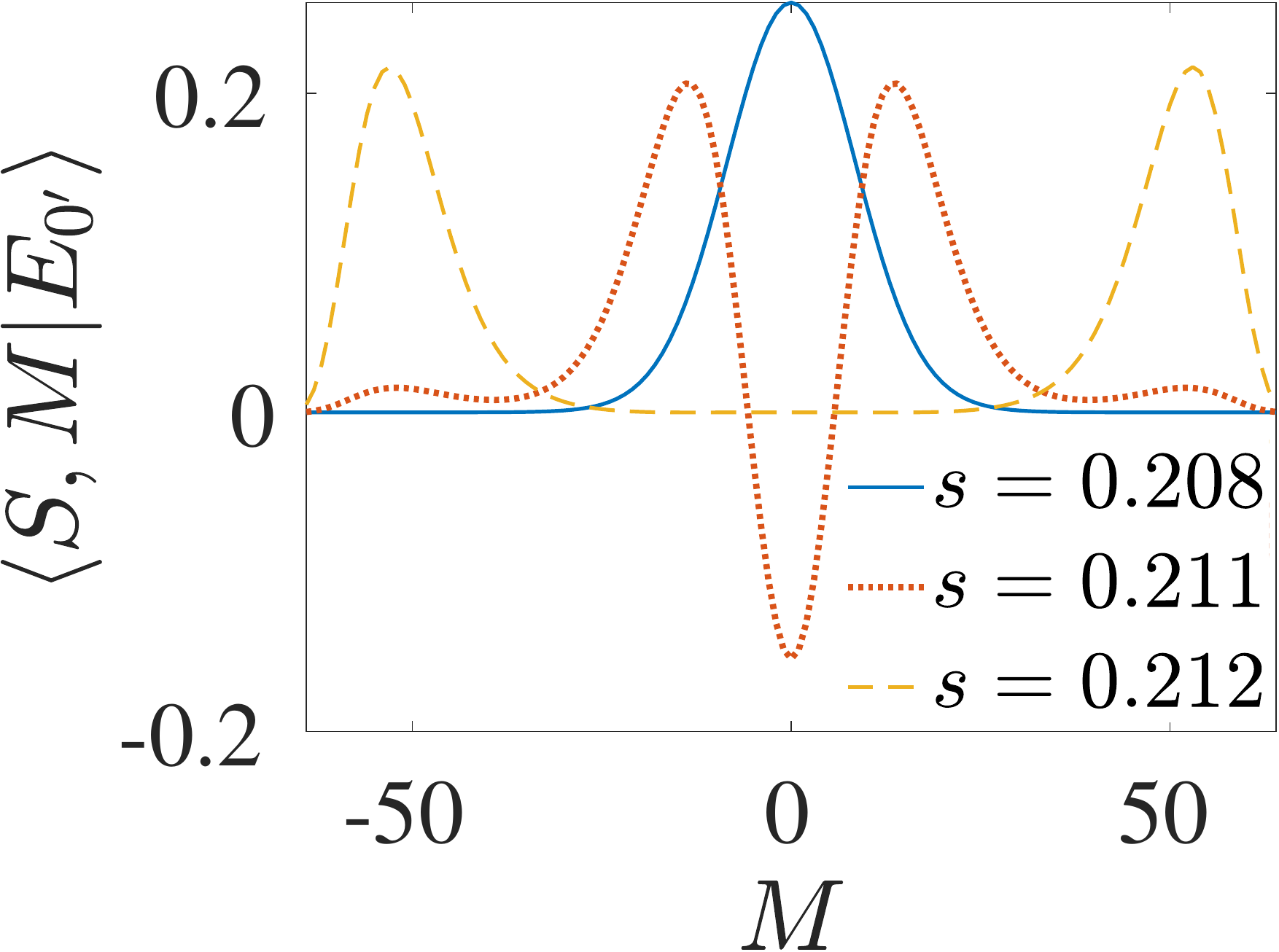} \label{fig:MF_GS1B}}
               \caption{(a) Energy gap $\Delta'$ between the ground state and first excited state within the subspace $\mathcal{S'}$ for the infinite-range ferromagnetic $(p = 6)$-spin model. (b) The ground state wavefunction $\ket{E_{0'}(s)}$ within the subspace $\mathcal{S'}$ in the Dicke basis for the same model.  Results shown are for $n=128$, and $\lambda= 2.425$.}
   \label{fig:MF_GS1}
\end{figure}

\begin{figure}[htbp]
   \centering
      \subfigure[]{\includegraphics[width=0.48\columnwidth]{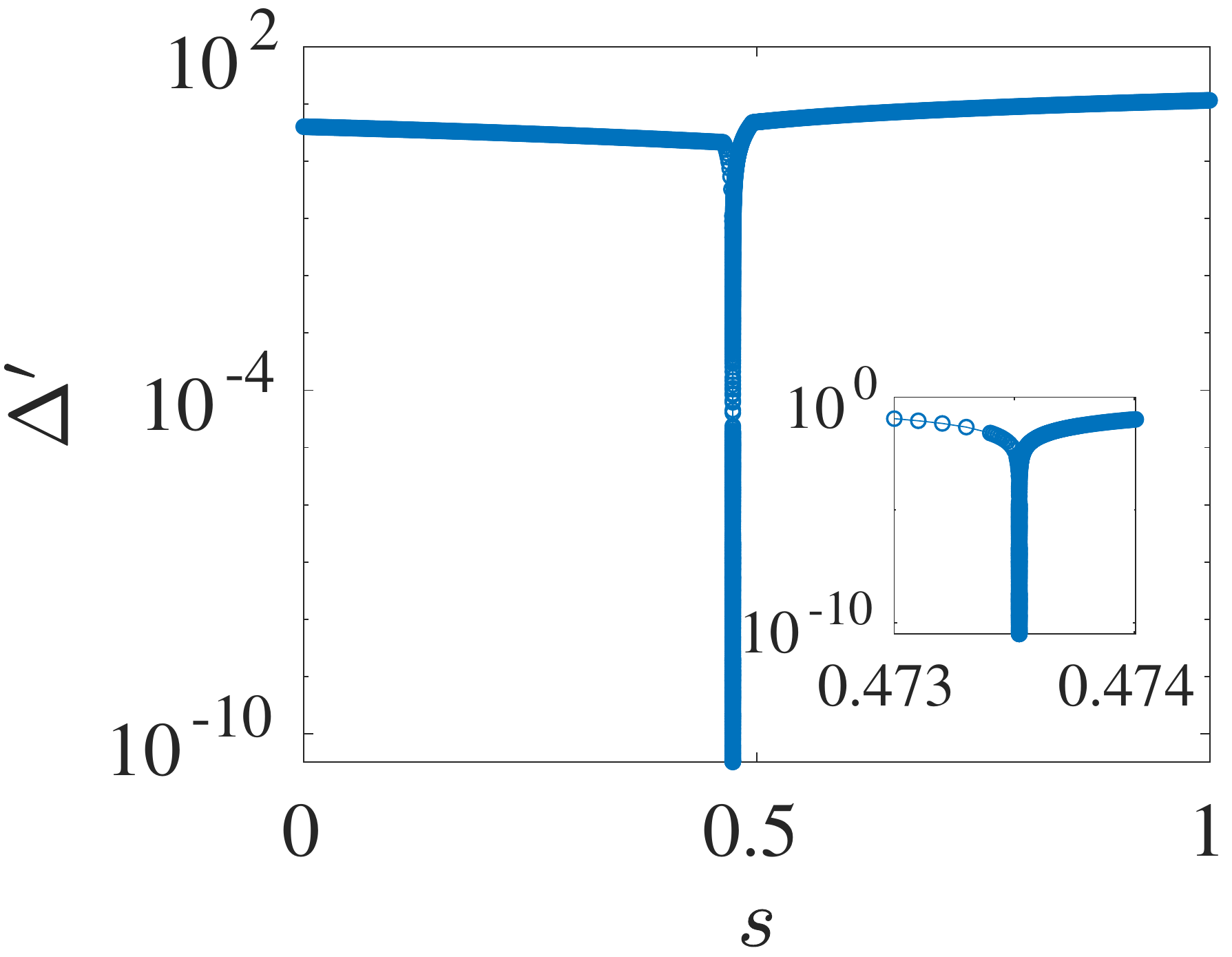} \label{fig:MF_GS2A}}
            \subfigure[]{\includegraphics[width=0.46\columnwidth]{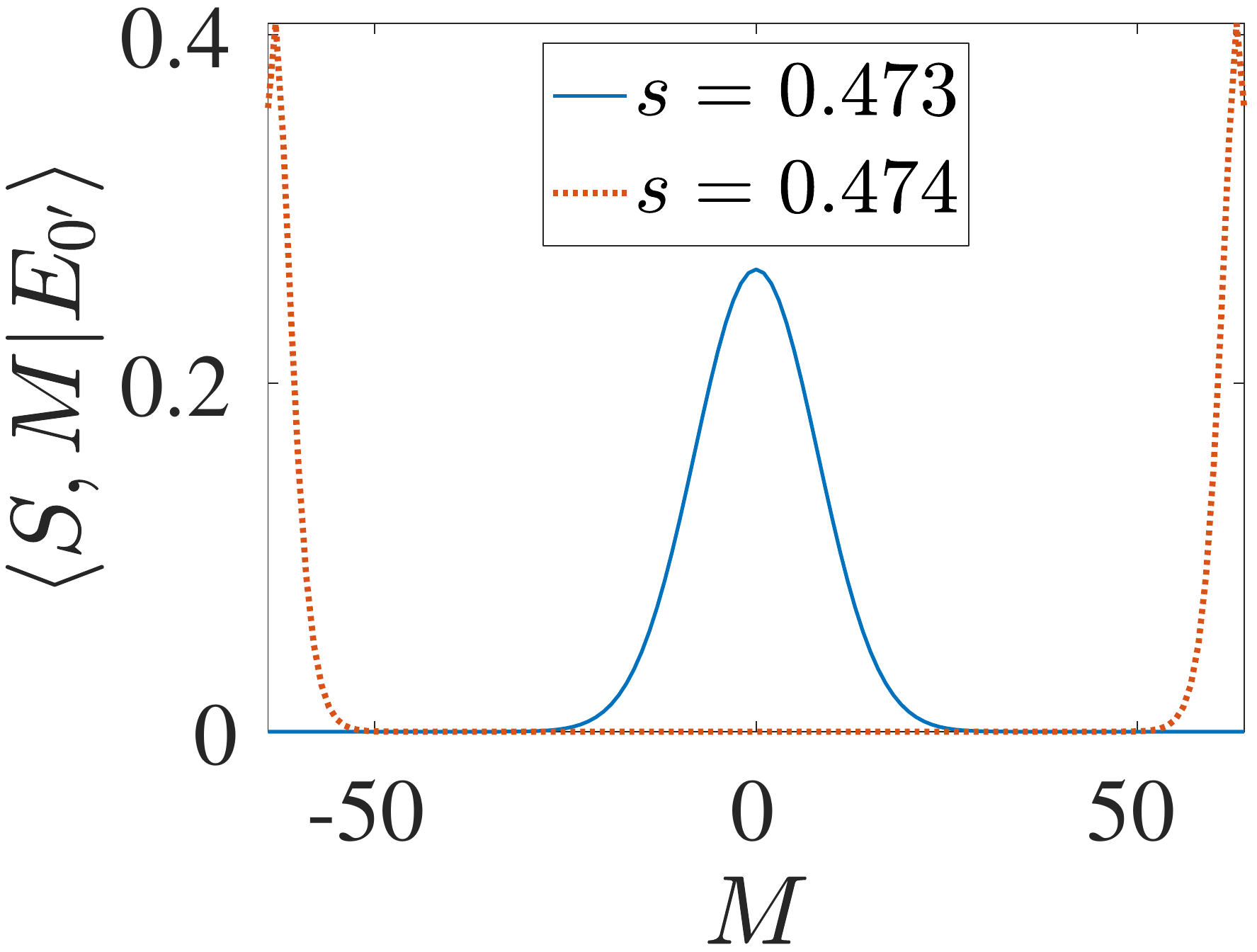} \label{fig:MF_GS2B}}
               \caption{(a) Energy gap $\Delta'$ between the ground state and first excited state within the subspace $\mathcal{S'}$ for the infinite-range ferromagnetic $(p = 6)$-spin model. (b) The ground state wavefunction $\ket{E_{0'}(s)}$ within the subspace $\mathcal{S'}$ in the Dicke basis for the same model.  Results shown are for $n=128$, and $\lambda= 0$.}
   \label{fig:MF_GS2}
\end{figure}

In order to understand this dramatic difference in scaling behavior, it is useful to study how the ground state gap and wavefunction evolve during the interpolation in the $\mathcal{S}'$ subspace.  In Fig.~\ref{fig:MF_GS1A}, we observe the appearance of multiple local minima in the gap along the interpolation.  More local minima appear in the case of unoptimized $\lambda$, as we show in Appendix \ref{App:Unoptimized}. The local minima are associated with the addition of pairs of nodes to the ground state wavefunction as we show in Fig.~\ref{fig:MF_GS1B}. Nodes must be added in pairs because the energy eigenstates within the subspace $\mathcal{S}'$ must remain symmetric about $M = 0$ for $p$ even.  For example, at the first local minimum, the ground state changes from having zero nodes to having two nodes, and at the second local minimum, the ground state changes from having two nodes to four nodes (the extra nodes around $M = 0$ are not visible in Fig.~\ref{fig:MF_GS1B}).  As the interpolation continues towards $s = 1$, additional pairs of nodes are added to the wavefunction as it approaches the ground state of the $p$-spin Hamiltonian, which for $p$ even corresponds to peaks at $M = -n/2$ and $M = n/2$.   These multiple but incremental changes in the ground state are to be contrasted to what happens when the gap closes exponentially:  for sufficiently small $\lambda$ in the non-stoquastic case or generally in the stoquastic case, the ground state changes dramatically as it crosses the unique minimum gap (Fig.~\ref{fig:MF_GS2}).

%As shown in Fig.~\ref{fig:MF_GS1A}, the ground state of $\mathcal{S}'$ deviates from that of $\mathcal{S}$ multiple times as a function of $s$; the number of times grows linearly with $n$.  These deviations are associated with energy level crossings in the $\mathcal{S}$ subspace, whereby a $P=-1$ state becomes lower in energy than the current $P=+1$ ground state.  For example, at the first deviation, the ground state in the subspace $\mathcal{S}$ changes from having zero nodes ($P=+1$) to having a single node ($P=-1$) (Fig.~\ref{fig:MF_GS1B}).  The deviation vanishes when the ground states of $\mathcal{S}$ and $\mathcal{S'}$ merge again when the two-node solution ($P=+1$) becomes energetically favored over the one-node solution. Each subsequent deviation in the ground state of the two subspaces occurs when the addition of a single node results in an odd number of nodes in the ground state wavefunction.
%%
%
% The transition from zero nodes to two-nodes occurs in the $\mathcal{S'}$ subspace and is associated with a local minimum in the ground state gap in that subspace (Fig.~\ref{fig:MF_GS2}).  Each local minimum in the gap is associated with an addition of two nodes to the ground state of the $\mathcal{S'}$ subspace.  
 
% 

\begin{figure}[htbp]
   \centering
      \subfigure[]{\includegraphics[width=0.48\columnwidth]{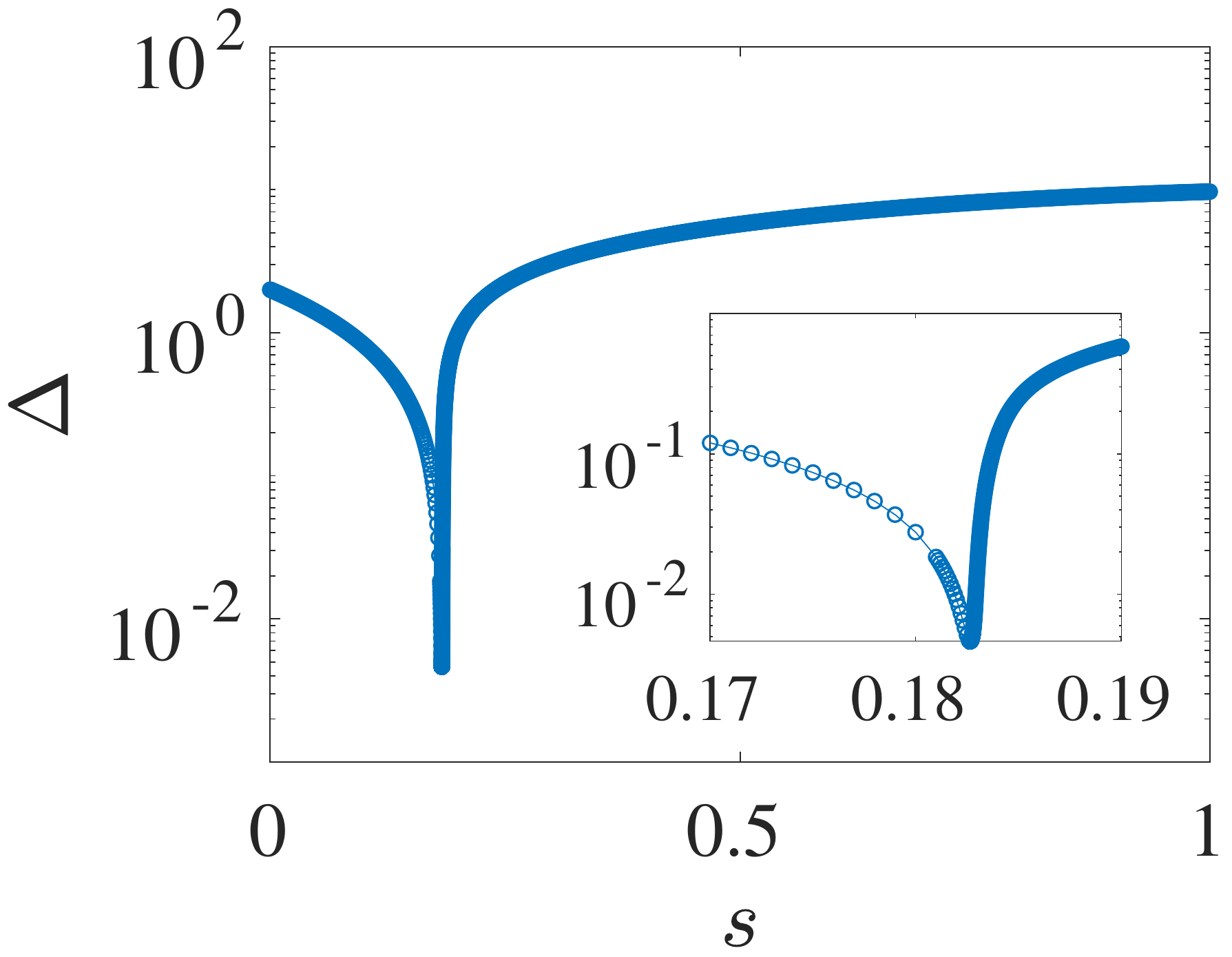} \label{fig:MF_GS3A}}
            \subfigure[]{\includegraphics[width=0.46\columnwidth]{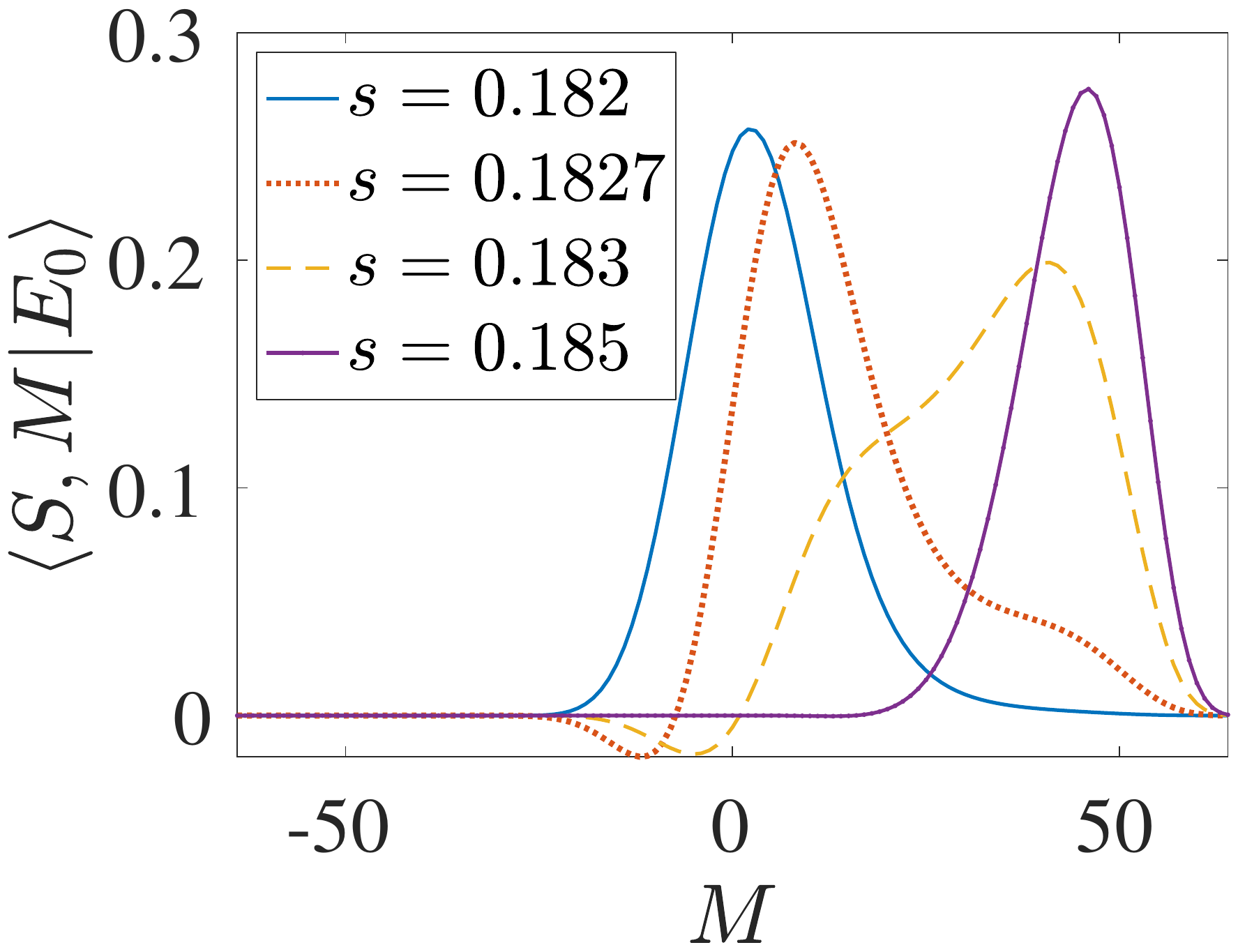} \label{fig:MF_GS3B}}
               \caption{(a) Energy gap $\Delta$ between the ground state and first excited state within the subspace $\mathcal{S}$. (b) The ground state wavefunction $\ket{E_{0}(s)}$ within the subspace $\mathcal{S}$ evaluated in the Dicke basis.  Results are for $p=5$, $n=128$, and $\lambda= 2.75$.}
   \label{fig:MF_GS3}
\end{figure}

In the case of $p$ odd, the Hamiltonian (Eq.~\eqref{eqt:pSpinH}) is not invariant under $P$, and the evolution is restricted to the $\mathcal{S}$ subspace.  This changes the qualitative behavior of the ground state wavefunction along the interpolation in that the local minima in the gap are associated with the addition of a single node as opposed to a pair of nodes, as we show in Fig.~\ref{fig:MF_GS3}.  Similarly to the $p$ even case, multiple local minima are evident in the case of unoptimized $\lambda$, as we show in Appendix \ref{App:Unoptimized}. As the interpolation continues towards $s = 1$, additional single nodes are added to the wavefunction as it approaches the ground state of the $p$-spin Hamiltonian, which for $p$ odd corresponds to a peak at $M = n/2$.  We provide additional comparisons between the even and odd $p$ cases in Appendix \ref{App:evenVSodd}.

\section{Infinite-range 2-local large-spin tunneling example}
%%%%%%%%%%%%%%%%%%%%%%%%%%%%
%
We now present a new example that exhibits a similar exponential advantage for non-stoquastic catalysts over stoquastic catalysts but where the interactions are only 2-local.  We consider the following interpolating Hamiltonian for the prototypical large-spin tunneling problem \cite{Farhi:02,oneOverF2}
\begin{eqnarray} \label{eqt:H2}
H_\lambda(s) &=& -2 (1-s) \left(S_1^x + S_2^x \right)  - s \left(  2 h_1 S_1^z - 2 h_2 S_2^z  \right. \nonumber \\
&& \left. + \frac{4}{n} \left((S_1^z)^2 +(S_2^z)^2 + S_1^z S_2^z\right) \right) \nonumber \\
&& + \frac{8\lambda  s (1-s)}{n}  S_1^x S_2^x \ , 
\end{eqnarray}
where $S_1^\alpha = \frac{1}{2} \sum_{i=1}^{n/2} \sigma_i^\alpha$, $S_2^\alpha = \frac{1}{2} \sum_{i=n/2 + 1}^n \sigma_i^\alpha$.  For simplicity, we restrict to the case where $n/2$ is an integer.  We take $h_1 = 1$ and $h_2 = 0.49$, such that the ground state has eigenvalues  $(+1,+1)$ under $\frac{2}{n} S_1^z$ and $\frac{2}{n} S_2^z$ respectively and the first excited state has eigenvalues $(+1,-1)$ under the same operators.  Unlike the $p$-spin model in Section \ref{Sec:pSpin}, there is no single-qubit transformation that makes the Hamiltonian stoquastic. 

The Hamiltonian is invariant under permutations of the two sets of qubits $\left\{k\right\}_{k=1}^{n/2}$ and $\left\{k\right\}_{k=n/2+1}^{n}$. Since the ground state at $s=0$ is symmetric under both permutations, the evolution under the Hamiltonian is restricted to a subspace $\tilde{\mathcal{S}}$ that is spanned by the product of Dicke states with total angular momentum $S_1 =  S_2 = n /4$, $\ket{S_1,M_1} \otimes \ket{S_2,M_2}$, which for brevity we denote by $\ket{M_1,M_2}$.

\begin{figure}[htbp]
   \centering
      \subfigure[]{\includegraphics[width=0.48\columnwidth]{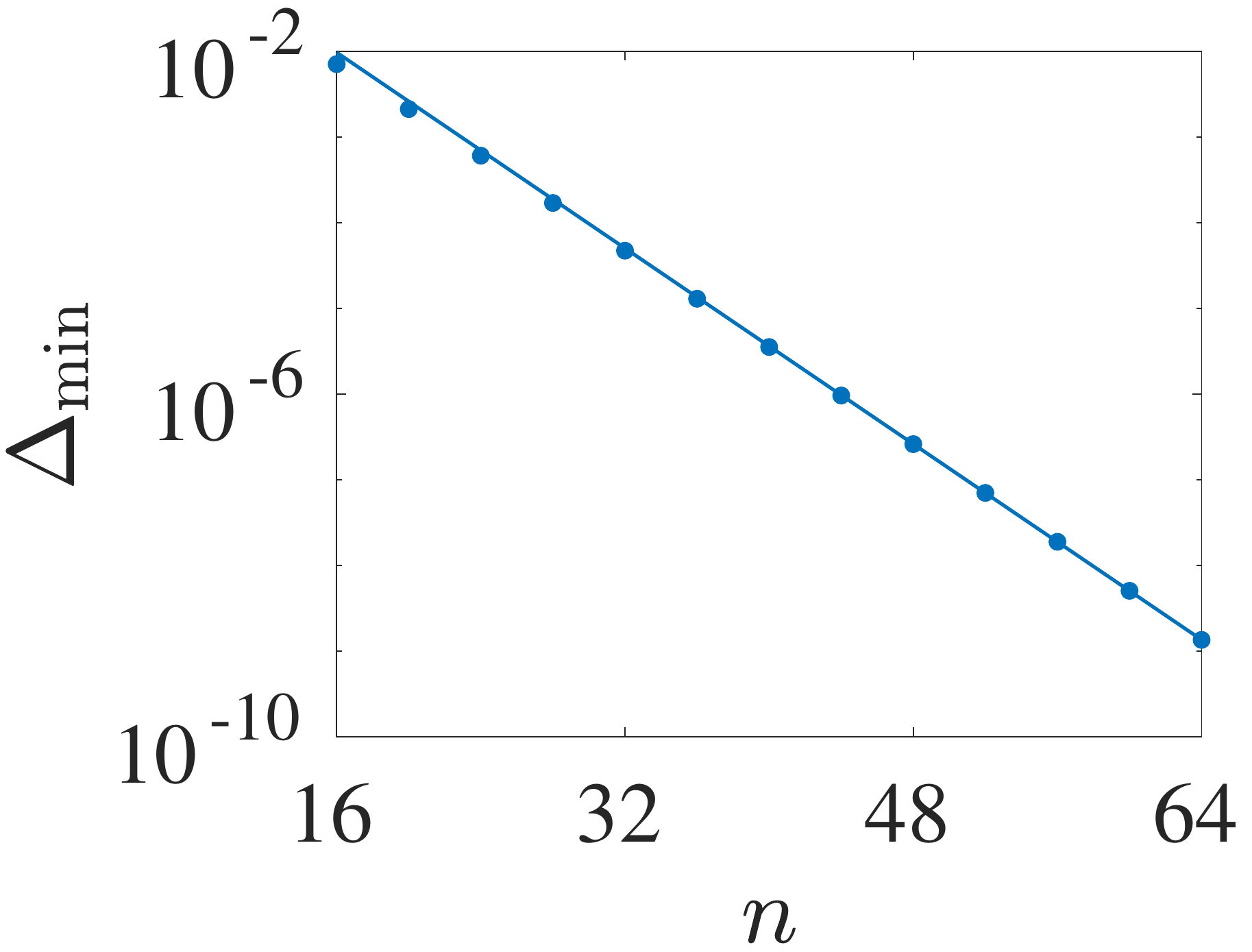} \label{fig:MF_GS4A}}
            \subfigure[]{\includegraphics[width=0.46\columnwidth]{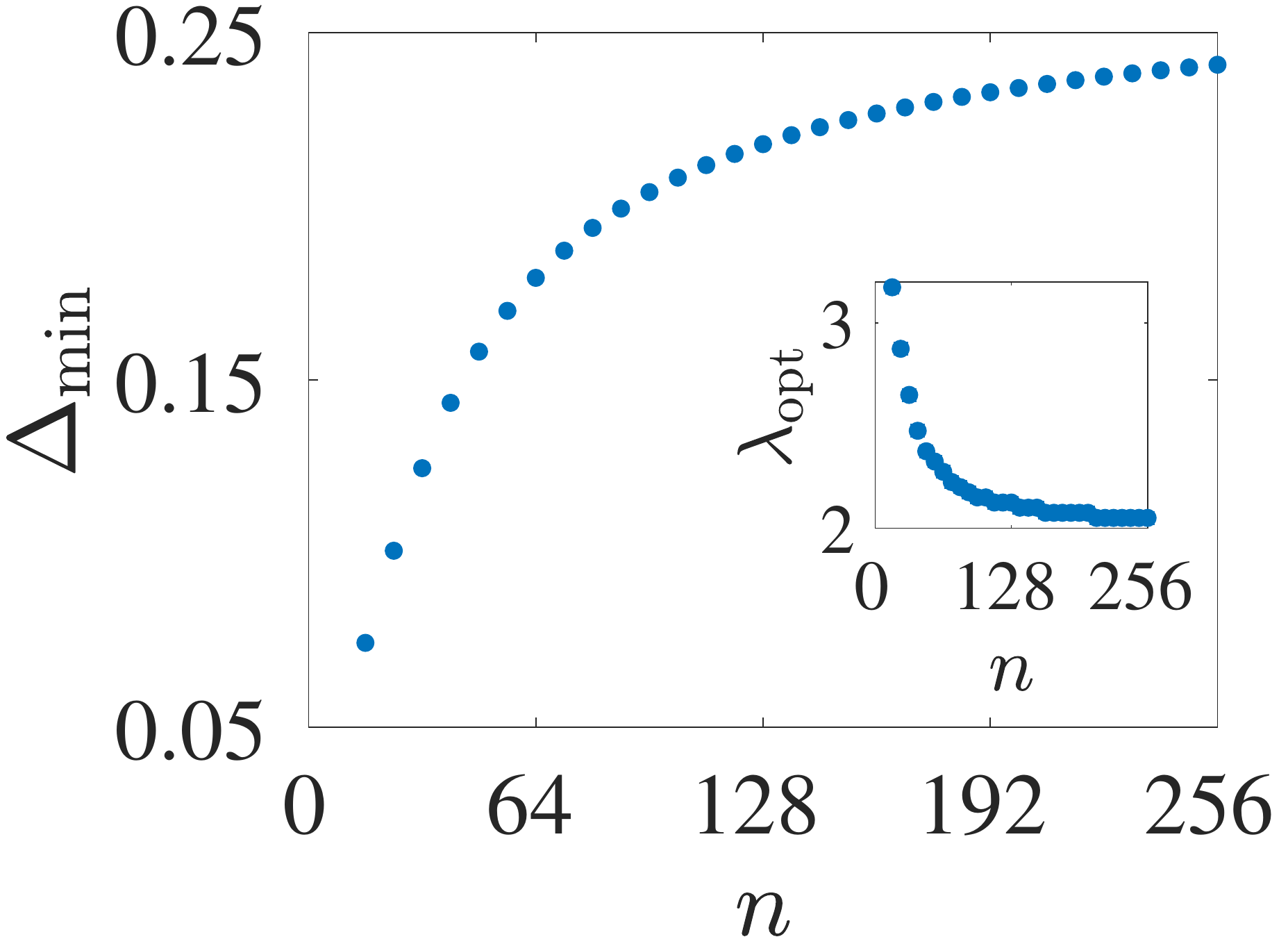} \label{fig:MF_GS4B}}
               \caption{Scaling of the minimum gap within the subspace $\mathcal{S}$ for (a) $\lambda = 0$ and (b) the optimized $\lambda$ for the infinite-range 2-local large-spin tunneling problem.  In (a), the solid line is the fit to $\sim~\exp(-0.33 n)$.  In (b), the inset shows the optimized values for $\lambda$ used.  The error bars, which are not visible because they are on the size of the data points, correspond to our uncertainty in the exact optimum value of $\lambda$.}
   \label{fig:MF_GS4}
\end{figure}

For $\lambda = 0$, the spectrum exhibits an exponentially closing minimum gap (shown in Fig.~\ref{fig:MF_GS4A}) with system size $n$ associated with the tunneling of approximately $n/2$ spins \cite{oneOverF2}.  In contrast, using an optimized $\lambda$ value the minimum gap approaches a constant with increasing $n$ as shown in Fig.~\ref{fig:MF_GS4B}.

\begin{figure}[htbp]
   \centering
      \subfigure[]{\includegraphics[width=0.48\columnwidth]{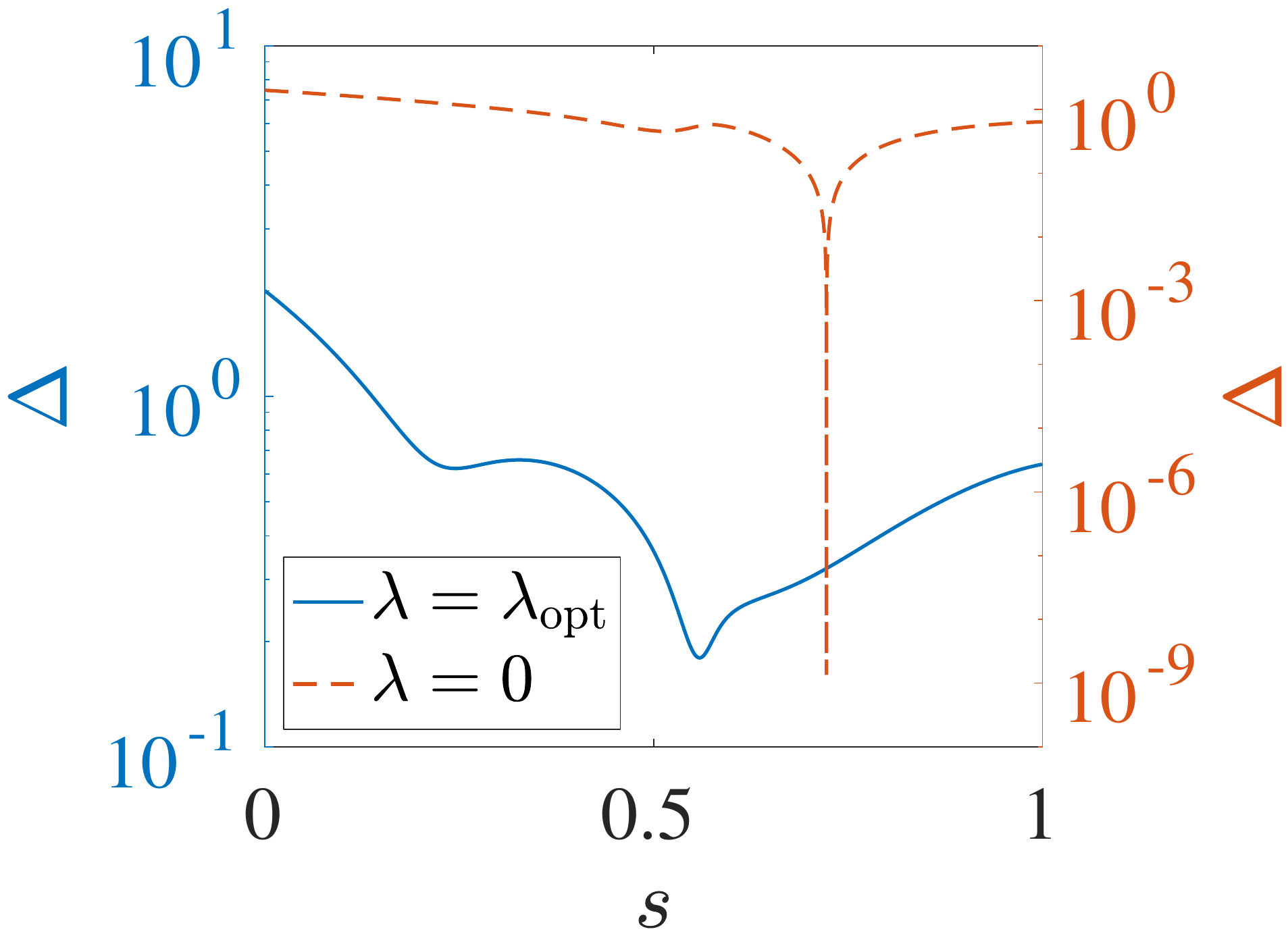} \label{fig:MF_GS5A}}
            \subfigure[]{\includegraphics[width=0.45\columnwidth]{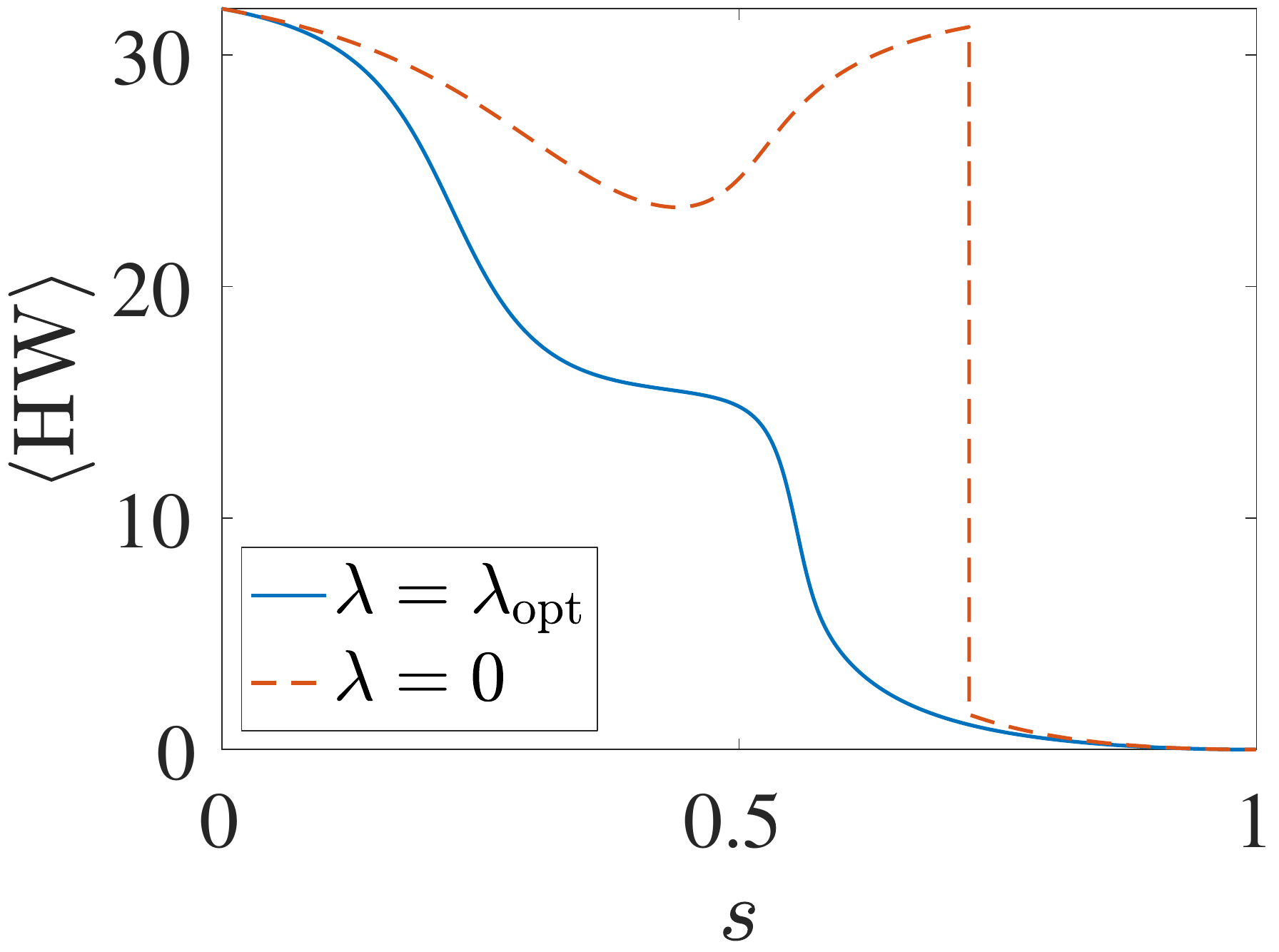} \label{fig:MF_GS5B}}
               \caption{(a) Ground state gap within the subspace $\mathcal{S}$ along the interpolation schedule and (b) the ground state expectation value of the Hamming weight operator for the infinite-range 2-local large-spin tunneling problem.  Results shown are for $n=64$ with  $\lambda = 0$ and $\lambda_{\mathrm{opt}} = 2.275$.}
   \label{fig:MF_GS5}
\end{figure}

In order to better understand why the non-stoquastic catalyst with an optimized $\lambda$ helps avoid the exponentially closing gap, we consider again the behavior of the gap and ground state along the interpolation, as shown in Fig.~\ref{fig:MF_GS5}.  In the stoquastic case, the exponentially closing gap is associated with a sharp change in the ground state expectation value of the Hamming weight operator, $\langle \mathrm{HW} \rangle= \frac{1}{2} \sum_{i=1}^n \left( 1  - \bra{E_0} \sigma_i^z \ket{E_0} \right)$. In contrast to the stoquastic case, the value of $\langle \mathrm{HW} \rangle$ decreases monotonically and gradually along the interpolation schedule when using the optimized non-stoquastic catalyst. We therefore observe a similar incremental change to the ground state wavefunction in the subspace of the evolution as in the $p$-spin model.

\begin{figure}[htbp]
   \centering
         \subfigure[]{\includegraphics[width=0.48\columnwidth]{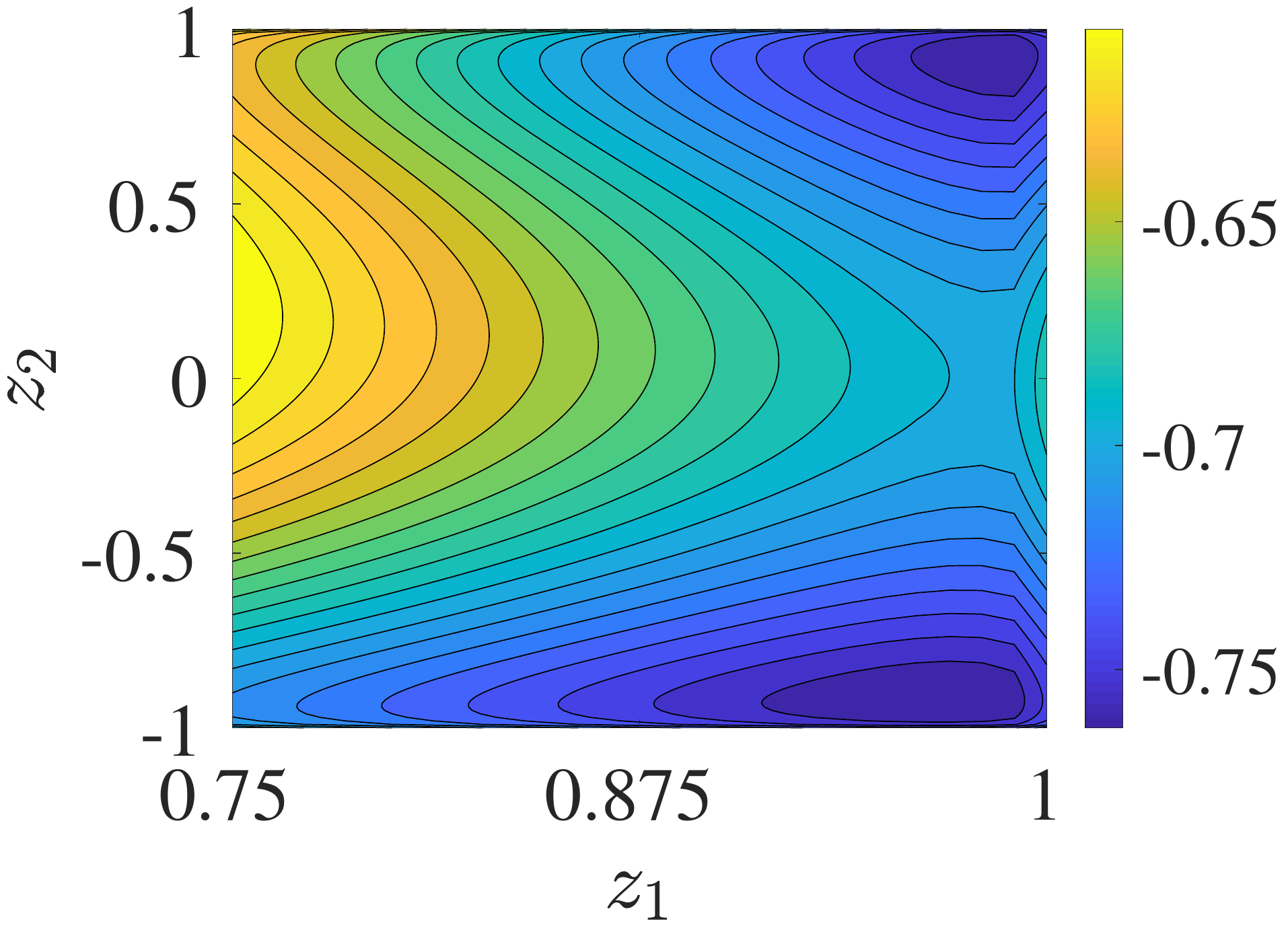} \label{fig:MF_GS6A}}
           \subfigure[]{\includegraphics[width=0.48\columnwidth]{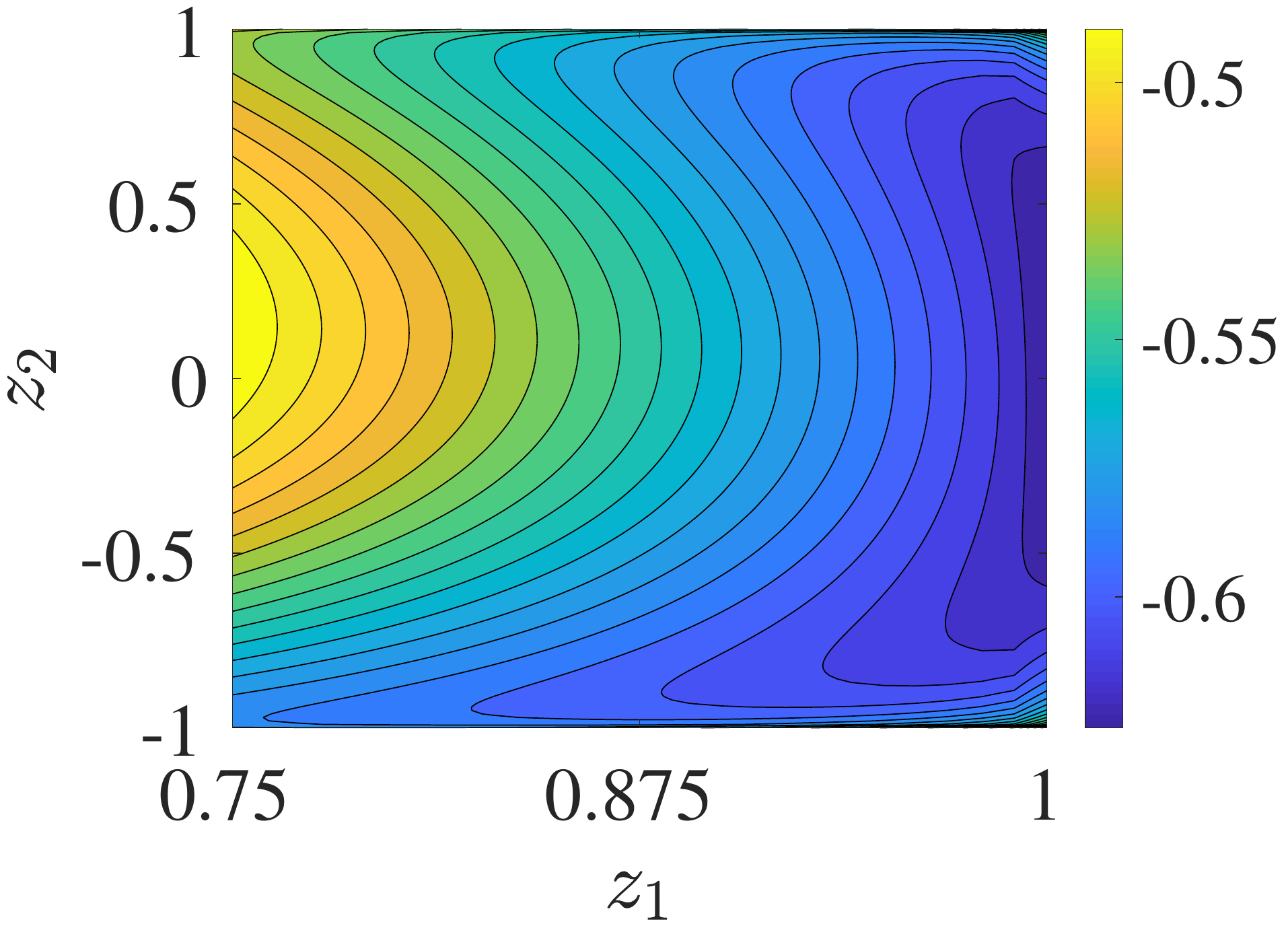} \label{fig:MF_GS5B}}
      \subfigure[]{\includegraphics[width=0.48\columnwidth]{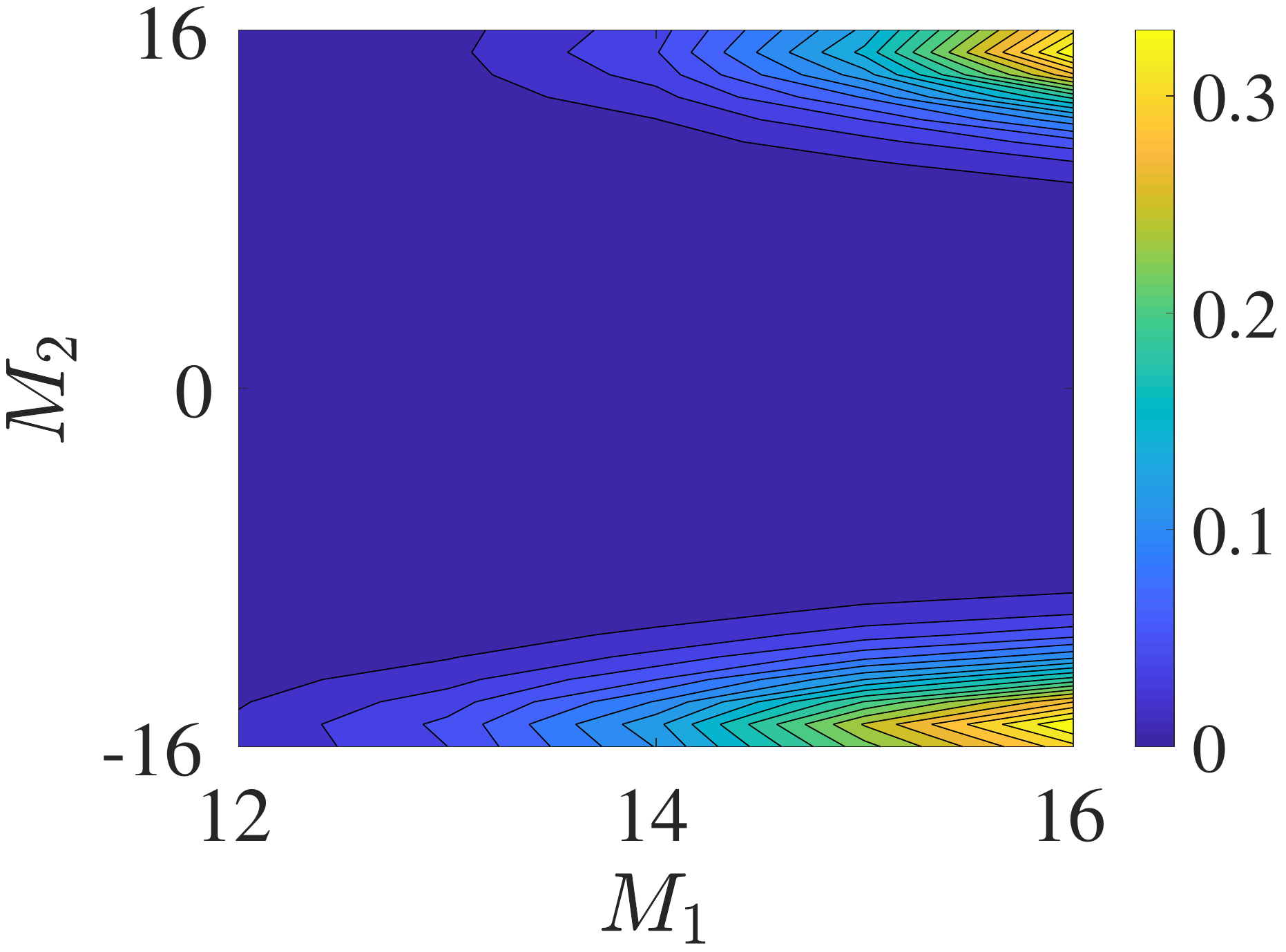} \label{fig:MF_GS6A}}
           \subfigure[]{\includegraphics[width=0.48\columnwidth]{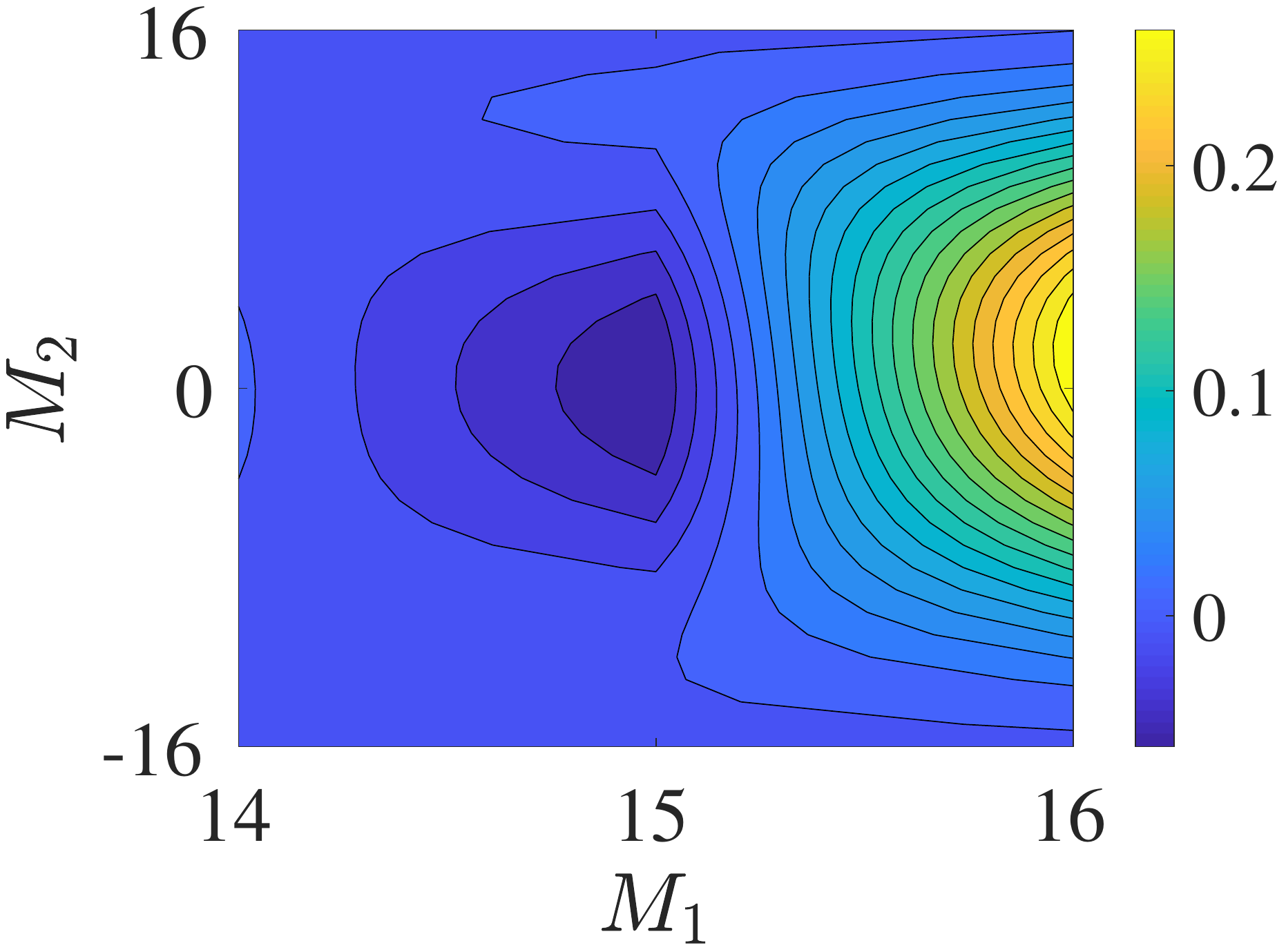} \label{fig:MF_GS5B}}
               \caption{(a,b) The semiclassical potential defined in Eq.~\eqref{eqt:V} for the infinite-range 2-local large-spin tunneling problem with $\lambda = 0$ and $\lambda = 2.275$ near their respective minimum gaps.  We evaluate the potential at $s \approx 0.722$ and $s = 0.5$ respectively.  (c,d) The ground state wavefunction $\braket{M_1,M_2}{E_0}$ for $\lambda = 0$ and $\lambda = 2.275$ near their respective minimum gaps for $n = 64$.}
   \label{fig:MF_GS6}
\end{figure}

It is also useful to consider the semiclassical potential derived in the Villain representation~\cite{Villain:1974,Enz:1986,2002quant.ph..8189B,PhysRevA.68.062321,oneOverF2,2018arXiv180607602D}:
\begin{eqnarray} \label{eqt:V}
V(z_1,z_2) & =& - \frac{1}{2} (1-s) \left(\sqrt{1-z_1^2} + \sqrt{1 - z_2^2} \right) \nonumber \\
& & - s \left(  \frac{1}{2} \left(h_1 z_1 - h_2 z_2 \right) +  \frac{1}{4} \left(z_1^2 + z_2^2 + z_1 z_2 \right)\right) \nonumber \\
&& + \frac{\lambda s(1-s)}{2} \sqrt{(1-z_1^2)(1-z_2^2)} \ .
\end{eqnarray}
  Near the minimum gap, the semiclassical potential for the the stoquastic and non-stoquastic cases has important differences.  For the stoquastic case, there is an energy barrier separating the degenerate minima of the potential through which the system must tunnel, whereas for the non-stoquastic case, there is a single wide minimum.  This then leads to very different behaviors for the ground state wavefunction at this point.  These features are depicted in Fig.~\ref{fig:MF_GS6}, where we note in particular the appearance of negative amplitude in the ground state wavefunction for the non-stoquastic case, in a similar way to the $(p=5)$-spin model in Fig.~\ref{fig:MF_GS3B}.

\begin{figure}[htbp]
   \centering
      \subfigure[]{\includegraphics[width=0.48\columnwidth]{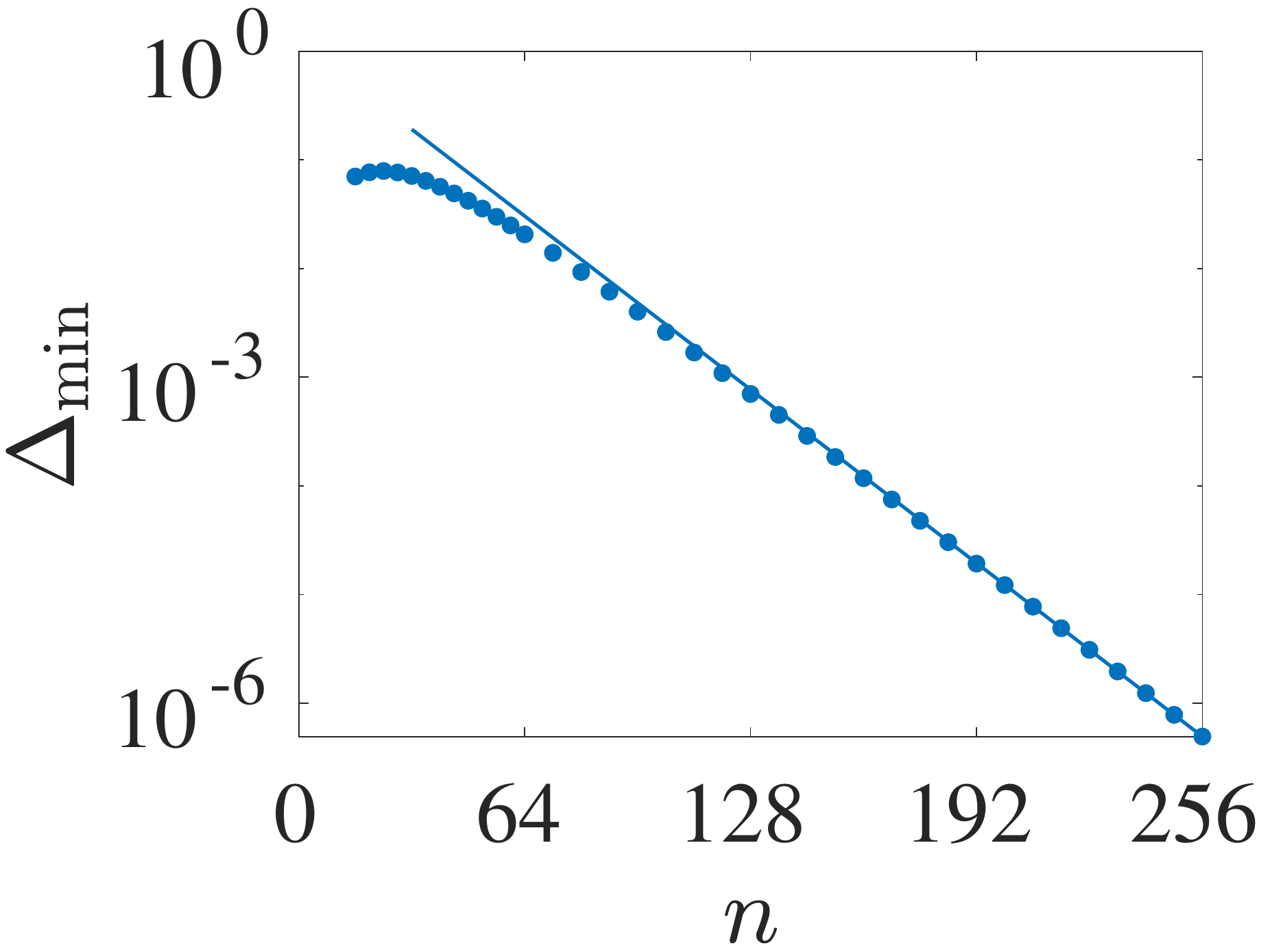} \label{fig:MF_GS8A}}
            \subfigure[]{\includegraphics[width=0.48\columnwidth]{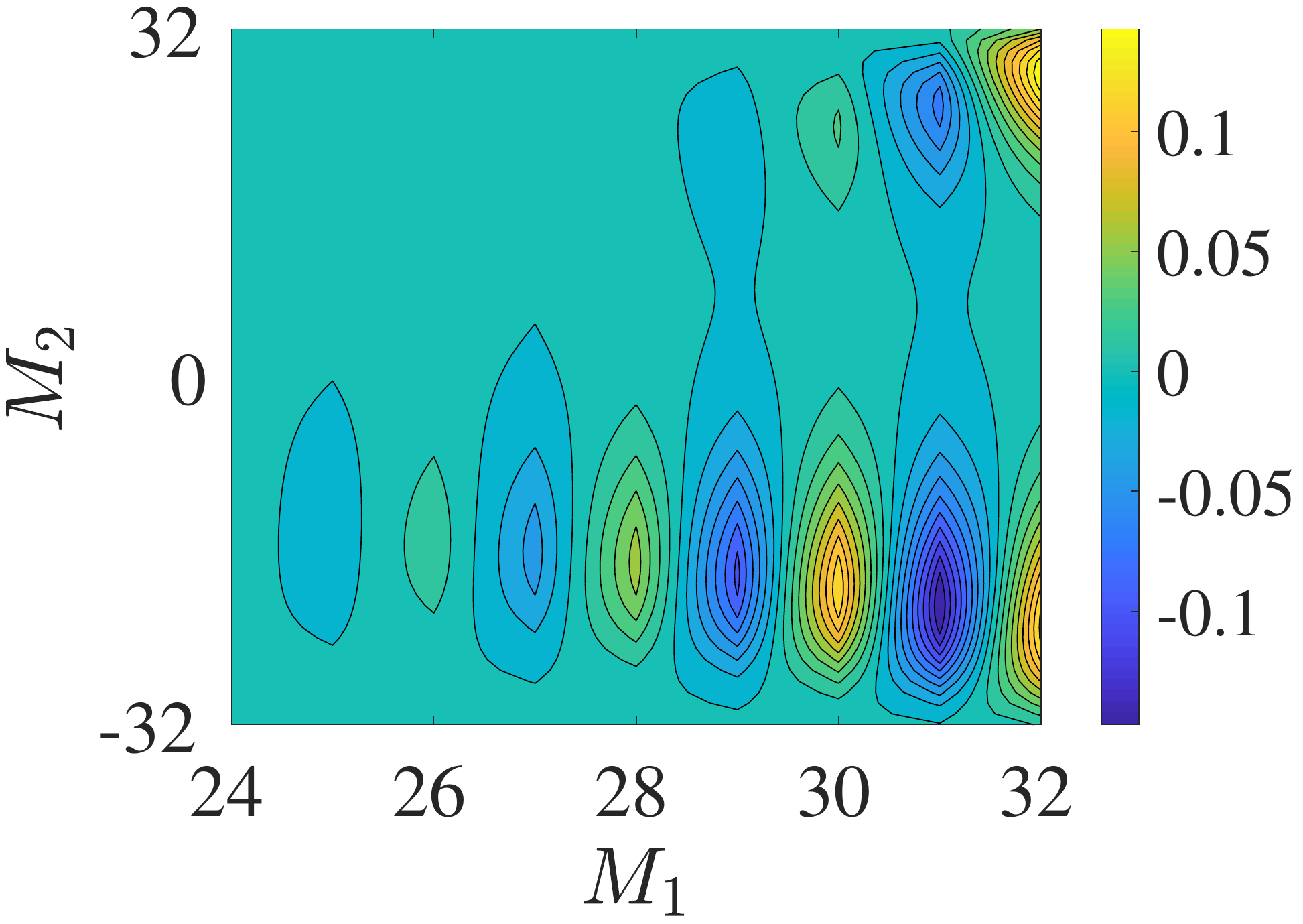} \label{fig:MF_GS8B}}
               \caption{(a) Scaling of the minimum gap within the subspace $\tilde{\mathcal{S}}$ for the infinite-range 2-local large-spin tunneling problem with $\lambda = 4$.  Solid line is the fit to $\sim \exp(-0.06 n)$. (b) Ground state wavefunction $\braket{M_1,M_2}{E_0}$ for $\lambda = 4, n=128$ evaluated at $s \approx 0.697$, near the location of the minimum gap.}
   \label{fig:MF_G87}
\end{figure}

Unlike the $p$-spin model we studied in Section~\ref{Sec:pSpin}, choosing a value of $\lambda$ that is too large does not retain the exponential advantage observed for $\lambda_{\mathrm{opt}}$.  As we show in Fig.~\ref{fig:MF_GS8A}, picking $\lambda$ too large causes the minimum gap to eventually scale exponentially with $n$ for sufficiently large $n$.  This arises because if $\lambda$ is too large, then the semiclassical potential becomes similar to that of the stoquastic case with an energy barrier separating the two local minima, although at finite size the ground state wavefunction exhibits a lot more structure, as shown in Fig.~\ref{fig:MF_GS8B}.  Nevertheless the exponential scaling is less severe than in the stoquastic case, so there is still an advantage even with this non-optimal choice for $\lambda$.

\section{Geometrically local Ising example} \label{Sec:LocalIsing}
%{\bf Geometrically local Ising example}.---
We now construct a geometrically local example with 2-local interactions that also exhibits an advantage for non-stoquastic catalysts over stoquastic catalysts. This example will not enjoy the permutation symmetry enjoyed by the previous two infinite-range models, so simulations much larger than $24$ qubits is computationally prohibitive.  In order to ensure that the stoquastic cases already exhibit their worse-case behavior at these sizes, we construct our problem to have a `perturbative crossing', which is a well-established QAO bottleneck \cite{Amin2009}.
 
% In order to demonstrate  %many of the same features as the infinite-range $p$-spin model.  
%We consider the class of Ising Hamiltonians shown in Fig.~\ref{fig:RingFigure}.
%, which have the all-zero and all-one bit-strings as their ground state (we pick $\sigma^z \ket{0} = \ket{0}$.  
%The instances are composed of two rings, each with $n/2$ (even) spins, connected to each other at the opposite ends of the ring.  
%
\begin{figure}[htbp] %  figure placement: here, top, bottom, or page
   \centering
   \includegraphics[width=3in]{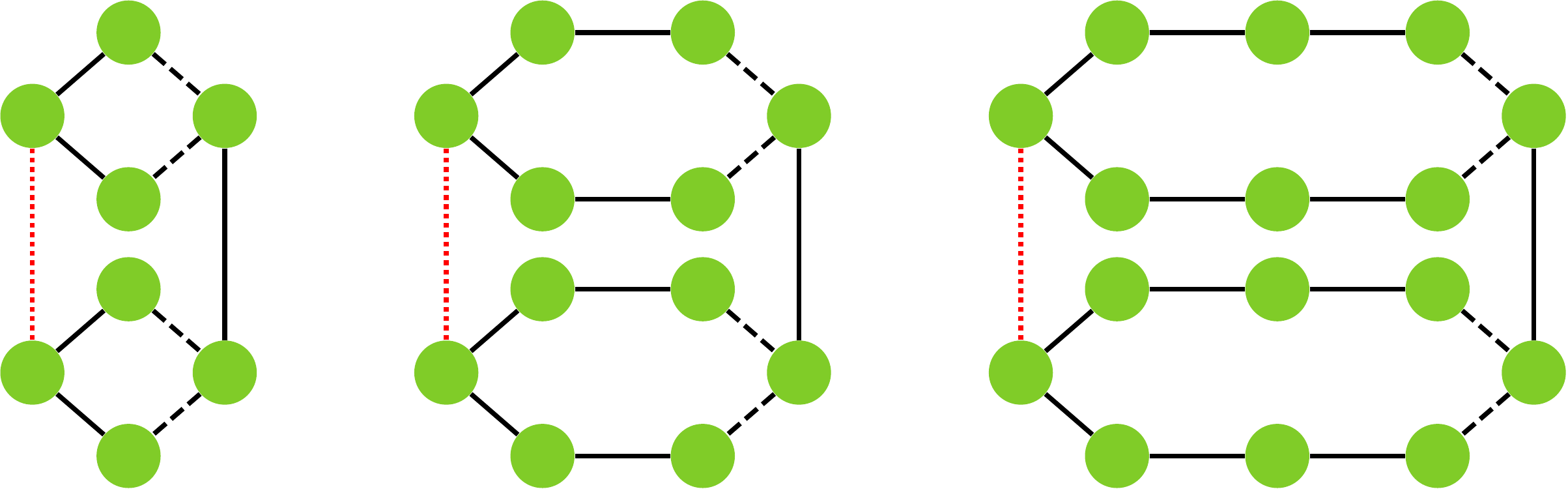} 
   \caption{Ising Hamiltonians of size $n=8$, $12$, and $16$ for the Hamiltonian in Eq.~\eqref{eqt:H2local}.  Spins are given by the green disks, and the lines between them correspond to Ising spin-spin interactions.  Solid black lines correspond to a ferromagnetic coupling of magnitude $1$, dashed black lines to a ferromagnetic coupling of magnitude $1/2$, and red dotted lines correspond to antiferromagnetic couplings of magnitude $1-1/6$.  The Ising Hamiltonian scales by introducing spins in the upper and lower rings at their centers in an alternating manner.} %  The minimum in the gap occurs at $s \approx 0.36$ at n=128.}
   \label{fig:RingFigure}
\end{figure}
\begin{figure}[t] %  figure placement: here, top, bottom, or page
   \centering
   \subfigure[]{\includegraphics[width=0.48\columnwidth]{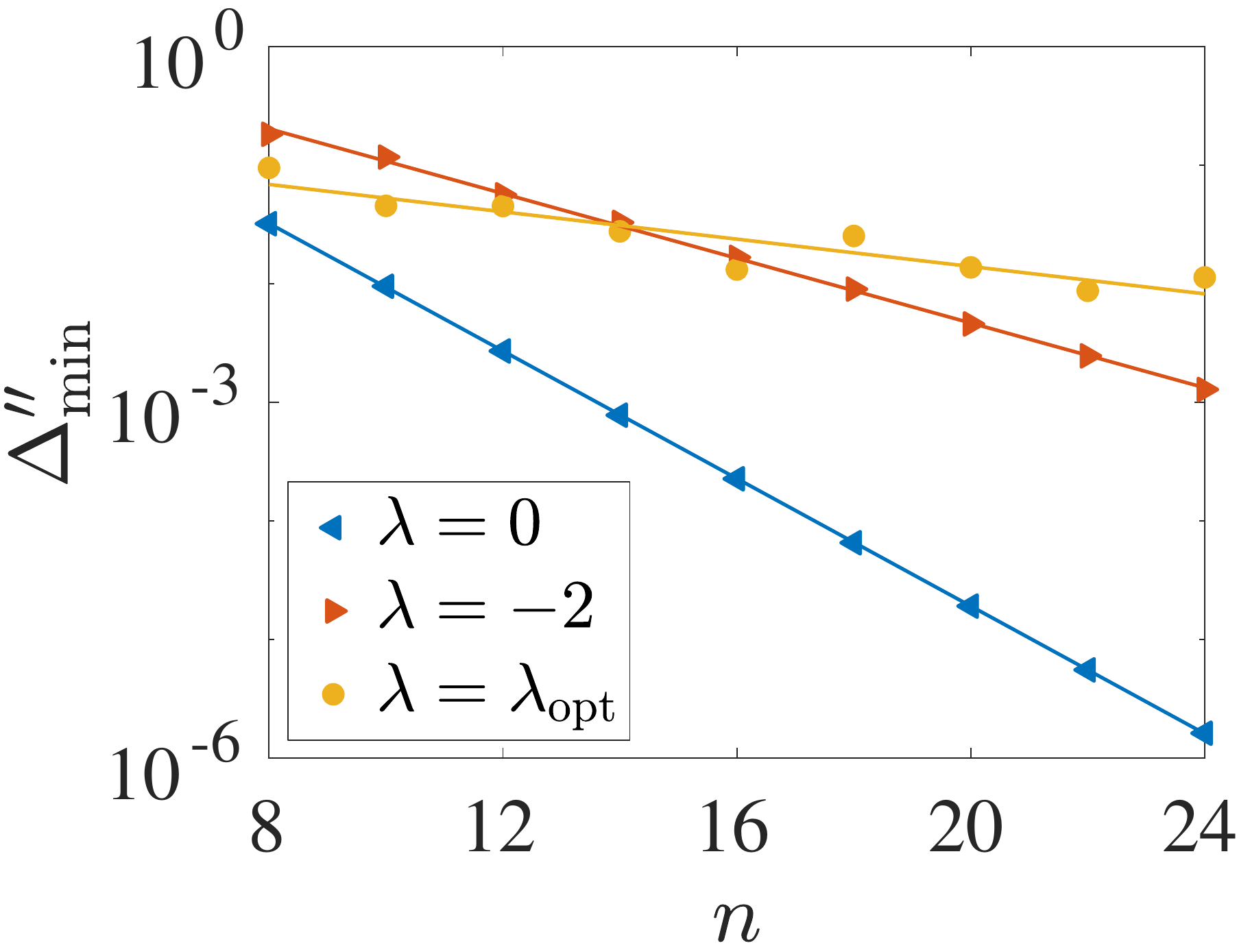} \label{fig:Ising_GapScaling}}
   \subfigure[]{\includegraphics[width=0.48\columnwidth]{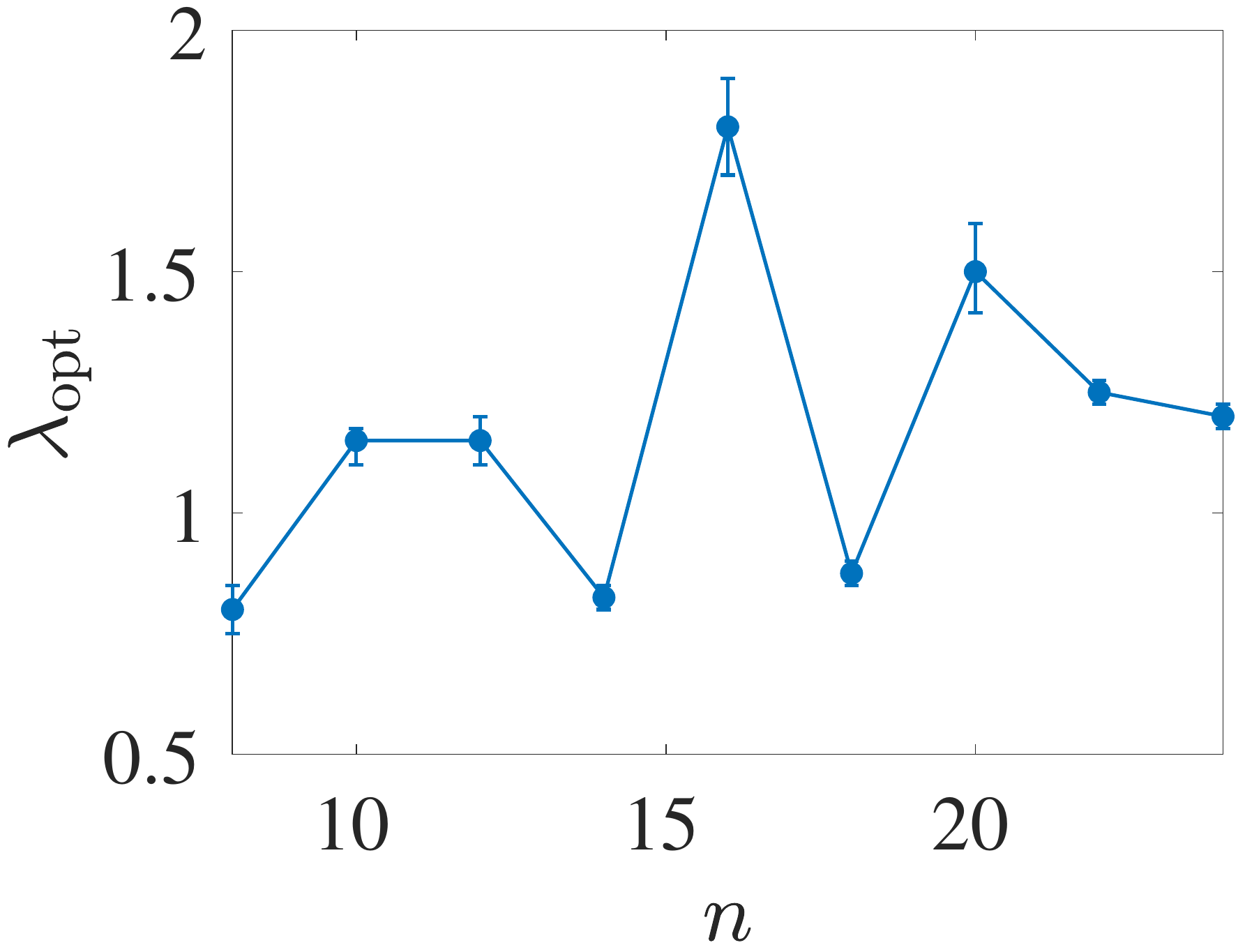}  \label{fig:Ising_Spectrum}}
   \caption{(a) Scaling of the minimum gap $\Delta_\mathrm{min}''$ within the subspace $\mathcal{S}''$ for the geometrically local Ising example with no catalyst ($\lambda= 0$), a stoquastic catalyst ($\lambda = -2$), and a non-stoquastic catalyst with optimized $\lambda$ ($\lambda = \lambda_{\mathrm{opt}}$).  Solid lines correspond to the fits of $\sim \exp(-0.62n)$, $\exp(-0.31n)$, and $\exp({-0.133n})$ respectively. (b) Optimized values of $\lambda$ used with the non-stoquastic catalyst. The error bars correspond to our uncertainty in the exact optimum value of $\lambda$.}
   \label{fig:Ising_Results}
\end{figure}

We consider an interpolation Hamiltonian for the QAO algorithm of the form:
\begin{eqnarray} \label{eqt:H2local}
H_\lambda(s) &=& -(1-s) \sum_i \sigma_i^x + s \sum_{\langle i, j \rangle} J_{ij} \sigma_i^z \sigma_j^z \nonumber \\
&& + \lambda s(1-s) \sum_{\langle i, j \rangle} \sigma_i^x \sigma_j^x \ ,
\end{eqnarray}
where $\left\{J_{ij} \right\}$ are the Ising interactions depicted in Fig.~\ref{fig:RingFigure}.
% and $R$ denotes the maximum magnitude coupling of the Ising interactions. 
The catalyst term has the same connectivity $\langle i,j \rangle$ as the Ising interactions.  The Hamiltonian $H_\lambda(s)$ is invariant under $P$ and under the interchange of the top and bottom rings of qubits. 
%This latter symmetry is analogous to the permutation symmetry of the infinite-range $p$-spin models. 
 Because the ground state at $s = 0$ is the uniform superposition state, the evolution under $H_\lambda(s)$ is restricted to the subspace $\mathcal{S}''$ with eigenvalue $1$ under the transformations associated with both these symmetries.  
 %
%For sufficiently large $R$ (we take $R = 6$), 
For our choice of Ising parameters (see Fig.~\ref{fig:RingFigure}), $H_\lambda(s)$ exhibits an exponentially closing minimum gap along the interpolation for the stoquastic cases ($\lambda \leq 0$) even for small system sizes $n \leq 24$ as shown in Fig.~\ref{fig:Ising_GapScaling}. In contrast, the closing of the minimum gap for the non-stoquastic case with a catalyst strength of the same order is significantly milder, and is equally well fit by a polynomial or a mild exponential.  It is difficult to distinguish the two possibilities at the sizes we consider.  We provide a more up-close comparison of the two fits in Appendix \ref{App:fits}.

\begin{figure}[htbp] %  figure placement: here, top, bottom, or page
   \centering
   \subfigure[]{\includegraphics[width=0.45\columnwidth]{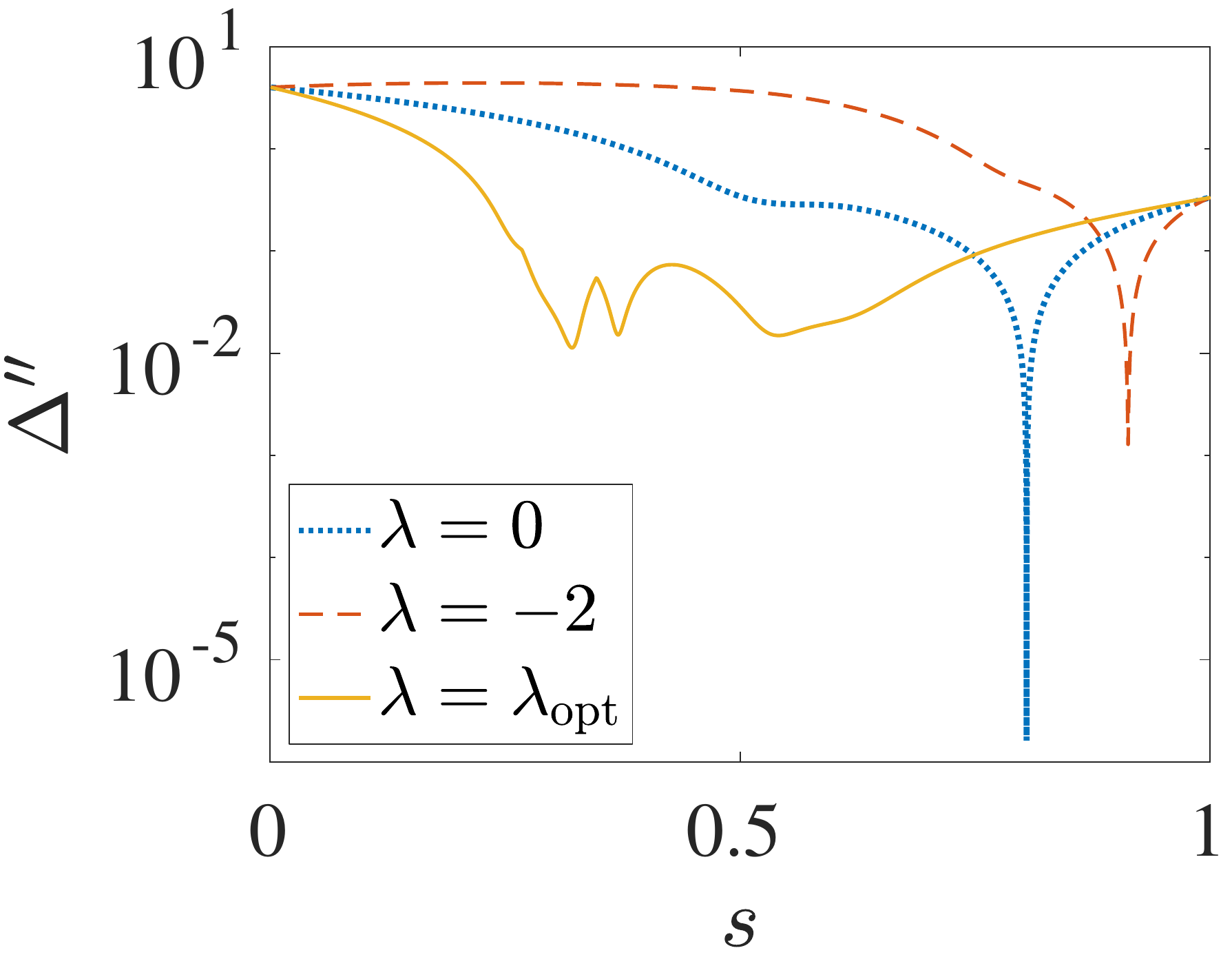} \label{fig:RingGap1}}
   \subfigure[]{\includegraphics[width=0.48\columnwidth]{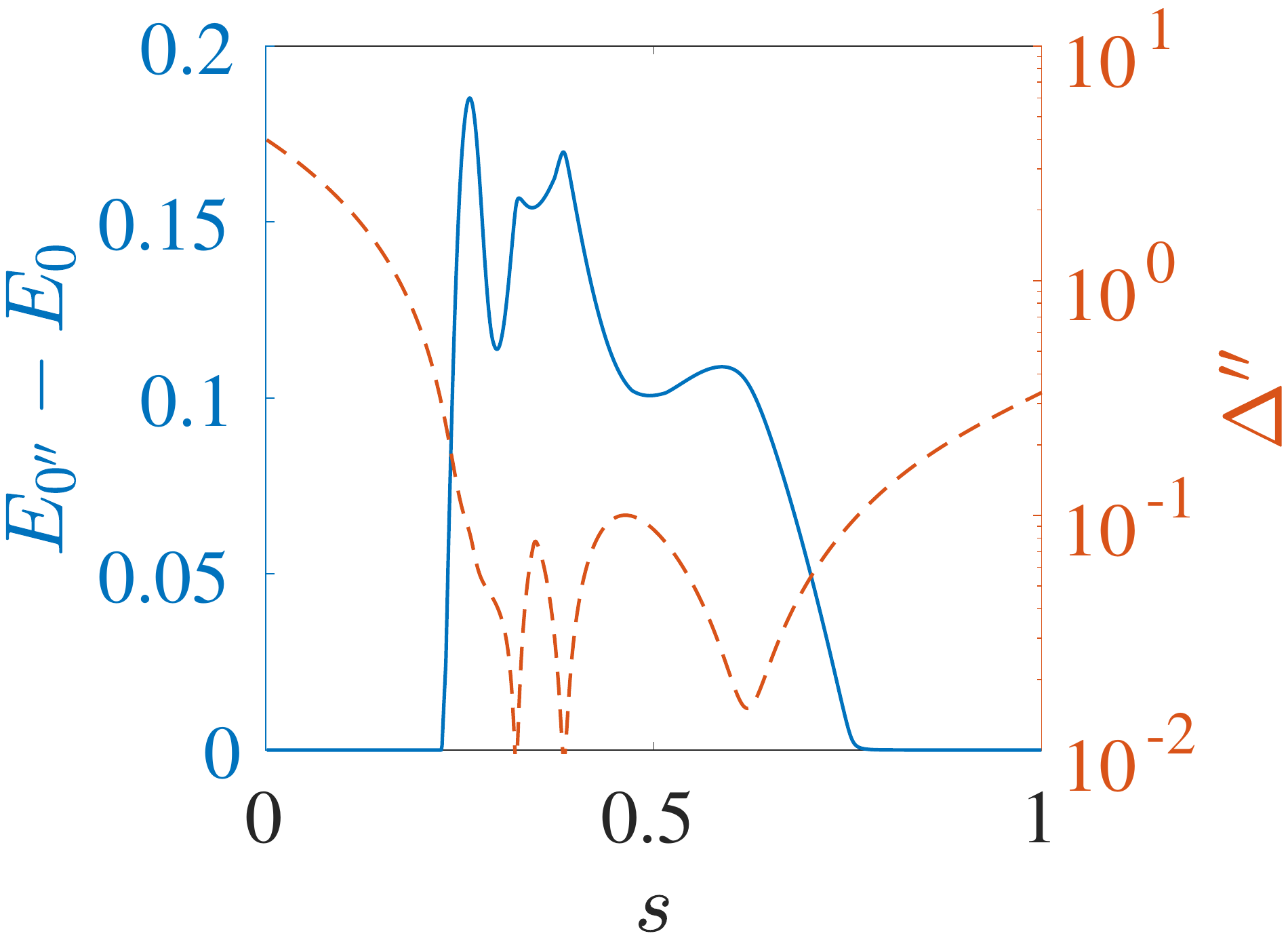} \label{fig:RingGap2}}
   \caption{(a) Comparison of the gap $\Delta^{\prime \prime}$ within the subspace $\mathcal{S}''$ for for $n=22$ without a catalyst ($\lambda = 0$) with a stoquastic catalyst ($\lambda = -2$), and with a non-stoquastic catalyst ($\lambda = 1.25$). (b) Energy difference $E_{0''} - E_0$ between the ground state of $H_\lambda$ within the subspace $\mathcal{S}''$ and the global ground state (left axis), and the ground state energy gap $\Delta''$ within the subspace $\mathcal{S}''$ along the interpolating path for $n=22$ using the non-stoquastic catalyst with $\lambda = 1.25$ (right axis). }
   \label{fig:RingGap}
\end{figure}

A key feature of the spectrum in the presence of the non-stoquastic catalyst is the absence of the perturbative crossing near $s = 1$.  We find that instead of the single local minimum associated with the perturbative crossing in the stoquastic case, we have multiple local minima that are much milder as shown in Fig.~\ref{fig:RingGap1}. We also show in Fig.~\ref{fig:RingGap2} that the ground state of $H_\lambda(s)$ within the subspace $\mathcal{S}''$ deviates from the global ground state, which is also what we observed for the $p$-spin model when $p$ is even (shown in Appendix \ref{App:evenVSodd}) .
%In our fully-connected examples, these local minima were associated with incremental changes in the ground state wavefunction, but the absence of the permutation symmetry here makes it impossible to make this comparison more concrete.  
%We also show in Fig.~\ref{fig:RingGap2} how the ground state of $H_\lambda$ within the subspace $\mathcal{S}''$ deviates from the global ground state, which resembles what we observed for the $p$-spin model when $p$ is even.

%features that resemble those uncovered in the previous two models studied.  We find that instead of the single local minimum associated with the perturbative crossing in the stoquastic case, we have multiple local minima that are much milder as shown in Fig.~\ref{fig:RingGap1}.  In our fully-connected examples, these local minima were associated with incremental changes in the ground state wavefunction, but the absence of the permutation symmetry here makes it impossible to make this comparison more concrete.  We also show in Fig.~\ref{fig:RingGap2} how the ground state of $H$ within the subspace $\mathcal{S}''$ deviates from the global ground state, which resembles what we observed for the $p$-spin model when $p$ is even.
%

We can relate the absence of the perturbative crossing near $s=1$ to the behavior of the low-lying energy spectrum of the Ising Hamiltonian, i.e. the spectrum at $s =1$ for Eq.~\eqref{eqt:H2local}, as the transverse field and catalyst are perturbatively turned on.  For concreteness, let us take the case of $n=8$ and label the states as $\ket{x_1,\dots x_4, x_5 \dots x_8}$ going around the top ring first followed by the bottom ring.  We take $x_i = \left\{0,1\right\}$ to denote the positive and negative eigenvalue of $\sigma_i^z$ respectively.  The doubly-degenerate ground states of the Ising Hamiltonian are given by $\ket{0000 \ 0000}$ and $\ket{1111 \ 1111}$, and there is a unique combination that is within the subspace $\mathcal{S}''$:%we write them in terms of eigenstates of the operator $P$:
\beq
 \ket{\mathbf{g}} \equiv  \frac{1}{\sqrt{2}} \left( \ket{0000 \ 0000} + \ket{1111 \ 1111}\right) \ . 
\eeq
Similarly, the 6-fold degenerate first excited states of the Ising Hamiltonian are given by $\ket{0000 \ 1111}$, $\ket{0000 \  1101}$, $\ket{0010 \ 1111}$ and their $P$-transformed complements, and there are two linear combinations that are within the subspace $\mathcal{S}''$:
\bes
\begin{align}
 \ket{\mathbf{a}} \equiv & \frac{1}{2} \left( \ket{0000 \ 1101} + \ket{1111 \ 0010} \right. \nonumber \\
& \left. + \ket{0010 \ 1111} + \ket{1101 \ 0000}\right) \ , \\
\ket{\mathbf{b}} \equiv & \frac{1}{\sqrt{2}} \left( \ket{0000 \ 1111} + \ket{1111 \ 0000}\right) \ .
   \end{align}
\ees
We now consider the effect of moving away from $s = 1$ on the Hamiltonian spectrum.  Since both the transverse field and catalyst are of the same order in $(1-s)$, they both contribute at first order in perturbation theory.  For simplicity, we will take the perturbation operator to be given by $V = - \sum_i \sigma_i^x + \lambda \sum_{\langle i,j \rangle} \sigma_i^x \sigma_j^x$, with either $\lambda = -1$ for the stoquastic case or $\lambda = 1$ for the non-stoquastic case. Since the ground state is at least $n/2 - 1$ in Hamming distance away from the first excited states, the  states $\ket{\mathbf{g}}$ remains unaffected at first order in perturbation theory.  However, the degeneracy of the first excited is broken at this order.  For the case of $\lambda =-1$, the state $\ket{\mathbf{e}} = \frac{1}{\sqrt{3}} \left( \sqrt{2} \ket{\mathbf{a}} + \ket{\mathbf{b}} \right)$ is lowered in energy, and it has eigenvalue $-2$ under $V$.  This eigenvalue determines the rate with which the energy of this state decreases with $(1-s)$.  As $s$ continues to decrease from $1$, the energy of $\ket{\mathbf{e}}$ eventually crosses that of $\ket{\mathbf{g}}$ at first order in perturbation theory, resulting in an avoided-level crossing. 
% The gap is then given by the tunneling matrix element of the effective two level system \cite{Amin2009}, which is proportional to the overlap $\langle \mathbf{g} |  V_1^{k_1} V_2^{k_2} | \mathbf{e} \rangle$, where $V_1$ is the transverse field operator and $V_2$ is the catalyst operator.  The integers $k_1$ and $k_2$ correspond to the number of these operators that we need to `connect' the states $\ket{\mathbf{g}}$ and $\ket{\mathbf{e}}$, and the gap scales exponentially in their sum.  Because these two states have an effective Hamming distance of $n/2 - 1$, the energy gap at this level crossing scales exponentially with $n$. 
%

For $\lambda = 1$, the degeneracy of the first excited state is broken differently. (The ground state remains unchanged at first order in perturbation theory.)  The degeneracy of the first excited states within the subspace $\mathcal{S}''$ is broken such that the state $\ket{\mathbf{e'}} = \frac{1}{\sqrt{3}}\left( \ket{\mathbf{a}} + \sqrt{2}\ket{\mathbf{b}} \right)$ is lowered in energy, but its eigenvalue under $V$ is only $-1$.  Therefore, the rate at which its energy is lowered is smaller relative to the stoquastic case.  Therefore, the avoided level crossing in principle happens at smaller $s$ values, where the driver and catalyst Hamiltonians are stronger.   This may partly explain why the gap associated with the perturbative crossing is softened in the presence of the non-stoquastic catalyst.
\section{Discussion and Conclusions}
%{\bf Discussion and Conclusions}.---
We analyzed the efficiency of the QAO algorithm for solving three different classes of problems using stoquastic and non-stoquastic catalyst Hamiltonians.
We first revisited the infinite-range $p$-spin model, and our finite size results corroborate the results of Refs.~\cite{Seki:2012,Seoane:2012uq,10.3389/fict.2017.00002,PhysRevA.95.042321}: we find that for a sufficiently strong non-stoquastic catalyst the relevant minimum gap along the interpolation path decreases only polynomially with system size $n$, whereas the stoquastic catalyst has the minimum gap decreasing exponentially with $n$.  We note that our choice of interpolation Hamiltonian in Eq.~\eqref{eqt:pSpinH} differs from the one used in Ref.~\cite{Seki:2012}, but similar results are obtained using either interpolating Hamiltonian, as we show in Appendix \ref{App:DifferentInterpolation}. Thus, the QAO algorithm with the non-stoquastic catalyst runs exponentially faster than with the stoquastic catalyst.

We restricted our study to the cases of $p=5$ and $p=6$.  We expect that similar results to be found for $p \geq 4$.  The case of $p=3$ is different because the mean-field potential always exhibits a discontinuous jump in its global minimum \cite{Seki:2012}.  A careful treatment of this model shows that it still exhibits the exponential advantage for the non-stoquastic catalyst \cite{2018arXiv180607602D}, but it requires an optimized $\lambda$ that grows with system size unlike the examples we studied here.  We show this in Appendix \ref{App:p=3}.

We then showed that a similar advantage is enjoyed by another infinite range model that only uses 2-local interactions.  In a similar manner to the $p$-spin model, the mean-field potential associated with the stoquastic Hamiltonian exhibits a discontinuous jump in the global minimum associated with a large tunneling event, whereas the potential associated with the non-stoquastic Hamiltonian avoids this discontinuity if the catalyst strength is chosen appropriately.  For this optimized choice of catalyst strength, the minimum gap asymptotes to a constant.

An important lesson that we derive from this example is that the interaction terms in the catalyst Hamiltonian can be crucial in determining whether an advantage can be had with a non-stoquastic catalyst.  In the Hamiltonian of Eq.~\eqref{eqt:H2}, the catalyst is taken to be proportional to $S_1^x S_2^x$ and not $(S_1^x + S_2^x)^2$.  We find that the latter case does not exhibit the exponential advantage that the former does; its scaling is similar to that without a catalyst as we show in Appendix \ref{App:DifferentCatalyst}.  This indicates that for a given connectivity graph defined by the optimization problem Hamiltonian, the catalyst should not always share the same connectivity to give the best results.  

Our finite $n$ analysis provides a different way to understand why the non-stoquastic catalyst can provide such a dramatic improvement over its stoquastic counterpart.  The non-stoquastic catalyst allows for multiple incremental changes to the ground state wavefunction, as opposed to the single large change that occurs for the stoquastic catalyst.  Qualitatively, we can interpret this as `spreading' the computational effort over a larger range of the interpolation as opposed to a narrow region only.

We have also constructed a geometrically local Ising example that exhibits many of the same qualitative features as the previous two examples.  We observe a growing advantage for the non-stoquastic catalyst over the stoquastic catalyst, and we attribute this to the non-stoquastic catalyst effectively softening or possibly even eliminating the perturbative level crossing that plagues the stoquastic case.  While we expect that the exponential scaling of the gap will continue for the stoquastic case, we cannot rule out that the non-stoquastic case may transition to another scaling at larger sizes. The lack of the permutation symmetry prevents us from performing a similar analysis as was done for the infinite-range models at larger system sizes. While there are other methods for eliminating perturbative crossings that do not rely on catalyst Hamiltonians \cite{Dickson_2011}, we hope this example in conjunction with the other examples presented in this work may help shed more light on the viability of non-stoquastic catalysts to give an advantage over their stoquastic counterparts.

For several of the cases we study, the ground state of the subspace in which the evolution occurs does not correspond to the global ground state of the non-stoquastic Hamiltonian because of energy-level crossings in the spectrum.  In the closed system setting, this does not pose a problem since adiabaticity will still allow us to reach the desired final ground state.  However, in the open system setting, thermal relaxation to the global ground state may actually hinder the QAO algorithm. In this case, the QAO algorithm would not necessarily have the robustness to thermal decoherence that it typically does~\cite{Farhi:01,2002quant.ph.11152K,PhysRevA.71.032330,PhysRevA.75.062313,ashhab:052330,amin_decoherence_2009,Sarovar:2013kx,Albash:2015nx,childs_robustness_2001,PhysRevLett.95.250503,Aberg:05,PhysRevA.72.042317,TAQC,PhysRevA.80.022303,Vega:2010fk,oreshkov_adiabatic_2010,Qiang:13,2002quant.ph.11152K}. Under what conditions the exponential advantage can still be maintained is an important issue to be addressed.

Our examples rely heavily on symmetries in the Hamiltonian to facilitate the analysis, and we should not expect this to be the typical situation for optimization problems.  Furthermore, any implementation of the QAO algorithm on a physical device will inherently have implementation errors \cite{childs_robustness_2001,Hauke:2012,scirep15:Martin-Mayor_Hen,Young:2013fk,analogErrors}, which would in turn break these symmetries. For the case of the infinite-range $p$-spin models, the case of $p$ odd breaks the symmetry associated with the operator $P$ and yet retains its exponential advantage.  The advantage is also retained for certain Hopfield models \cite{Seki:2015}.  However, we show in the Supplementary Materials that the introduction of implementation errors that break all the symmetries can change the energy gaps of the models we study, and it is not clear whether the advantage is retained in this situation.  %More generally, it remains unknown to what extent the presence of symmetries is required for a non-stoquastic catalyst to have an advantage over its stoquastic counterpart.  

%also changes the scaling of the non-stoquastic minimum gap from polynomial to exponential, indicating that the advantage of the non-stoquastic catalyst relies sensitively on the symmetries of the problem.  It remains unclear whether more generally an advantage for a non-stoquastic catalyst necessitates the presence of symmetries.  

We stress that the optimization problems considered here are trivial, so our work does not address whether non-stoquastic catalysts can help the QAO algorithm achieve true quantum advantages over classical algorithms. Nevertheless, our geometrically local Ising example suggests that physically implementable examples can be constructed and studied, both on a small scale using classical simulation but hopefully also on a large scale using quantum simulators with sufficiently rich programmable interactions. This opens up the possibility of better addressing this question in the near future as next generation experimental quantum information processing devices become available.\\

\begin{acknowledgments}
{\bf Acknowledgements}.---
We thank Itay Hen and Hidetoshi Nishimori for useful discussions. We also thank Gabriel Durkin for useful comments on the manuscript. Computation for the work described in this paper was supported by the University of Southern California's Center for High-Performance Computing (hpc.usc.edu) and by ARO grant number W911NF1810227.
The research is based upon work partially supported by the Office of the Director of National Intelligence (ODNI), Intelligence Advanced Research Projects Activity (IARPA), via the U.S. Army Research Office contract W911NF-17-C-0050. The views and conclusions contained herein are those of the authors and should not be interpreted as necessarily representing the official policies or endorsements, either expressed or implied, of the ODNI, IARPA, or the U.S. Government. The U.S. Government is authorized to reproduce and distribute reprints for Governmental purposes notwithstanding any copyright annotation thereon.
\end{acknowledgments}
%\bibliography{refs}

\appendix
%
%%%%%%%%%%%%%%%%%%%%%%%%%%%%%
\section{Unoptimized $\lambda$ in the infinite-range ferromagnetic $p$-spin model} \label{App:Unoptimized}
\begin{figure}[htbp]
   \centering
      \subfigure[]{\includegraphics[width=0.48\columnwidth]{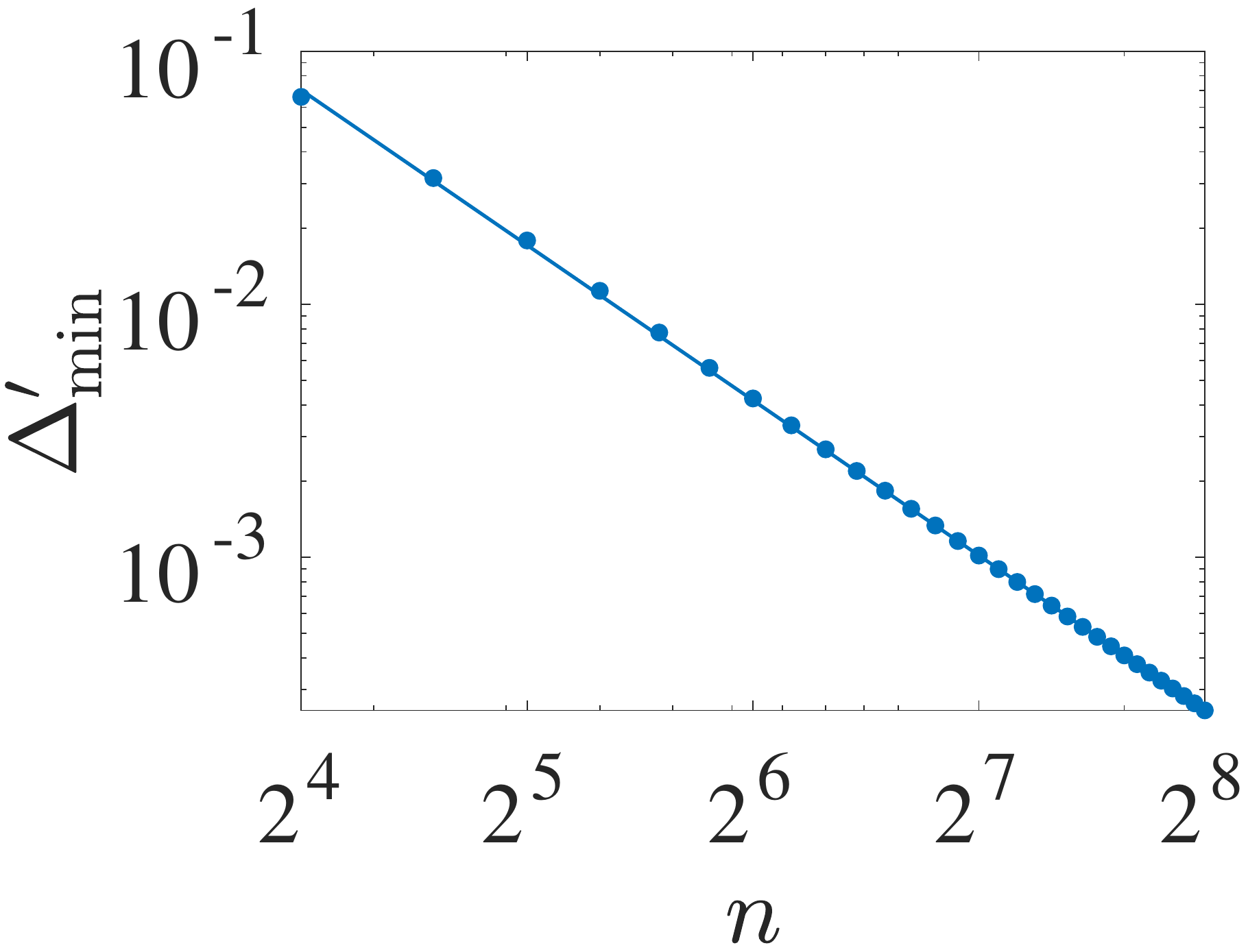} \label{fig:MF_GSS1A}}
            \subfigure[]{\includegraphics[width=0.46\columnwidth]{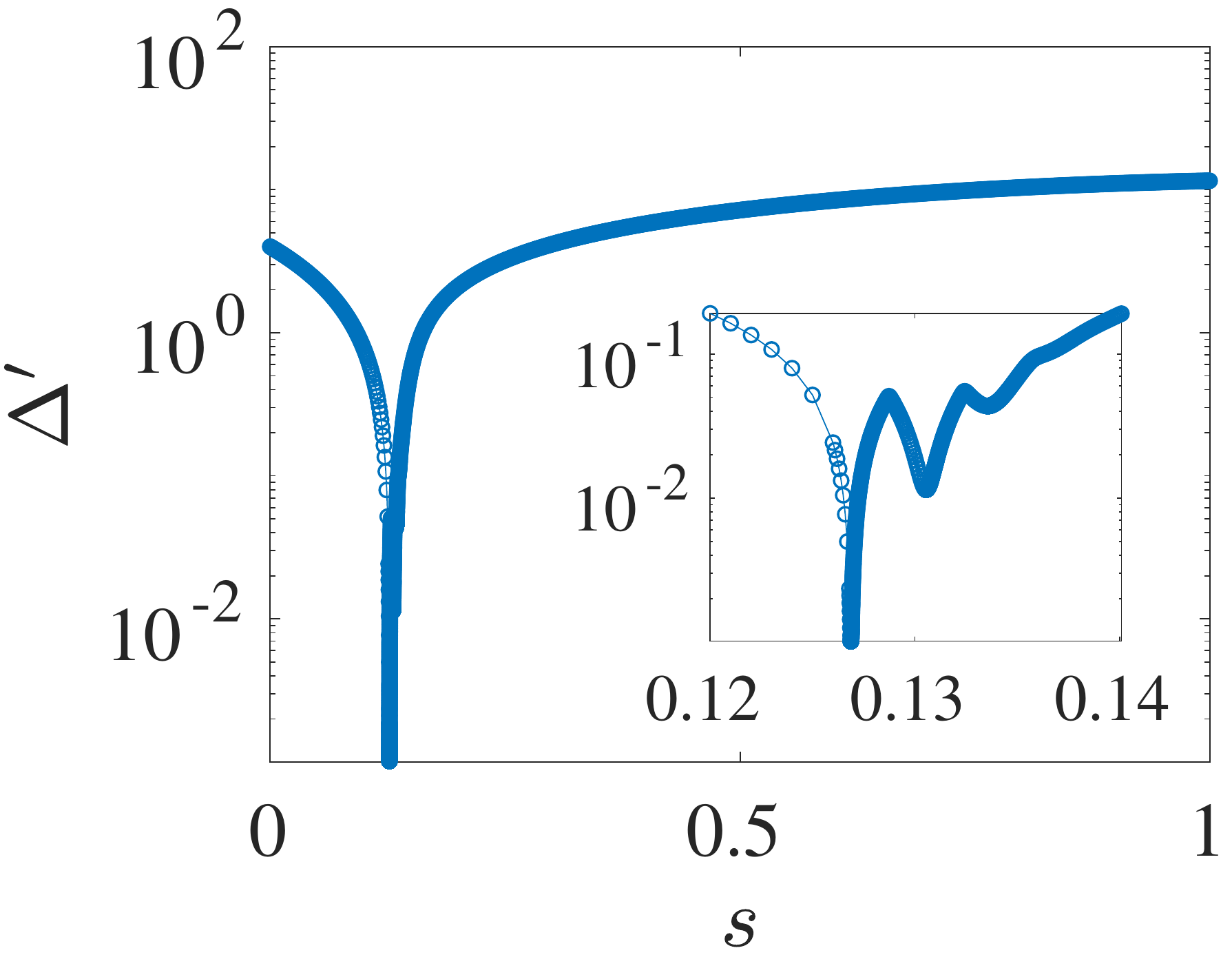} \label{fig:MF_GSS1B}}
               \caption{(a) Scaling of the minimum gap within the subspace $\mathcal{S}'$ for the infinite-range ferromagnetic $(p = 6)$-spin model with a non-stoquastic catalyst and $\lambda = 4$. The solid line corresponds to a fit of $\sim n^{-2.04}$. The inset shows a close-up of the region of interest. (b) Energy gap $\Delta'$ between the ground state and first excited state in $\mathcal{S'}$ for $p=6$, $n=128$ and $\lambda = 4$.  The inset shows a close-up of the region of interest.}
   \label{fig:MF_GSS1}
\end{figure}

%%%%%%%%%%%%%%%%%%%%%%%%%%%%%
In Section \ref{Sec:pSpin} we focused on the case where $\lambda$ is picked to maximize the minimum gap encountered during the interpolation.  In the case of the $p$-spin model, we can pick a single sufficiently large $\lambda > 0$ for all problem sizes and reproduce the polynomial scaling (Fig.~\ref{fig:MF_GSS1A}). Of particular interest though is that in this unoptimized case more local minima in the gap are apparent (compare Fig.~\ref{fig:MF_GSS1B} and Fig.~\ref{fig:MF_GS2A}).  

\begin{figure}[htbp]
   \centering
      \subfigure[]{\includegraphics[width=0.48\columnwidth]{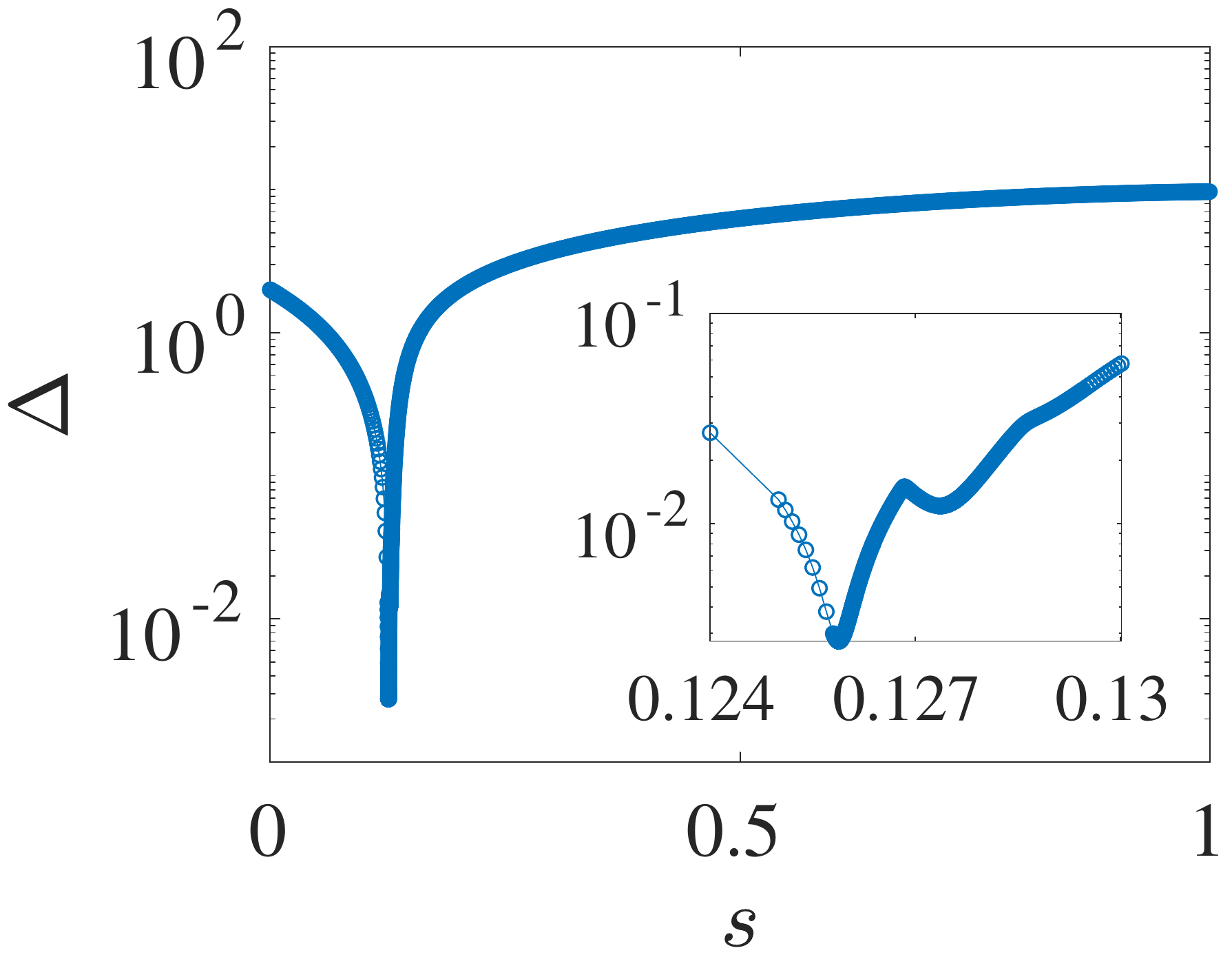} \label{fig:MF_GSS2A}}
            \subfigure[]{\includegraphics[width=0.48\columnwidth]{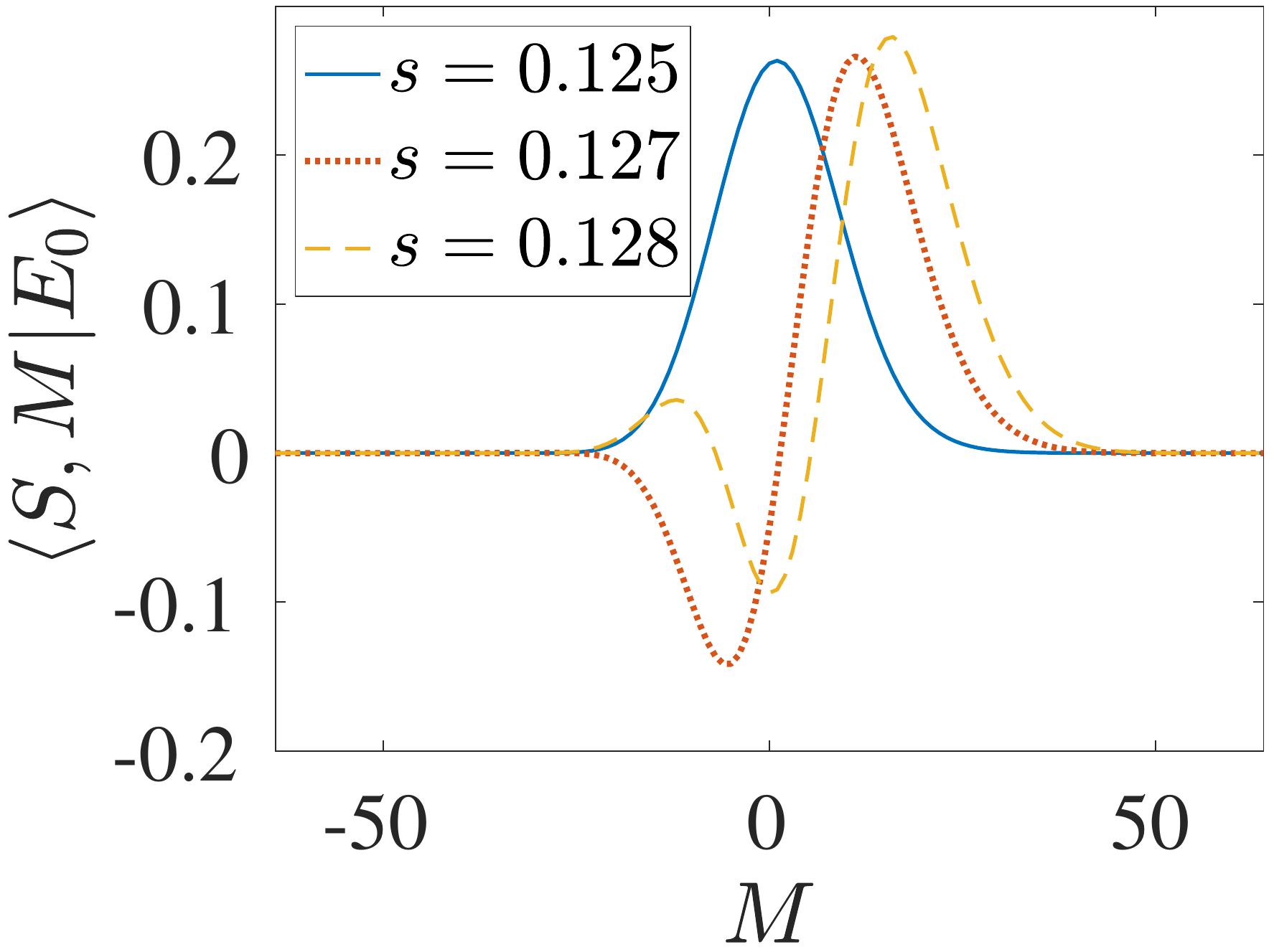} \label{fig:MF_GSS2B}}
              \caption{(a) Energy gap $\Delta$ between the ground state and first excited state within the subspace $\mathcal{S}$ for the infinite-range ferromagnetic $(p = 5)$-spin model. (b) The ground state wavefunction $\ket{E_{0}(s)}$ within the subspace $\mathcal{S}$ evaluated in the Dicke basis.  Results shown are for $n=128$ and $\lambda= 4$.}
   \label{fig:MF_GSS2}
\end{figure}

Similar results are observed for the case of $p$ odd, as shown in Fig.~\ref{fig:MF_GSS2}.
%
%%%%%%%%%%%%%%%%%%%%%%%%%%%%%
\section{Comparing $p$ even and $p$ odd in the infinite-range ferromagnetic $p$-spin model} \label{App:evenVSodd}
\begin{figure}[htbp]
   \centering
    \subfigure[]{\includegraphics[width=0.46\columnwidth]{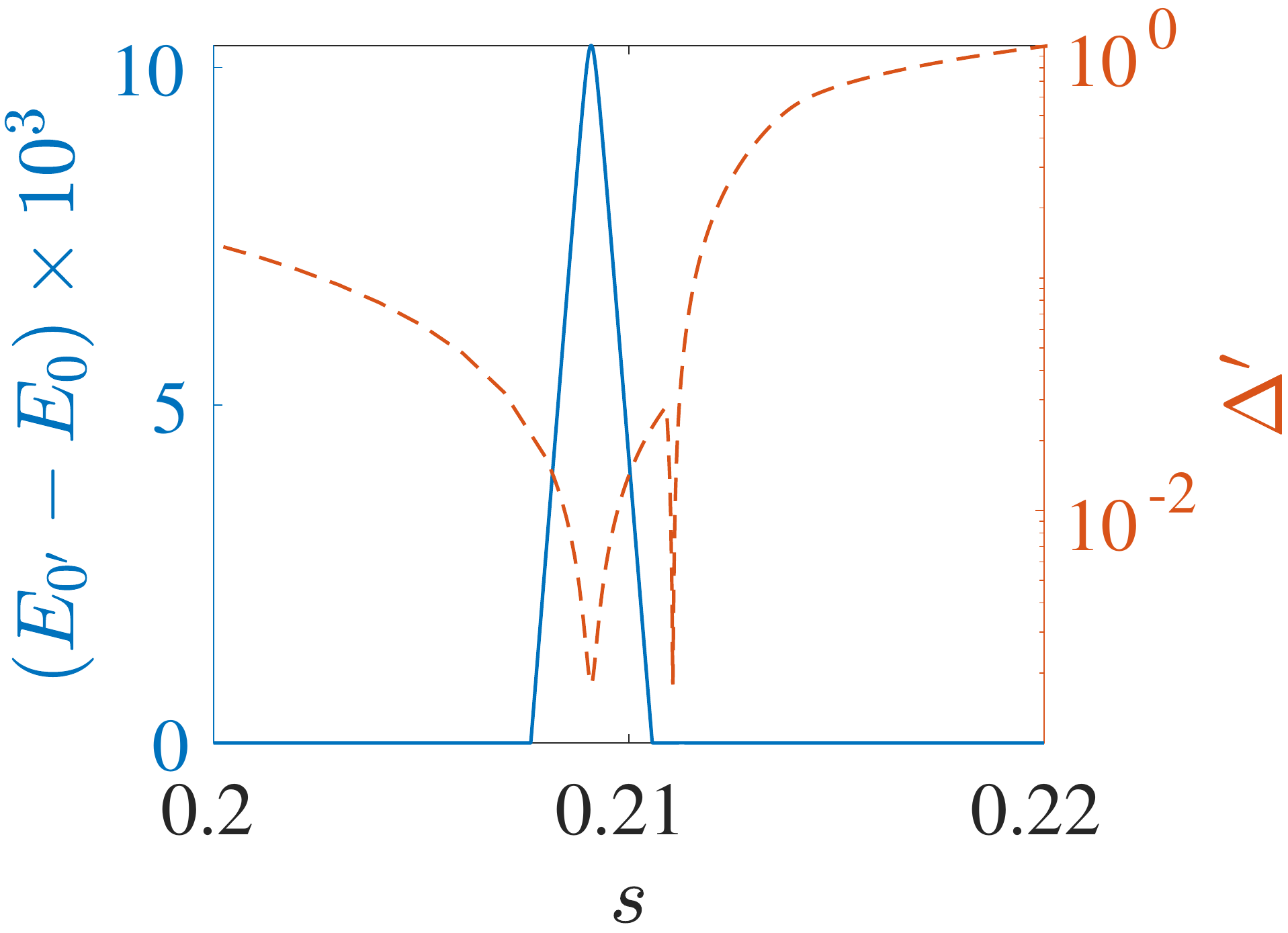} }
    \subfigure[]{\includegraphics[width=0.46\columnwidth]{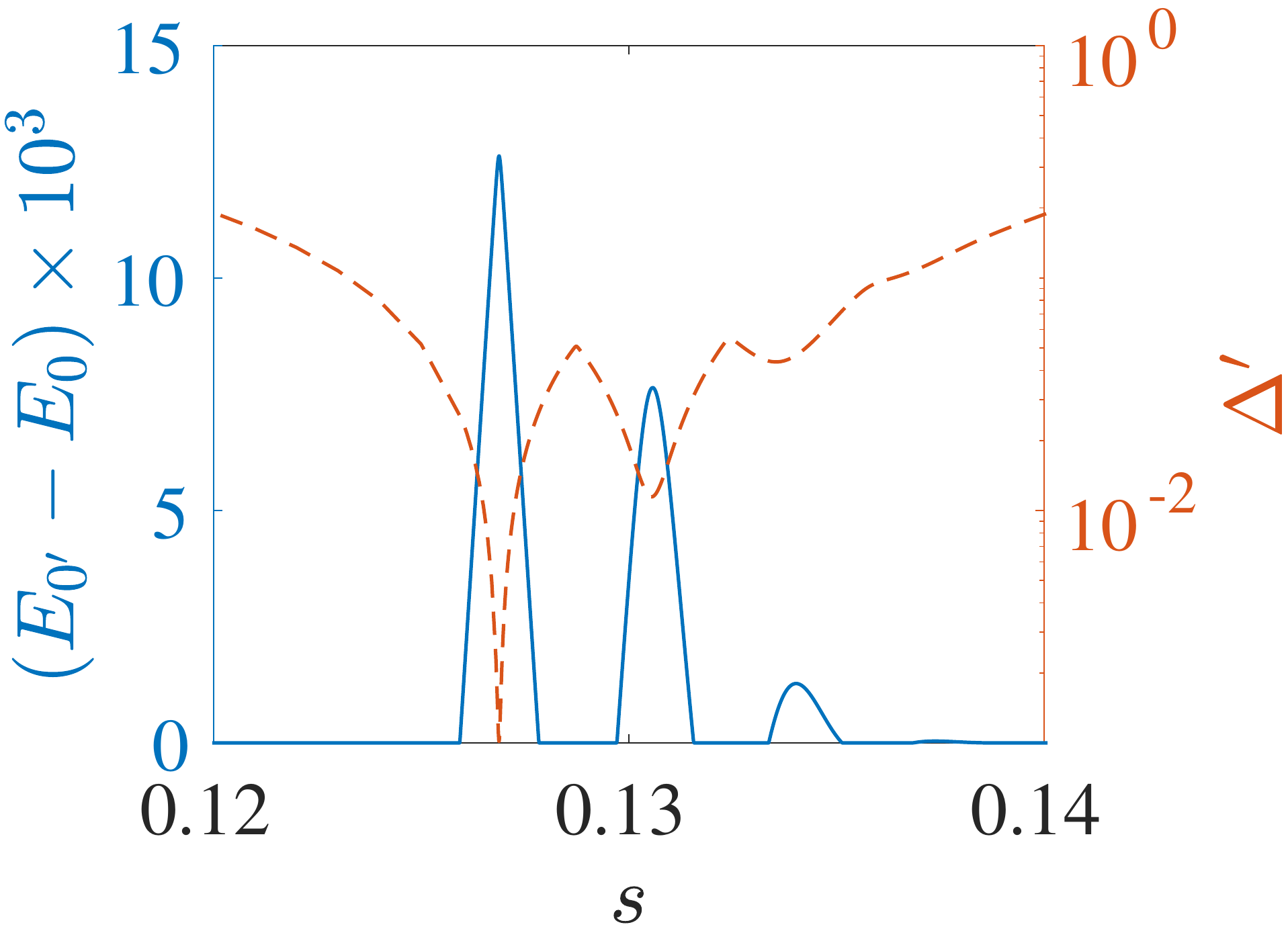} }
  %  \subfigure[]{\includegraphics[width=0.48\columnwidth]{MeanField_Z2SymmetricSubspaceGSWfn_n=128_p=6_lambda=0dot90_NonStoq_LargeFont} }
   \caption{The energy difference $E_{0'} - E_0$ between the ground states of the subspaces $\mathcal{S'}$ and $\mathcal{S}$  (solid line) and the ground state energy gap $\Delta'$ within the subspace $\mathcal{S}'$ (dashed line) for the infinite-range ferromagnetic $(p = 6)$-spin model with $n=128$ and (a) $\lambda =2.425$ and (b) $\lambda = 4$.}
   \label{fig:MF_GSS3}
\end{figure}

%%%%%%%%%%%%%%%%%%%%%%%%%%%%%
%
In the main text, we emphasized that due to the additional symmetry associated with $p$ even, the evolution subspaces $\mathcal{S}'$ and $\mathcal{S}$ associated with the two cases of $p$ even and $p$ odd are different. For $p$ even, ground state wavefunctions with an odd number of nodes are not present within the subspace $\mathcal{S}'$, but we do find that this does not necessarily correspond to the global ground state during the entire interpolation. As shown in Fig.~\ref{fig:MF_GSS3}, the ground state of $H_\lambda$ within the subspace $\mathcal{S}'$ deviates from that of the subspace $\mathcal{S}$ as a function of $s$; the number of times this occurs follows the number of local minima in the gap.  These deviations are associated with energy level crossings within the subspace $\mathcal{S}$ subspace, whereby a $P=-1$ state becomes lower in energy than the current $P=+1$ ground state. For example, at the first deviation, the ground state within the subspace $\mathcal{S}$ changes from having zero nodes ($P=+1$) to having a single node ($P=-1$).  The deviation vanishes when the ground states of the subspaces $\mathcal{S}$ and $\mathcal{S'}$ merge again when the two-node solution ($P=+1$) becomes energetically favored over the one-node solution. Each subsequent deviation in the ground state of $H_\lambda$ within the two subspaces occurs when the addition of a single node results in an odd number of nodes in the ground state wavefunction.

\begin{figure}[htbp]
   \centering
    \subfigure[]{\includegraphics[width=0.48\columnwidth]{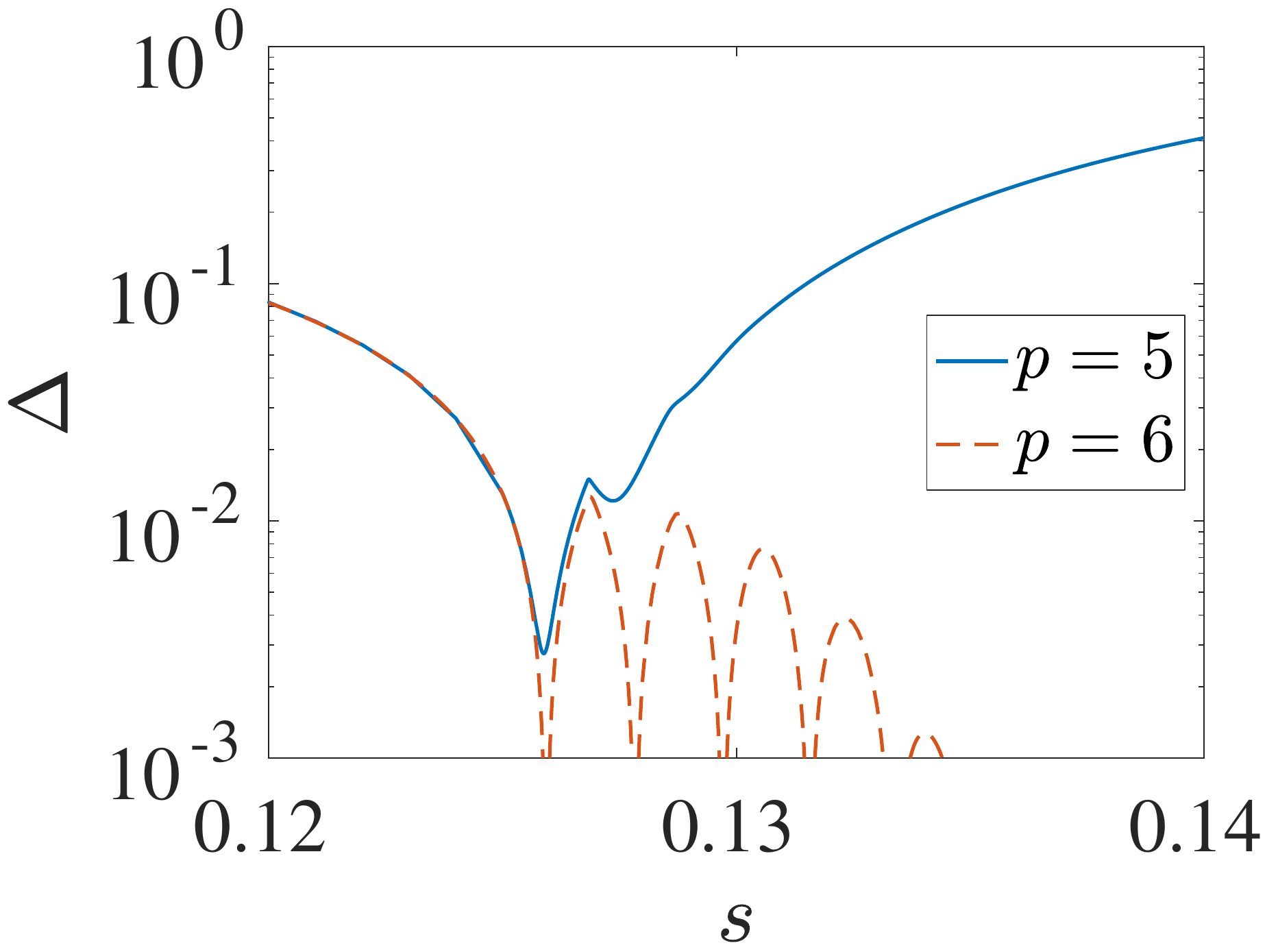} \label{fig:MF_GSS4A}}
    \subfigure[]{\includegraphics[width=0.48\columnwidth]{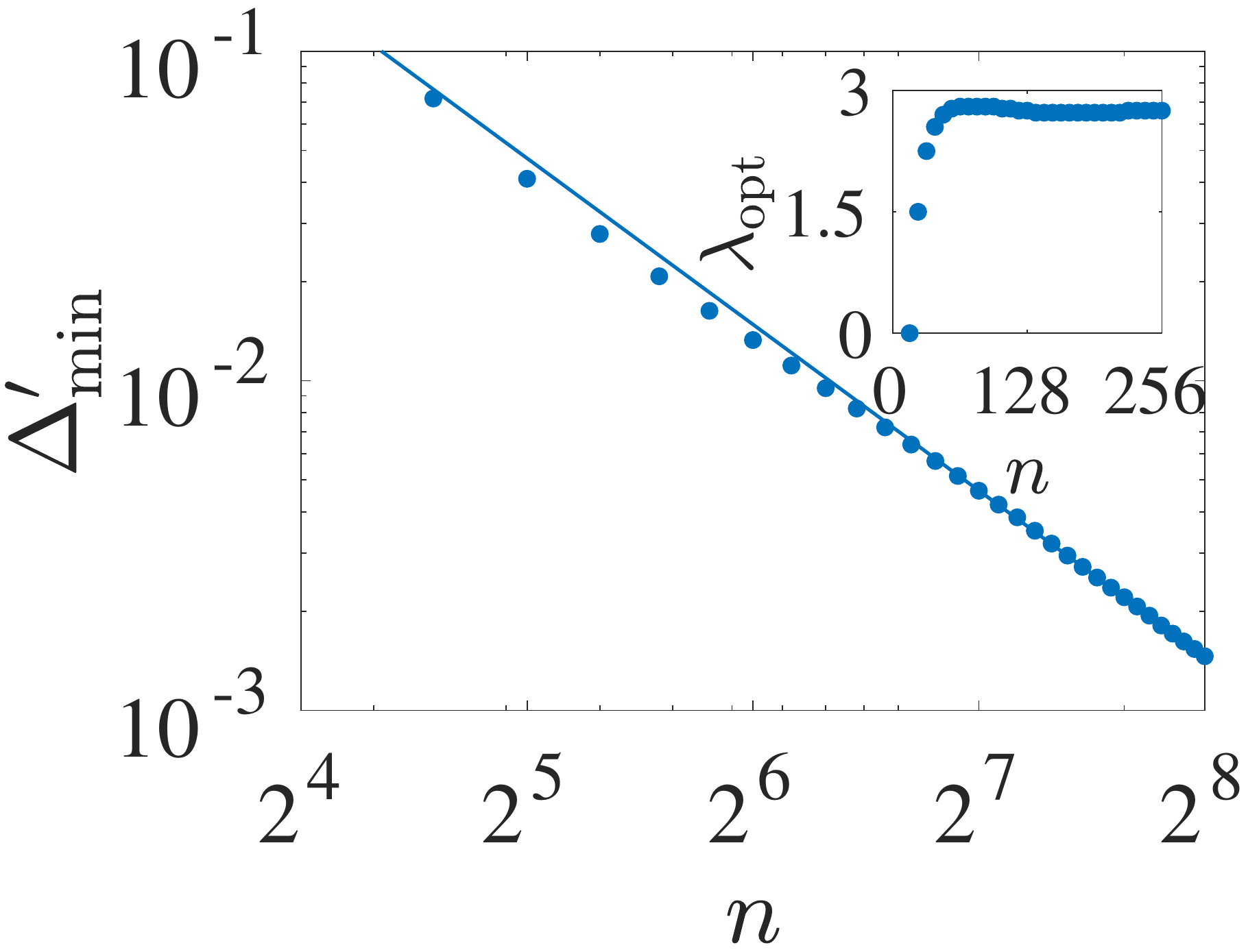} \label{fig:MF_GSS4B}}
   \caption{(a) Comparison between the ground state gap $\Delta$ within the subspace $\mathcal{S}$ for the infinite-range ferromagnetic $p$-spin model with $p=5$ and $p=6$ for $\lambda = 4, n=128$. (b) Scaling of the minimum gap for the infinite-range ferromagnetic $(p=5)$-spin model. Solid line corresponds to the fit of $\sim n^{-1.68}$.  The inset shows the optimized $\lambda$ values.}
   \label{fig:MF_GSS4}
\end{figure}

The true energy-level crossings that occur in the $\mathcal{S}$ subspace for $p$ even are replaced by avoided level crossings (Fig.~\ref{fig:MF_GSS4A}), and the resulting multiple local minima in the gap are each associated with an addition of a single node to an even-node ground state wavefunction.  Because these increments are smaller than in the $p$ even case, we find that the polynomial scaling of the minimum gap is now even milder, as shown in Fig.~\ref{fig:MF_GS3B}.
%
%
%
%%%%%%%%%%%%%%%%%%%%%%%%%%%%%%
%\section{Another example}
%%%%%%%%%%%%%%%%%%%%%%%%%%%%%%
%%
%We consider the same Hamiltonian of Eq.~\eqref{eqt:H2local} in the main text, but now with the Ising Hamiltonian represented in Fig.~\ref{fig:RingFigure2}.  The 
%%
%\begin{figure}[htbp] %  figure placement: here, top, bottom, or page
%   \centering
%   \includegraphics[width=2in]{EvenGadget7_cropped} 
%   \caption{Ising Hamiltonians of size $n=8$ for the Hamiltonian in Eq.~\eqref{eqt:H2local}.  Spins are given by the disks, and the lines between them correspond to Ising spin-spin interactions.  Solid black lines correspond to a ferromagnetic coupling of magnitude $1$, dashed black lines to a ferromagnetic coupling of magnitude $1/2$.  A red disk corresponds to a spin with a local field of $3/4$, energetically favoring the spin-down state, and the blue disk corresponds to spin with a local field of $-1$, energetically favoring the spin-up state.  The Ising Hamiltonian scales by introducing spins in the upper and lower parts of the rings at their centers in an alternating manner.} %  The minimum in the gap occurs at $s \approx 0.36$ at n=128.}
%   \label{fig:RingFigure2}
%\end{figure}
%%%%%%%%%%%%%%%%%%%%%%%%%%%%%
\section{Comparing exponential and polynomial fits for the geometrically local Ising example} \label{App:fits}
\begin{figure}[htbp] %  figure placement: here, top, bottom, or page
   \centering
   \includegraphics[width=0.48\columnwidth]{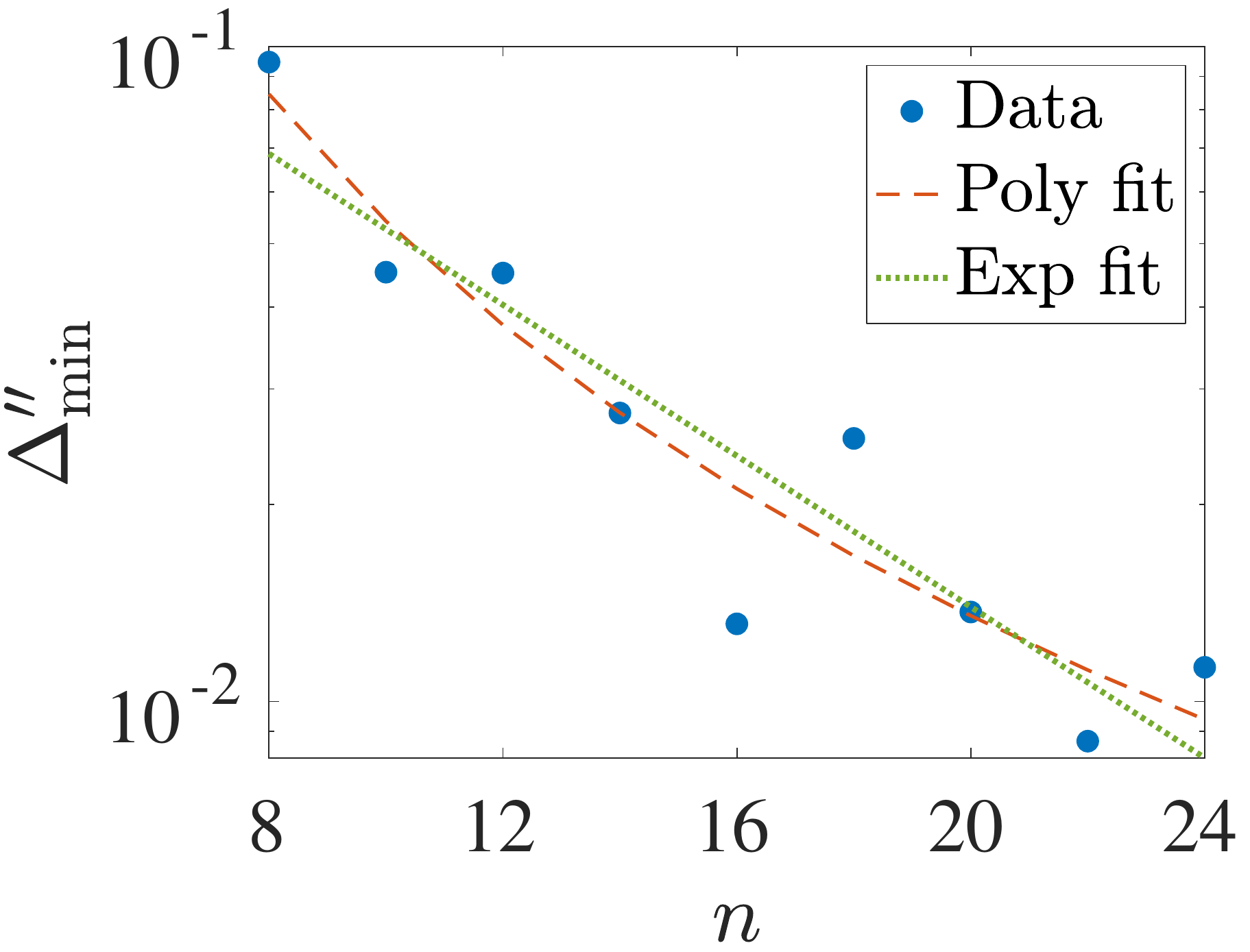} 
   \caption{The minimum gap $\Delta_\mathrm{min}''$ within the subspace $\mathcal{S}''$ for the geometrically local Ising example with a non-stoquastic catalyst with optimized $\lambda$ ($\lambda = \lambda_{\mathrm{opt}}$).  The dashed line corresponds to a polynomial fit of $\sim n^{-2.001}$, and the dotted line corresponds to an exponential fit of $\sim \exp(-0.133n)$.} %  The minimum in the gap occurs at $s \approx 0.36$ at n=128.}
   \label{fig:polyExp}
\end{figure}

%%%%%%%%%%%%%%%%%%%%%%%%%%%%%
%
We show in Fig.~\ref{fig:polyExp} the exponential and polynomial fits to the minimum gaps of the geometrically local Ising example in Section~\ref{Sec:LocalIsing}.  Because we are restricted to small sizes, both fits reasonably capture the data.
%
%%%%%%%%%%%%%%%%%%%%%%%%%%%%%
\section{Different interpolations for the infinite-range ferromagnetic $p$-spin model} \label{App:DifferentInterpolation}
\begin{figure}[htbp] %  figure placement: here, top, bottom, or page
   \centering
  \subfigure[]{ \includegraphics[width=0.46\columnwidth]{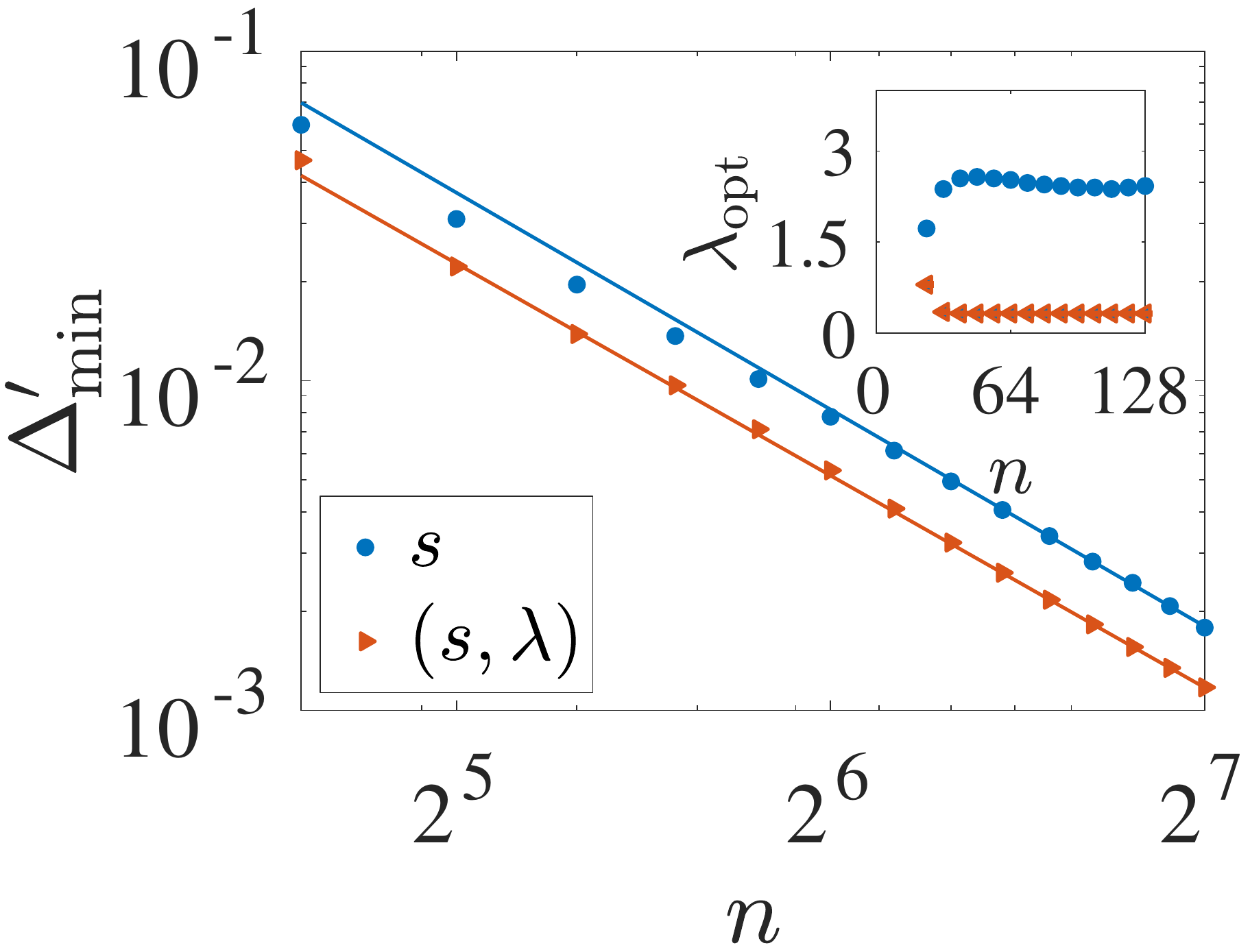}   \label{fig:Schedule1}}
%  \subfigure[]{ \includegraphics[width=0.46\columnwidth]{Figure09b}  \label{fig:Local_NoisyExample}}
   \caption{Scaling of the minimum gap within the subspace $\mathcal{S}''$ with system size using the interpolating Hamiltonians in Eq.~\eqref{eqt:pSpinH} (denoted `$s$.') and in Eq.~\eqref{eqt:pSpinH2} (denoted `$(s,\lambda)$') for the infinite-range $(p=6)$-spin model.  The solid lines correspond to best fits of $\sim n^{-2.19}$ and $\sim n^{-2.14}$ respectively.  The inset shows the optimized $\lambda$ values for both schedules.}
\label{fig:schedues}
\end{figure}

%%%%%%%%%%%%%%%%%%%%%%%%%%%%%
%
In Eq.~\eqref{eqt:pSpinH} of the main text, we used the conventional interpolation schedule for the catalyst Hamiltonian \cite{crosson2014different}.  This was not the interpolation used in Ref.~\cite{Seki:2012}, which had an interpolating Hamiltonian of the form:
\begin{eqnarray} \label{eqt:pSpinH2}
H_\alpha(s,\lambda) &=& -(1-s) \sum_{i=1}^n \sigma_i^x - \frac{s \lambda}{n^{p-1}} \left(\sum_{i=1}^n \sigma_i^z \right)^p \nonumber \\
&& + \alpha \frac{s (1-\lambda)}{n} \left( \sum_{i=1}^n \sigma_i^x \right)^2 \ ,
\end{eqnarray}
whee $\alpha = 0,-1$ for stoquastic catalysts and $\alpha = 1$ for non-stoquastic catalysts.  Unlike the choice to keep $\lambda$ fixed during the interpolation with $s$ in Eq.~\eqref{eqt:pSpinH}, both $\lambda$ and $s$ can vary during the interpolation with this choice.  For simplicity we consider an interpolating path $H_1(0,0) \rightarrow H_1(0, \lambda_\ast) \rightarrow H_1(1,\lambda_\ast) \rightarrow H_1(1,1)$, where we optimize the value of $\lambda_\ast$ to maximize the minimum gap crossed, and compare to the results using the schedule in the main text.  We show in Fig.~\ref{fig:schedues} that for the case of $p=6$ the large $n$ scaling is essentially identical, with the only significant difference being an overall constant shift in the minimum gap encountered.
%
%%%%%%%%%%%%%%%%%%%%%%%%%%%%%
\section{Results for the $(p=3)$-spin model} \label{App:p=3}
\begin{figure}[htbp] %  figure placement: here, top, bottom, or page
   \centering
   \subfigure[]{\includegraphics[width=0.48\columnwidth]{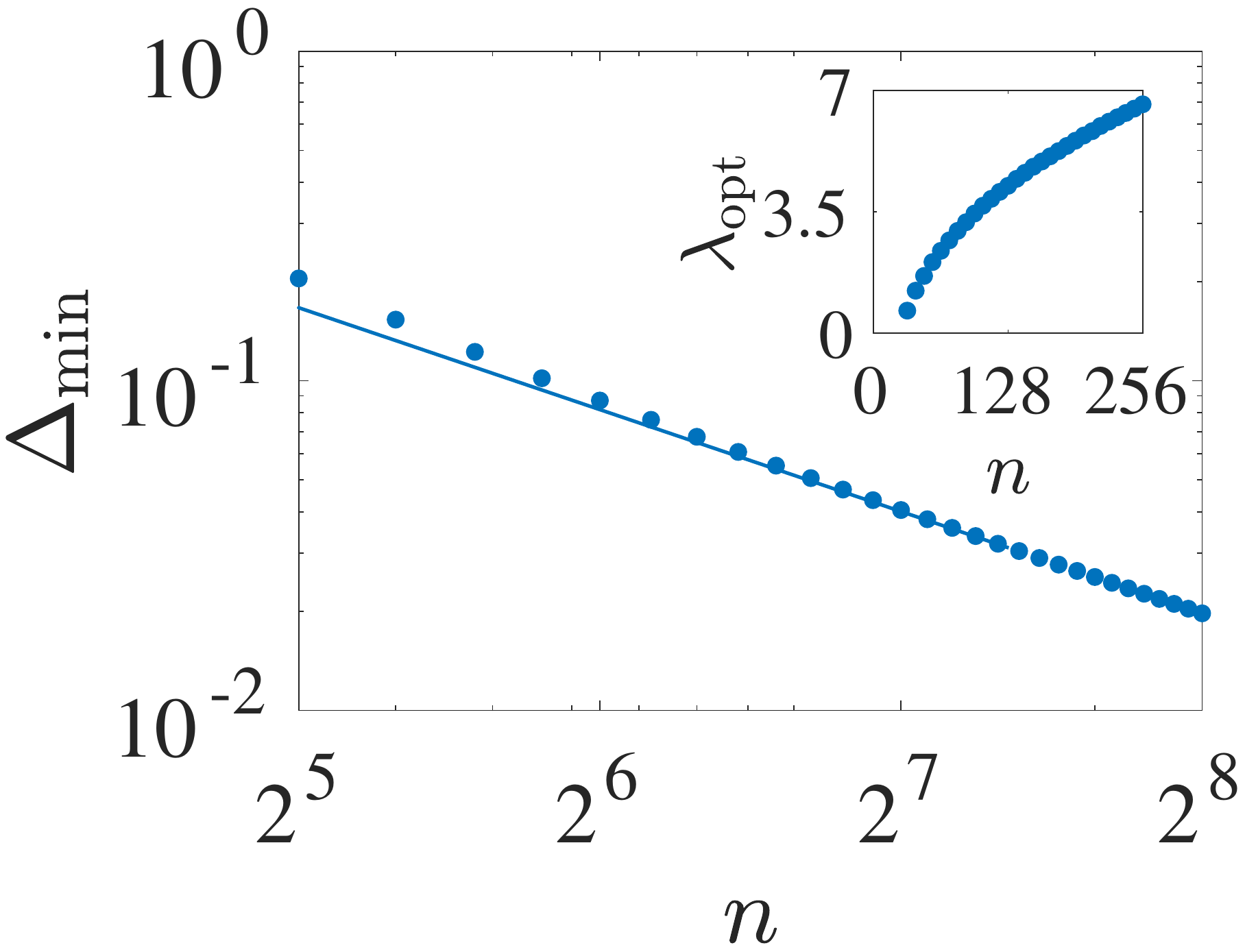} \label{fig:p=3a}}
   \subfigure[]{\includegraphics[width=0.48\columnwidth]{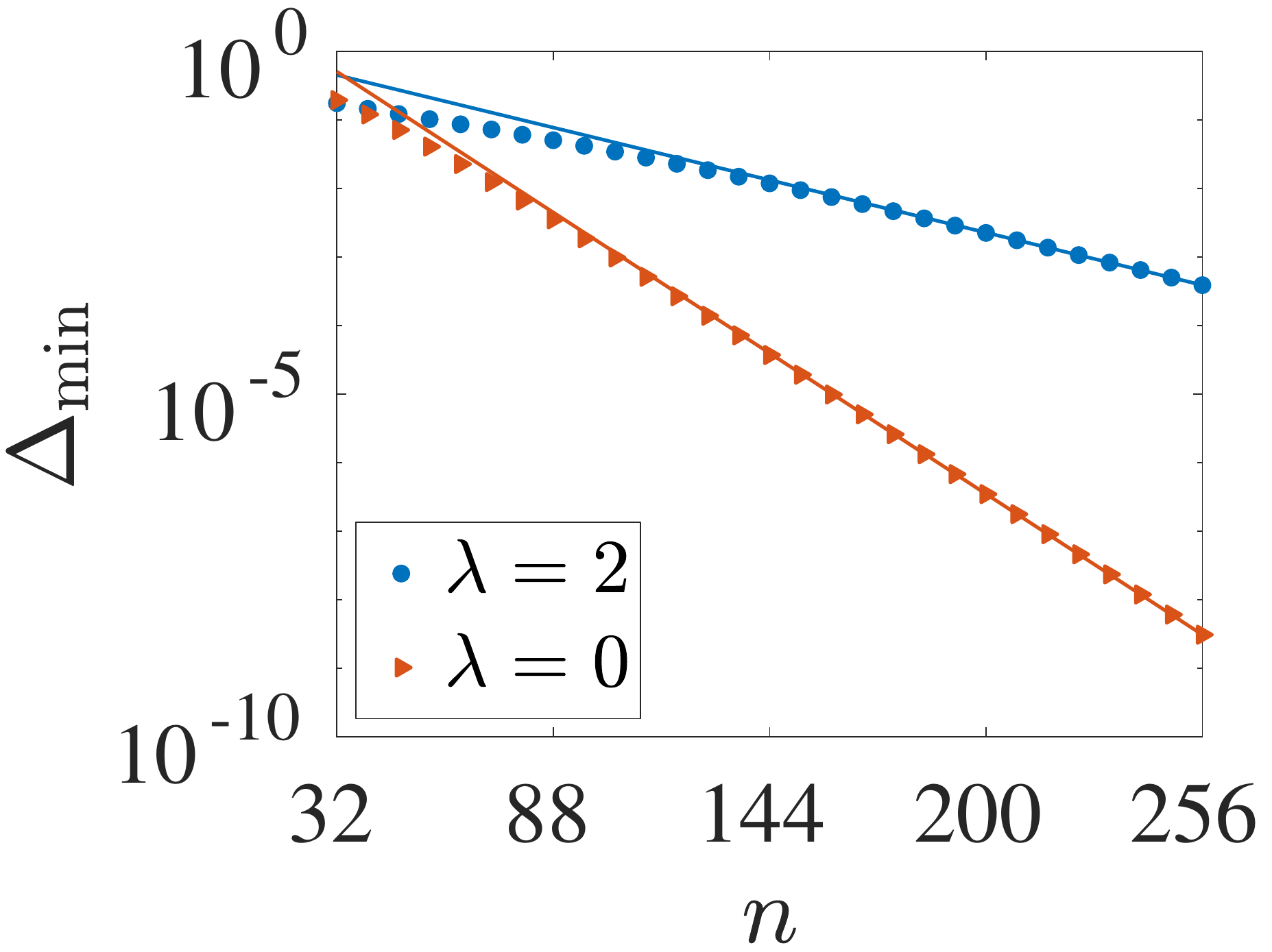} \label{fig:p=3b}}
   \caption{(a) The minimum gap $\Delta_\mathrm{min}$ within the subspace $\mathcal{S}$ for the infinite-range $(p=3)$-spin model with a non-stoquastic catalyst and optimized $\lambda$ ($\lambda = \lambda_{\mathrm{opt}}$).  The dashed line corresponds to a polynomial fit of $\sim n^{-1.03}$.  The inset shows the optimized values for $\lambda$ used.  The error bars, which are not visible because they are on the size of the data points, correspond to our uncertainty in the exact optimum value of $\lambda$ (b) Comparison of the minimum gap for $p=3$ with a non-stoquastic catalyst with fixed $\lambda = 2$ and no catalyst ($\lambda = 0$).} 
   \label{fig:p=3}
\end{figure}

%%%%%%%%%%%%%%%%%%%%%%%%%%%%%
%
We show in Fig.~\ref{fig:p=3a} results for the minimum gap for the infinite-range ferromagnetic $(p=3$)-spin model.  Our results are consistent with the conclusions of Ref.~\cite{2018arXiv180607602D}, whereby the exponential advantage of the non-stoquastic catalyst over the stoquastic catalyst is only maintained if the relative strength of the non-stoquastic catalyst to the problem Hamiltonian grows with system size.  For a fixed non-stoquastic catalyst strength, the scaling returns to an exponential scaling but it is milder than the stoquastic case, as shown in Fig.~\ref{fig:p=3b}.  Therefore, while the exponential advantage is not maintained in this case, there is still an advantage that can be had with a non-stoquastic catalyst.
%
%%%%%%%%%%%%%%%%%%%%%%%%%%%%%%
\section{Infinite-range 2-local large-spin tunneling example with a different catalyst} \label{App:DifferentCatalyst}
\begin{figure}[htbp] %  figure placement: here, top, bottom, or page
   \centering
 \includegraphics[width=0.46\columnwidth]{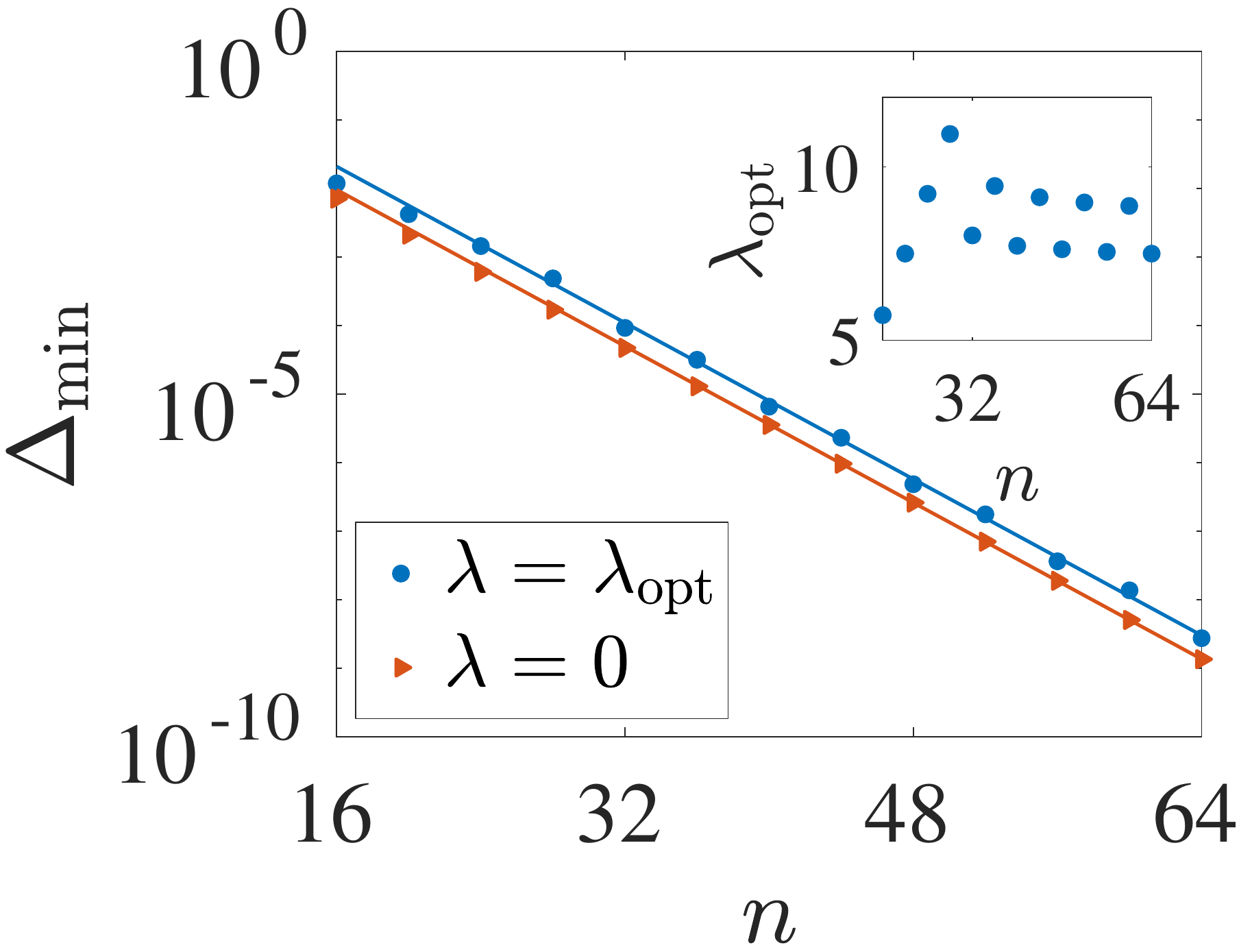}   
 %\label{fig:Schedule1}
%  \subfigure[]{ \includegraphics[width=0.46\columnwidth]{Figure09b}  \label{fig:Local_NoisyExample}}
   \caption{Scaling of the minimum gap within the subspace $\mathcal{S}$ for the infinite-range 2-local large-spin tunneling example with a different non-stoquastic catalyst ($\lambda = \lambda_{\mathrm{opt}}$) and without a catalyst ($\lambda = 0$).  The solid lines corresponds to a best fit of $\sim \exp(0.33)$.  The inset shows the optimized $\lambda$ values for the case with a catalyst.}
\label{fig:NewCatalyst}
\end{figure}

%%%%%%%%%%%%%%%%%%%%%%%%%%%%%%
%
We consider a different catalyst Hamiltonian for the 2-local large spin tunneling example in the main text.  Instead of Eq.~\eqref{eqt:H2}, we take the interpolating Hamiltonian to be given by
\begin{eqnarray} \label{eqt:H4}
H_\lambda(s) &=& -2 (1-s) \left(S_1^x + S_2^x \right)  - s \left(  2 h_1 S_1^z - 2 h_2 S_2^z  \right. \nonumber \\
&& \left. + \frac{4}{n} \left((S_1^z)^2 +(S_2^z)^2 + S_1^z S_2^z\right) \right) \nonumber \\
&& + \frac{4\lambda  s (1-s)}{n} ( S_1^x +S_2^x )^2 
\end{eqnarray}
where we have changed the catalyst Hamiltonian from $2 S_1^x S_2^x$ to $(S_1^x + S_2^x)^2$.  We find that this changes the scaling behavior of the minimum gap, as we show in Fig.~\ref{fig:NewCatalyst}.  We find that even with an optimized $\lambda$ (we find that the cases of $n/4$ even and odd give different asymptotic values for $\lambda_{\mathrm{opt}}$), the non-stoquastic catalyst does not exhibit an exponential improvement over the stoquastic case, as we saw in Fig.~\ref{fig:MF_GS4} of the main text.  Instead, the scaling of the non-stoquastic catalyst in this case is indistinguishable from that of the case with no catalyst.
\section{Implementation errors}
%%%%%%%%%%%%%%%%%%%%%%%%%%%%%
%
\begin{figure*}[htbp] %  figure placement: here, top, bottom, or page
   \centering
  \subfigure[]{ \includegraphics[width=0.46\columnwidth]{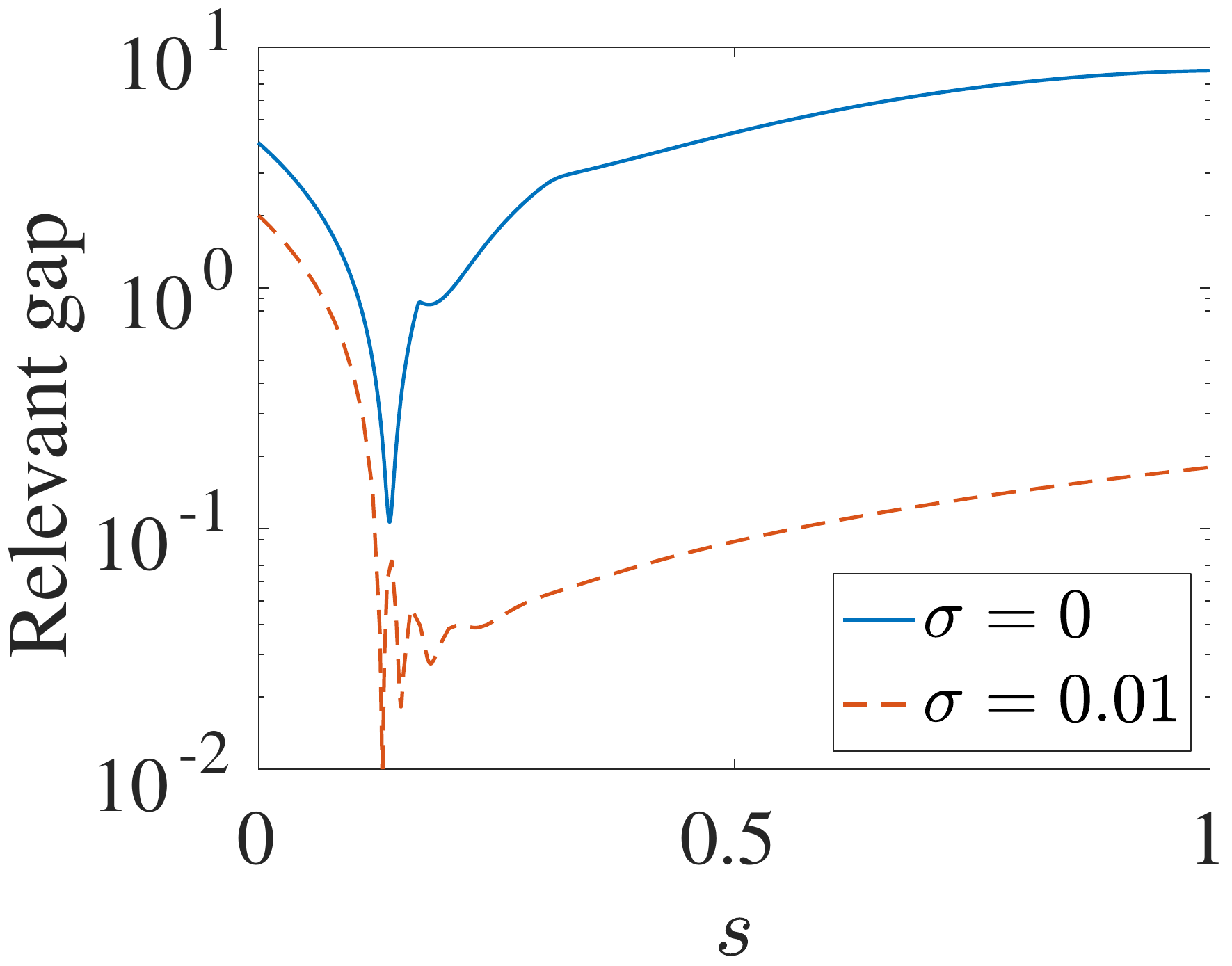}   \label{fig:MF_NoisyExample}}
    \subfigure[]{ \includegraphics[width=0.46\columnwidth]{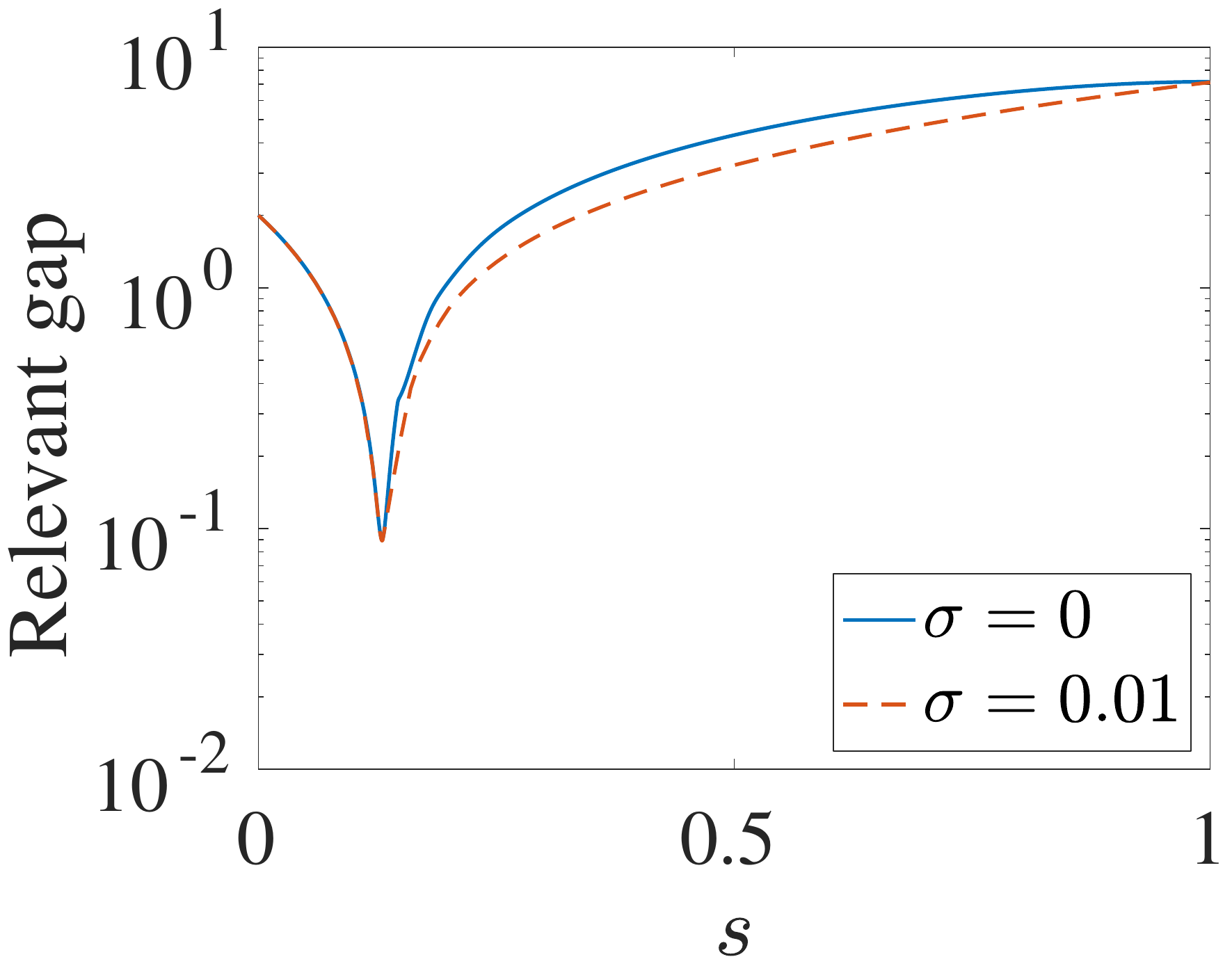}   \label{fig:MF2_NoisyExample}}
  \subfigure[]{ \includegraphics[width=0.46\columnwidth]{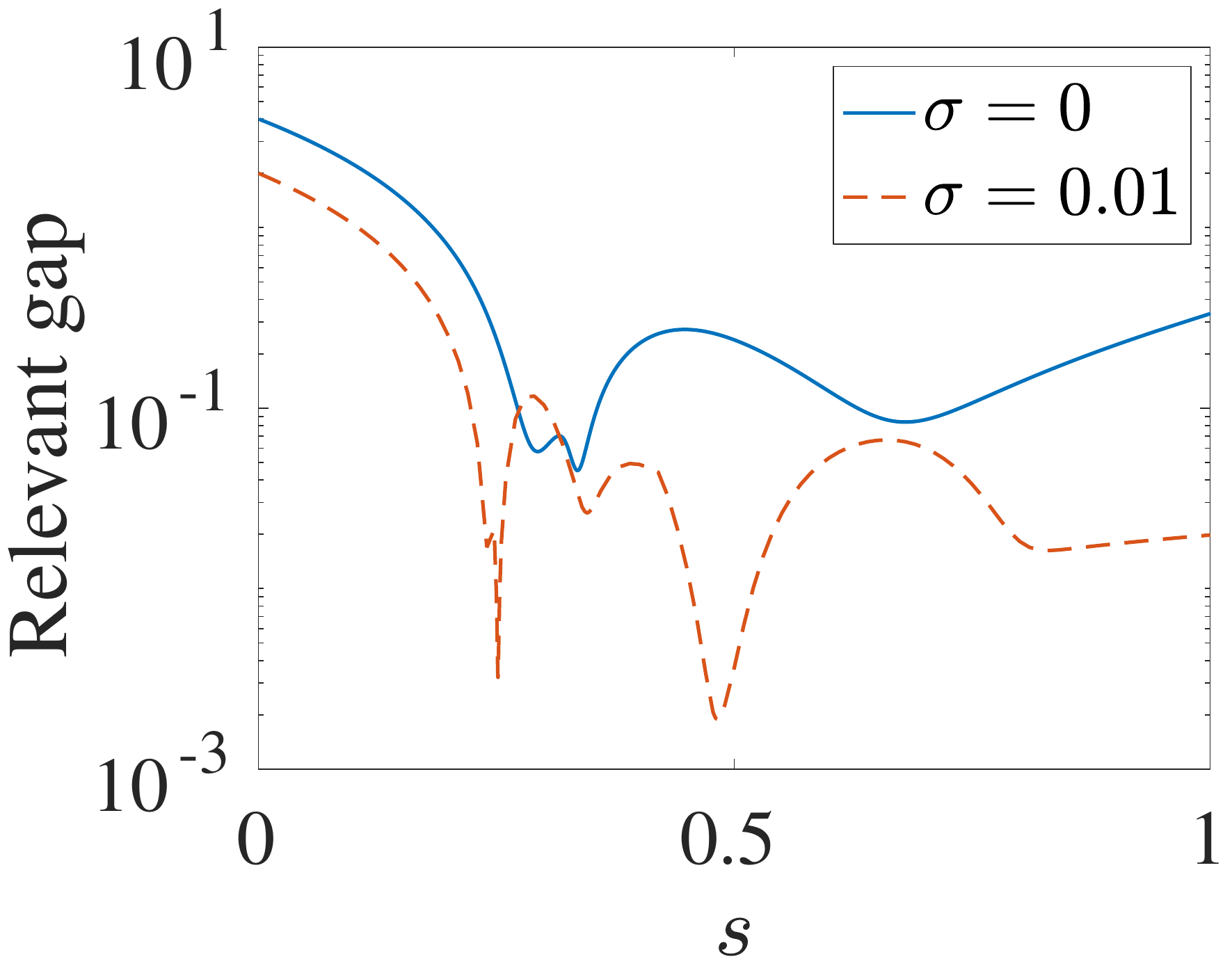}  \label{fig:Local_NoisyExample}}\\
       \subfigure[]{ \includegraphics[width=0.46\columnwidth]{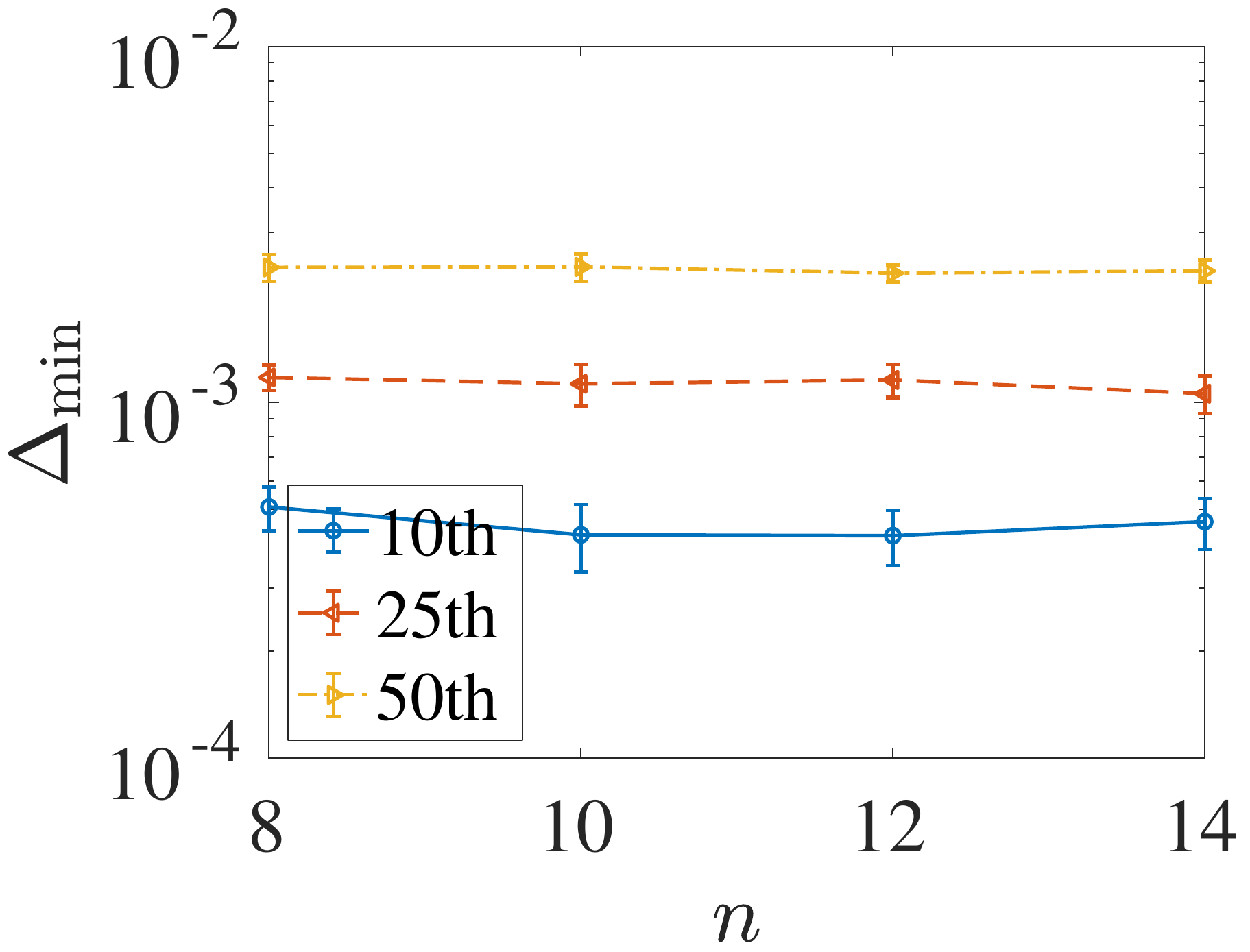}   \label{fig:MF_NoisyExample2}}
          \subfigure[]{ \includegraphics[width=0.46\columnwidth]{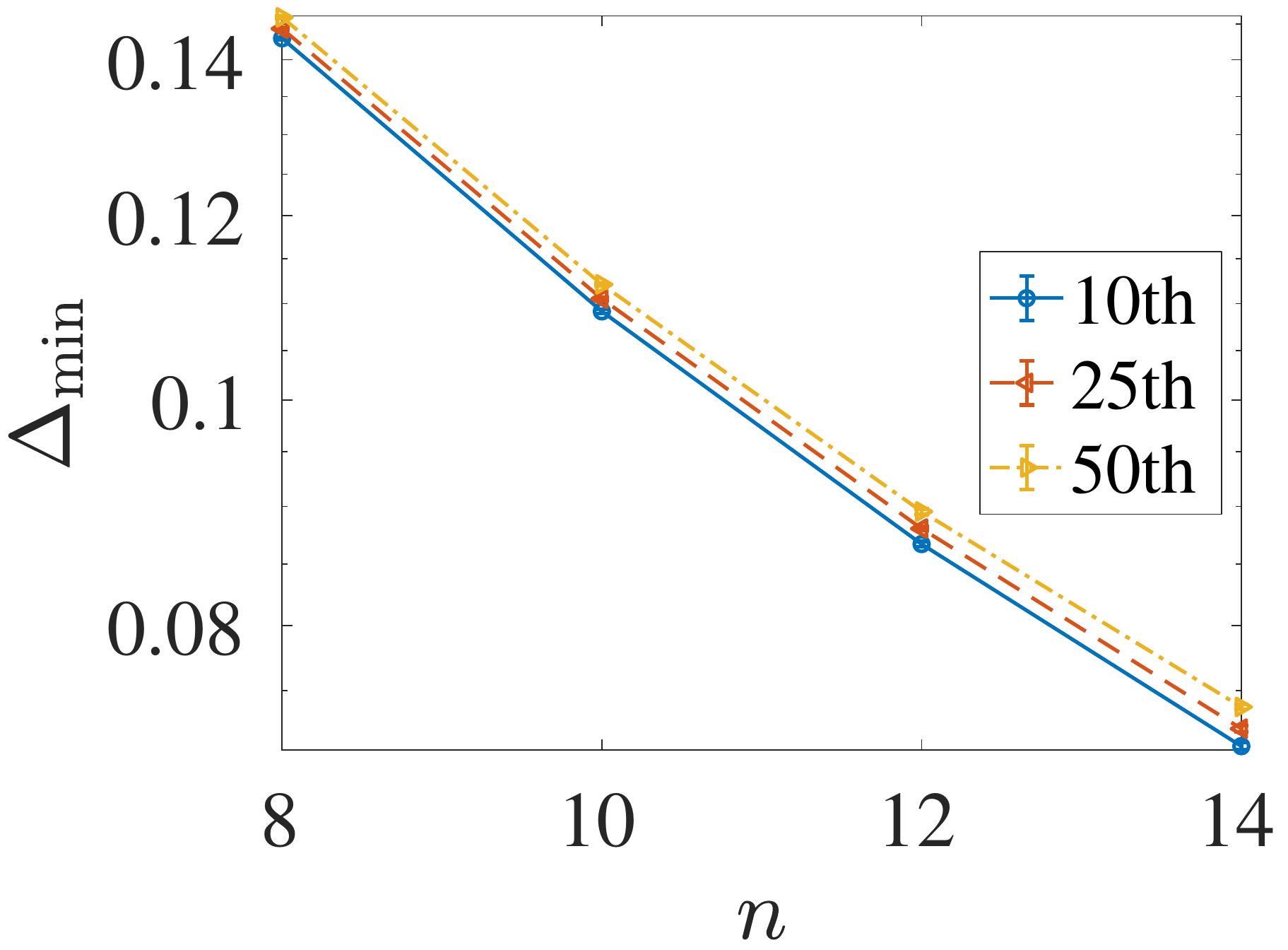}   \label{fig:MF2_NoisyExample2}}
    \subfigure[]{ \includegraphics[width=0.46\columnwidth]{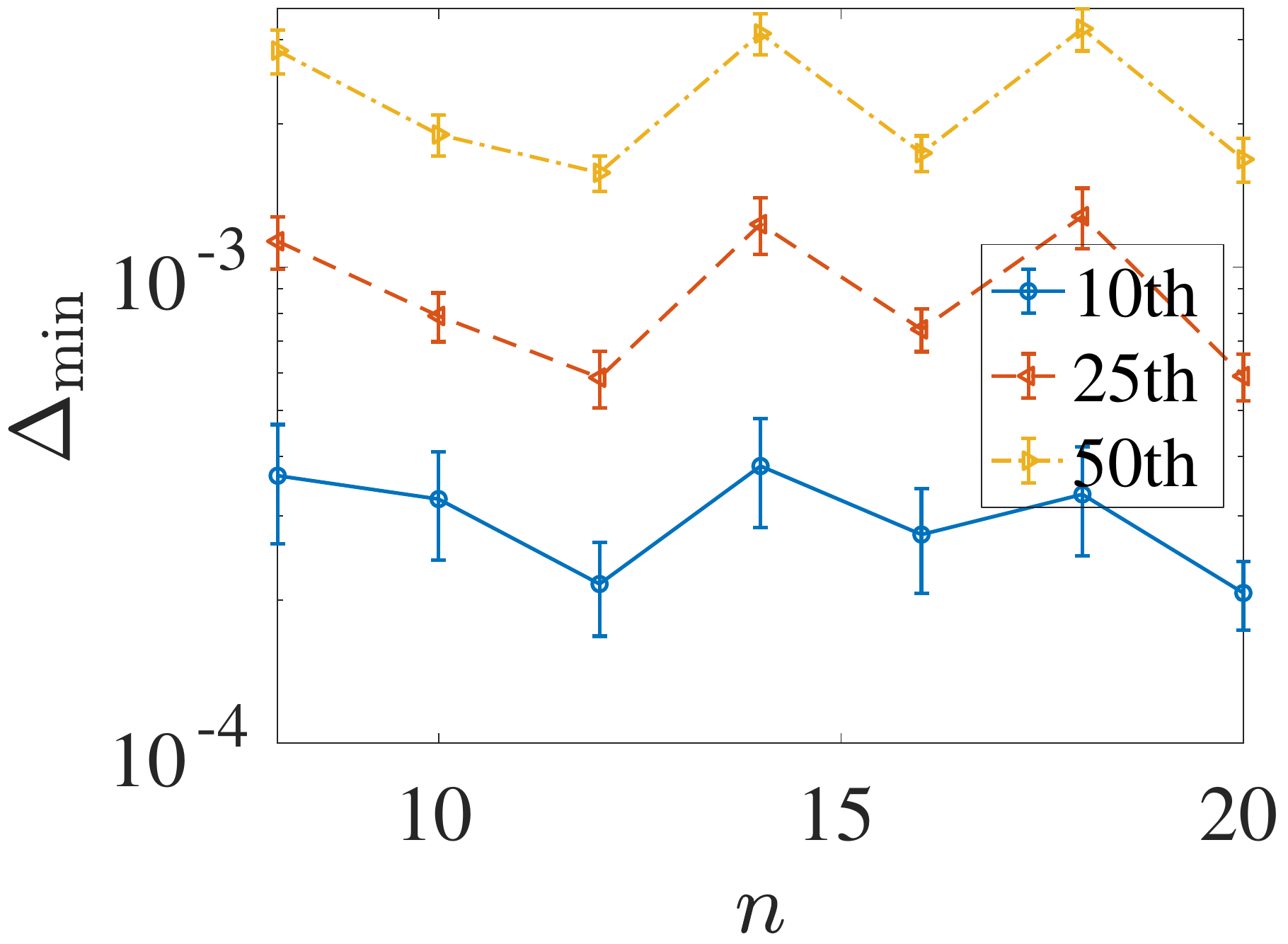}  \label{fig:Local_NoiseScaling2}}
   \caption{(a-c) Relevant ground state energy gap $\Delta$ for the noiseless case ($\sigma = 0$) and a noisy realization ($\sigma = 10^{-2}$) for (a) the infinite-range $(p=6)$-spin model with $n=12$ and $\lambda=4$, (b) the infinite-range $(p=5)$-spin model with $n=12$ and $\lambda=4$ and (c) the geometrically local Ising example with $n=12$ and $\lambda = 1.15$. (d-f) Percentiles of the minimum gap for $10^3$ noisy realizations ($\sigma = 10^{-2}$) at different sizes for (d) the infinite-range $(p=6)$-spin model with $\lambda=4$, (e) the infinite-range $(p=5)$-spin model with $\lambda=4$  and (f) the geometrically local Ising example with the optimized $\lambda$ values for the noiseless case.  The error bars represent 95\% confidence intervals ($2\sigma$) calculated using $10^3$ bootstraps of the noisy realizations. }
\label{fig:NoiseExamples}
\end{figure*}
%
%\begin{figure}[hb] %  figure placement: here, top, bottom, or page
%   \centering
%%
%       \caption{Percentiles of the minimum gap for $10^3$ noisy realizations ($\sigma = 10^{-2}$) at different sizes for (a) the infinite-range $p=6$ spin model with $\lambda=4$, (b) the infinite-range $p=5$ spin model with $\lambda=4$  and (c) the geometrically local example with the optimized $\lambda$ values for the noiseless case.  The error bars represent 95\% confidence intervals ($2\sigma$) calculated using $10^3$ bootstraps of the noisy realizations.}
%\label{fig:NoiseScaling}
%\end{figure}

In order to address how dependent the non-stoquastic advantage is on the presence of symmetries in the Hamiltonian, we consider introducing noise to the Hamiltonian defining the optimization problem.  For example, we replace the Hamiltonian of the  infinite-range $p$-spin model with
\begin{eqnarray}
\frac{1}{n^{p-1}}\left( \sum_i \sigma_i^z \right)^p  &\rightarrow& \frac{1}{n^{p-1}} \left( \sum_i \sigma_i^z \right)^p  + \sum_i \delta h_i \sigma_i^z \ ,
%\nonumber \\
%&& + \sum_{\langle i_1, \dots, i_p \rangle}  \delta J_{i_1, \dots i_p}  \sigma_{i_1}^z \dots \sigma_{i_p}^z
\end{eqnarray}
where 
%the sum is over the allowed unique combination of Pauli operators (including the identity).  We take $ \delta J_{i_1, \dots i_p}, 
$\delta h_i \sim  \mathcal{N}(0, \sigma^2)$.  Under this noise model, the time-dependent Hamiltonian (Eq.~\eqref{eqt:pSpinH} in the main text) is no longer invariant under permuting the qubits nor under the operator $P$ for $p$ even, so the evolution is not restricted to any obvious subspace.  We compare in Fig.~\ref{fig:MF_NoisyExample} and Fig.~\ref{fig:MF2_NoisyExample} the original (noiseless) relevant gap and the gap for one realization of the noise.  Significantly smaller minimum gaps now appear in the spectrum for the $p=6$  case but not for the $p=5$ case.  The smaller gaps result from true level-crossings in the spectrum becoming avoided level-crossings.

We can consider a similar noise model for our geometrically local example, and we show an example of the gap for one noise realization in Fig.~\ref{fig:Local_NoisyExample}.

It is difficult to ascertain how the minimum gap in the noisy models scales with problem size $n$.  As we show in Figs.~\ref{fig:MF_NoisyExample2}-\ref{fig:Local_NoiseScaling2}, while the introduction of random noise results in a drop in magnitude in the minimum gap for the cases where the Hamiltonian is invariant under $P$, no obvious scaling is seen for the small system sizes we study.

\end{document}